\documentclass[12pt]{iopart}
\usepackage{graphicx}
\usepackage{amssymb}
\usepackage[
    range-units=single,
]{siunitx}
\usepackage{lineno}
\usepackage{xspace}
\usepackage{subcaption}
\usepackage{latexsym}
\usepackage{amsmath}
\usepackage{amsfonts}
\usepackage{mathtools}
\usepackage[
    usenames,
    dvipsnames,
    x11names,
]{xcolor}
\usepackage[
    colorlinks=True,
    linkcolor=NavyBlue,
    citecolor=NavyBlue,
    urlcolor=PaleGreen4,
]{hyperref}
\usepackage[capitalise]{cleveref}
\crefname{figure}{figure}{figures}
\Crefname{figure}{Figure}{Figures}
\crefname{table}{table}{tables}
\Crefname{table}{Table}{Tables}
\crefname{section}{section}{sections}
\Crefname{section}{Section}{Sections}
\crefname{subsection}{section}{sections}
\Crefname{subsection}{Section}{Sections}
\crefname{appendix}{appendix}{appendices}
\Crefname{appendix}{Appendix}{Appendices}
\usepackage{ifthen}
\usepackage{array}
\usepackage{calc}
\usepackage{vmargin}
\usepackage{colortbl}
\usepackage{enumerate} % Makes it so you can format enumerate environments as you wish.
\usepackage{multirow}
\usepackage{cite}
\usepackage{lineno}
\usepackage{placeins}
\usepackage{orcidlink}
\makeatletter
\newrobustcmd{\fixappendix}{%
	\patchcmd{\l@section}{1.5em}{7em}{}{}%
	\patchcmd{\l@subsection}{2.3em}{7em}{}{}%
}
\makeatother

\makeatletter
\def\@fnsymbol#1{\ifcase#1\or
	\dagger\or
	\ddagger\or
	\S\or
	\|\or
	\P\or
	^{+}\or
	^{\tsty *}\or
	\sharp\or
	\dagger\dagger\or
	\ddagger\ddagger\or
	\S\S
	\else\@ctrerr
	\fi\relax}
\makeatother

\graphicspath{{../figures/}{./figures/}}

\DeclareSIUnit{\ppm}{ppm}
\DeclareSIUnit{\partspermillion}{ppm}
\DeclareSIUnit{\ppb}{ppb}
\DeclareSIUnit{\partsperbillion}{ppb}
\DeclareSIUnit{\pc}{pc}
\DeclareSIUnit{\parsec}{pc}
\DeclareSIUnit{\rtHz}{\big/\sqrt{Hz}}
\DeclareSIUnit{\torr}{torr}
\DeclareSIUnit{\year}{yr}
\DeclareSIUnit{\yr}{yr}
\DeclareSIUnit{\solarmass}{\text{\ensuremath{M_\odot}}}
\DeclareSIUnit{\deg}{deg}

\newread\paramfile
\openin\paramfile=data/parameters.txt
\def\DefVal#1: #2\relax{\expandafter\gdef\csname val-#1\endcsname{#2}}
{\endlinechar=-1
\loop
\ifeof\paramfile\else
\read\paramfile to \tmp
\ifx\tmp\empty\else
\expandafter\DefVal\tmp\relax
\fi
\repeat
}
\closein\paramfile
\providecommand{\Val}[1]{\csname val-#1\endcsname}

\usepackage{xparse}
\newlength{\nbvspace}
\newlength{\nbheight}
\newcommand{\Asharp}{A\textsuperscript{$\sharp$}}

\newcommand{\ValInt}[1]{\fpeval{round(\Val{#1}, 0)}}

\newcommand{\citealog}[2]
{\ifthenelse{\equal{#1}{lho}}{LHO aLOG \href{https://alog.ligo-wa.caltech.edu/aLOG/index.php?callRep=#2}{#2}}{}%
	\ifthenelse{\equal{#1}{llo}}{LLO aLOG \href{https://alog.ligo-la.caltech.edu/aLOG/index.php?callRep=#2}{#2}}{}%
}

\newif\ifshowcomments
\showcommentstrue % show comments
\ifshowcomments
\newcommand{\comments}[1]{\textcolor{blue}{[#1]}}
\else
\newcommand{\comments}[1]{}
\fi

\begin{document}
%\linenumbers

\title{LIGO \Asharp{}: Detector Design and Science Prospects Beyond A+}

\author{%
	% Leading authors (writing/editing/analysis/discussion)
	L~Sun\,\orcidlink{0000-0001-7959-892X}$^{1}$,
	K~Kuns\,\orcidlink{0000-0003-0630-3902}$^{2}$,
	B~J~J~Slagmolen\,\orcidlink{0000-0002-2471-3828}$^{1}$,
	P~Fritschel$^{2}$,
	P~Schmidt\,\orcidlink{0000-0003-1542-1791}$^{3}$,
	B~T~Lantz\,\orcidlink{0000-0002-7404-4845}$^{4}$,
	S~S~Y~Chua\,\orcidlink{0000-0001-8026-7597}$^{1}$,
	Divyajyoti\,\orcidlink{0000-0002-2787-1012}$^{5}$,
	S~W~Ballmer\,\orcidlink{0000-0003-2306-523X}$^{6}$,
	M~A~Barton\,\orcidlink{0000-0002-9948-306X}$^{7}$,
	A~V~Cumming\,\orcidlink{0000-0003-4096-7542}$^{7}$,
	K~L~Dooley\,\orcidlink{0000-0002-1636-0233}$^{5}$,
	J~C~Driggers\,\orcidlink{0000-0002-6134-7628}$^{8}$,
	A~Effler\,\orcidlink{0000-0001-8242-3944}$^{9}$,
	M~Evans\,\orcidlink{0000-0001-8459-4499}$^{2}$,
	B~Farr\,\orcidlink{0000-0002-2916-9200}$^{10}$,
	G~Gonz\'alez\,\orcidlink{0000-0003-0199-3158}$^{11}$,
	N~Lu\,\orcidlink{0000-0002-8861-9902}$^{1}$,
	D~J~Ottaway\,\orcidlink{0000-0001-6794-1591}$^{12}$,
	C~Palomba\,\orcidlink{0000-0002-4450-9883}$^{13}$,
	O~J~Piccinni\,\orcidlink{0000-0001-5478-3950}$^{14}$,
	G~Pratten\,\orcidlink{0000-0003-4984-0775}$^{3}$,
	S~Raja$^{15}$,
	A~P~Subhash\,\orcidlink{0009-0008-1458-3338}$^{1}$,
	P~J~Sutton\,\orcidlink{0000-0003-1614-3922}$^{5}$,
	K~Toland\,\orcidlink{0000-0001-9537-9698}$^{7}$,
	R~L~Ward\,\orcidlink{0000-0001-5503-5241}$^{1}$,
	% All other authors (opt-in merged with LSC Feb-2026 list, alphabetical)
	A~G~Abac\,\orcidlink{0000-0003-4786-2698}$^{16}$,
	I~Abouelfettouh$^{8}$,
	K~Ackley\,\orcidlink{0000-0002-8648-0767}$^{17}$,
	A~Adam$^{18}$,
	S~Adhicary\,\orcidlink{0009-0004-2101-5428}$^{19}$,
	D~Adhikari$^{20,21}$,
	R~X~Adhikari\,\orcidlink{0000-0002-5731-5076}$^{22}$,
	V~K~Adkins$^{11}$,
	S~Afroz\,\orcidlink{0009-0004-4459-2981}$^{23}$,
	M~Agathos\,\orcidlink{0000-0002-9072-1121}$^{24}$,
	N~Aggarwal$^{25}$,
	S~Aggarwal$^{26}$,
	O~D~Aguiar\,\orcidlink{0000-0002-2139-4390}$^{27}$,
	P~Ajith\,\orcidlink{0000-0001-7519-2439}$^{28}$,
	L~Albers$^{29}$,
	S~Al-Kershi$^{20,21}$,
	S~Al-Shammari$^{5}$,
	J~A~Alvarez$^{30}$
% ...
	A~B~Zimmerman\,\orcidlink{0000-0002-7453-6372}$^{86}$,
	M~E~Zucker\,\orcidlink{0000-0002-2544-1596}$^{2,22}$,
	% External authors
	R~Ar\`es\,\orcidlink{0000-0001-7697-9735}$^{159}$,
	J~J~Carter\,\orcidlink{0000-0001-8845-0900}$^{20,21}$,
	S~J~Madden\,\orcidlink{0000-0003-3146-285X}$^{1}$,
	K~McKenzie\,\orcidlink{0000-0002-1463-4595}$^{1}$,
	E~R~Rees\,\orcidlink{0000-0003-0112-6716}$^{1}$,
	B~Shapiro\,\orcidlink{0000-0002-0216-9534}$^{160}$
}

\address {$^{1}$OzGrav, Australian National University, Canberra, Australian Capital Territory 2601, Australia }
\address {$^{2}$LIGO Laboratory, Massachusetts Institute of Technology, Cambridge, MA 02139, USA }
\address {$^{3}$University of Birmingham, Birmingham B15 2TT, United Kingdom }
\address {$^{4}$Stanford University, Stanford, CA 94305, USA }
\address {$^{5}$Cardiff University, Cardiff CF24 3AA, United Kingdom }
\address {$^{6}$Syracuse University, Syracuse, NY 13244, USA }
\address {$^{7}$IGR, University of Glasgow, Glasgow G12 8QQ, United Kingdom }
\address {$^{8}$LIGO Hanford Observatory, Richland, WA 99352, USA }
\address {$^{9}$LIGO Livingston Observatory, Livingston, LA 70754, USA }
\address {$^{10}$University of Oregon, Eugene, OR 97403, USA }
\address {$^{11}$Louisiana State University, Baton Rouge, LA 70803, USA }
\address {$^{12}$OzGrav, University of Adelaide, Adelaide, South Australia 5005, Australia }
\address {$^{13}$INFN, Sezione di Roma, I-00185 Roma, Italy }
\address {$^{14}$IAC3--IEEC, Universitat de les Illes Balears, E-07122 Palma de Mallorca, Spain }
\address {$^{15}$RRCAT, Indore, Madhya Pradesh 452013, India }
\address {$^{16}$Max Planck Institute for Gravitational Physics (Albert Einstein Institute), D-14476 Potsdam, Germany }
\address {$^{17}$University of Warwick, Coventry CV4 7AL, United Kingdom }
\address {$^{18}$OzGrav, University of Western Australia, Crawley, Western Australia 6009, Australia }
\address {$^{19}$The Pennsylvania State University, University Park, PA 16802, USA }
\address {$^{20}$Max Planck Institute for Gravitational Physics (Albert Einstein Institute), D-30167 Hannover, Germany }
\address {$^{21}$Leibniz Universit\"{a}t Hannover, D-30167 Hannover, Germany }
\address {$^{22}$LIGO Laboratory, California Institute of Technology, Pasadena, CA 91125, USA }
\address {$^{23}$Tata Institute of Fundamental Research, Mumbai 400005, India }
\address {$^{24}$Queen Mary University of London, London E1 4NS, United Kingdom }
\address {$^{25}$University of California, Davis, Davis, CA 95616, USA }
\address {$^{26}$University of Minnesota, Minneapolis, MN 55455, USA }
\address {$^{27}$Instituto Nacional de Pesquisas Espaciais, 12227-010 S\~{a}o Jos\'{e} dos Campos, S\~{a}o Paulo, Brazil }
\address {$^{28}$International Centre for Theoretical Sciences, Tata Institute of Fundamental Research, Bengaluru 560089, India }
\address {$^{29}$Universit\"{a}t Hamburg, D-22761 Hamburg, Germany }
\address {$^{30}$University of California, Berkeley, CA 94720, USA }
% ...
\address {$^{86}$University of Texas, Austin, TX 78712, USA }
% ...
\address {$^{159}$Institut Interdisciplinaire d'Innovation Technologique (3IT), Universit\'e de Sherbrooke, Sherbrooke, Qu\'ebec J1K 0A5, Canada (external) }
\address {$^{160}$The Johns Hopkins University Applied Physics Laboratory, Laurel, MD 20723, USA }
\ead{ling.sun@ligo.org}

\vspace{10pt}
%\begin{indented}
%\item[]May 2025}
%\end{indented}

\begin{abstract}
We present the LIGO \Asharp{} detector concept, an upgrade for the LIGO observatories based on room-temperature interferometers beyond the fifth observing run (O5). Building on the A+ sensitivity, \Asharp{} targets broadband sensitivity improvements through heavier test masses, improved suspensions and seismic isolation, increased arm-cavity power, enhanced frequency-dependent squeezing, reduced coating thermal noise considering two scenarios, and improved control of mechanical motion and optical modes. We describe the principal design choices, projected noise performance, and corresponding astrophysical prospects. LIGO \Asharp{} substantially increases compact-binary detection rates, strengthens population inference, and improves both early-warning times and localization for binary neutron star mergers. The improved sensitivity enables more detailed studies of compact-binary coalescences, including higher-order multipoles, intermediate-mass black holes, remnant black hole ringdown, and the neutron star equation of state. It also broadens the discovery potential for new gravitational-wave sources such as continuous waves and bursts, should enable detection of the stochastic background from compact binary mergers if it remains undetected after O5, and strengthens the role of gravitational-wave detectors as probes of fundamental physics. We discuss key technical challenges and the role of \Asharp{} as both a major scientific upgrade for the 2030s and a technology pathfinder for next-generation gravitational-wave observatories, such as Cosmic Explorer.

%\textit{Draft revision: \IfFileExists{./gitID.txt}{git hash \input{gitID.txt}}{Missing gitID  (\today)}}
\end{abstract}

%\maketitle

\tableofcontents

\section{Introduction}
\label{sec:introduction}

Hundreds of compact binary coalescence events have been observed by the LIGO--Virgo--KAGRA detector network~\cite{LIGOScientific:2014pky,VIRGO:2014yos,KAGRA:2020tym,Prospects_network} in the first decade since the first direct gravitational-wave (GW) detection, establishing a powerful new observational channel for studying compact-object astrophysics and gravity in the strong-field regime~\cite{GWTC-4,GWTC-5,GWTC-4_intro,GWTC-4_methods,GWTC-5-intro,GWTC-5-method}. 
As the field matures, the scientific priorities are shifting from the initial discovery phase toward precision astrophysics, population studies, and multi-messenger observations. Achieving these goals requires sustained improvements in detector sensitivity beyond that of the current Advanced LIGO and A+ configurations~\cite{AplusASD}.

In the fifth observing run (O5), planned for the late 2020s to early 2030s~\cite{obs_plan}, Advanced LIGO will represent the culmination of the A+ upgrade program. LIGO's ultimate O5 performance, represented here by the A+ design sensitivity~\cite{AplusASD}, is expected to improve the strain sensitivity by nearly a factor of two relative to the fourth observing run (O4) across the full frequency band. With observing data collected over approximately three calendar years~\cite{obs_plan}, O5 will add significantly to the catalog of GW events. 
In the years beyond O5, LIGO's scientific output would be dramatically enhanced by taking the time to double the broadband sensitivity and reduce the low-frequency cutoff before observing again, rather than simply continuing to observe for an extended period at the final O5 sensitivity. 
This will significantly increase the number of observed events, particularly currently rare sources such as binary neutron star (BNS) coalescences, and will allow higher-mass black holes (BHs) to be observed. It will also provide more detailed measurements of observed coalescence signals, enabling studies of the nuclear equation of state and higher-order multipole GW emission.

To meet these challenges, the LIGO Scientific Collaboration has developed the \Asharp{} (pronounced ``A sharp'') design~\cite{Fritschel:2024postO5}, with the aim of installing and operating it in the LIGO facilities after the end of the O5 run in the early 2030s.
Conceived as a broadband upgrade that significantly extends LIGO's astrophysical reach, \Asharp{} is compatible with existing infrastructure and uses technologies that can be ready on the post-O5 timescale; the expected science outcomes are therefore shaped by the instrument sensitivity achievable through feasible upgrades.
The \Asharp{} design simultaneously targets improvements at low frequencies (below \qty{50}{\Hz}), mid-frequencies (\qtyrange{50}{300}{\Hz}), and high frequencies (above \qty{300}{\Hz}), enabling a broad spectrum of new science.
Detection rates for binary mergers are expected to increase by approximately a factor of four to eight relative to the A+ design, depending on the upgrade scenario. 
Improved low-frequency sensitivity enhances observations of higher-mass BBH systems and enables early warning for BNS mergers, while the broader sensitivity improvement increases the number of well-localized BNS events, improving the prospects for coordinated electromagnetic follow-up.
At mid and high frequencies, the sensitivity gains strengthen compact-binary population inference and enable more precise measurements of source properties.
The broadband sensitivity improvement also increases the possibility of observing new source classes and probing fundamental physics.

Additionally, \Asharp{} would provide an important scientific and technological bridge between A+ and the next-generation ground-based observatories. 
Scientifically, it would substantially expand LIGO's discovery potential in the period before next-generation detectors, Cosmic Explorer~\cite{evans2021CEHS,evans2023} and the Einstein Telescope~\cite{Maggiore2020,ET_Abac_2026}, become operational. 
Technologically, \Asharp{} core upgrades, including heavier test masses, improved suspensions, higher power operation, advanced test-mass coatings, and enhanced squeezing, directly overlap with technologies required for these future detectors. Implementing \Asharp{} within the current LIGO facilities therefore delivers major scientific gains in the coming decade while establishing and validating critical technologies for next-generation observatories.

In this paper, we present the \Asharp{} design and the scientific opportunities it enables. \Cref{sec:upgrades} describes the principal design choices, required technology developments, and projected noise performance. In \cref{sec:science}, we examine the scientific prospects of the \Asharp{} detectors. \Cref{sec:discussion} discusses alternative configuration options, key challenges, and \Asharp{}'s role within the broader global roadmap for GW astronomy.

\section{LIGO \Asharp{} design}
\label{sec:upgrades}

The \Asharp{} upgrades to the LIGO A+ detectors retain room-temperature operation, fused-silica test masses, and \qty{1064}{\nm} laser light. This leverages existing expertise and provides a natural path toward implementation in Cosmic Explorer detectors. 
The design changes can be divided into improvements targeting low frequencies (below  \qty{50}{\Hz}), mid frequencies (\qtyrange{50}{300}{\Hz}), and high frequencies (above \qty{300}{\Hz}).

\begin{figure}[h]
	\centering
	\includegraphics[width=0.8\textwidth]{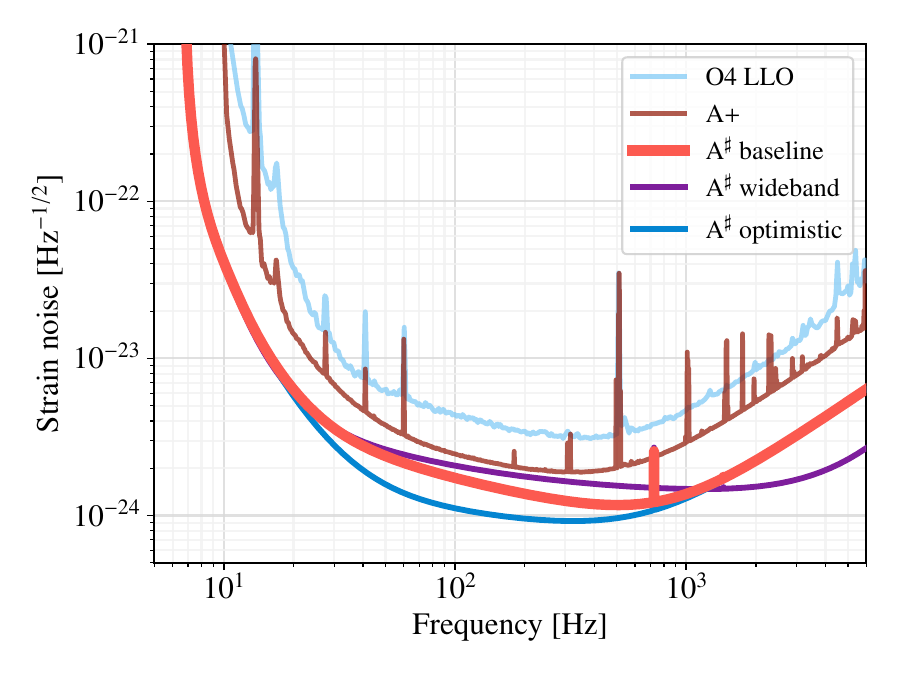}
	\caption[Overview of post-O5 sensitivity scenarios]
	{Design strain noise spectra for the LIGO \Asharp{} configurations considered in this work. The O4 LLO strain noise curve represents a typical sensitivity achieved at LIGO Livingston during O4 in September 2024, while the A+ curve corresponds to the projected sensitivity for the final stage of O5~\cite{AplusASD,HelmlingCornell:2025O4bSensitivity,HelmlingCornell:2025O4cSensitivity,PostO5SensitivityCurves}.}
	\label{fig:postO5-overview}
\end{figure}

\Cref{fig:postO5-overview} summarizes the strain sensitivity of the detector configurations considered in this work. The principal designs are as follows:

\begin{itemize}
	    \item \textbf{A+ design.} The A+ configuration builds on Advanced LIGO~\cite{LIGOScientific:2014pky} and corresponds to the projected sensitivity of the final stage of O5 (O5c)~\cite{AplusASD}. It incorporates a 30\% reduction in coating thermal noise relative to the current Advanced LIGO level and implements \qty{7}{\dB} of observed frequency-dependent squeezing~\cite{AplusASD}. 
	    Although the test masses will be recoated, the A+ suspension design remains unchanged from Advanced LIGO.
		In addition, balanced homodyne detection (BHD) is employed to further optimize the quantum-noise readout~\cite{Fritschel2014.OE,Heinze2022}.
	    The test masses are \qty{40}{\kg}, and the circulating laser power in each arm cavity is \qty{550}{\kW}. 
	
		\item \textbf{\Asharp{} baseline.} 
		Relative to A+, \Asharp{} baseline increases the test masses to \qty{105}{\kg}, redesigns the suspensions for the heavier optics and improved controllability, and raises the suspension-fiber stress by a factor of two to \qty{1.6}{\GPa}. The low-frequency design also includes upgraded seismic isolation sensors and a modest factor-of-two suppression of Rayleigh-wave Newtonian noise. At high frequencies, \Asharp{} baseline increases the arm-cavity circulating power to \qty{1.5}{\MW} and pushes the observed squeezing level to \qty{10}{\dB}. The coating thermal noise is reduced to 50\% of the Advanced LIGO level; 
		%this is a relatively modest reduction from the O5 level, but as discussed in \cref{subsec:aplus_coatings}, technically it appears to be quite feasible.
		pathways envisioned to obtain such a reduction are discussed in \cref{subsec:aplus_coatings}.
		
		\item \textbf{\Asharp{} wideband.} Identical to \Asharp{} baseline, but with the signal bandwidth broadened to \qty{3.4}{\kHz} (from \qty{450}{\Hz}) by increasing the reflectivity of the signal recycling mirror, together with corresponding adjustments to the finesse and detuning of the squeezer filter cavity. The \Asharp{} wideband configuration updates the A+ wideband tuning proposed in Ref.~\cite{Ganapathy2021}, using the improved \Asharp{} quantum-noise model described in \cref{subsec:wideband}.
	
		\item \textbf{\Asharp{} optimistic.} The \Asharp{} baseline configuration defined above is limited by coating thermal noise at mid-frequencies, so lower-thermal-noise coatings would improve the interferometer performance and increase the science reach. As an optimistic limiting case, we consider a strain curve in which the coating thermal noise is equal to the \Asharp{} quantum noise at \qty{100}{\Hz}; this corresponds to a coating thermal noise reduction of approximately 75\% relative to the Advanced LIGO level. The possibility of realizing this target with GaAs/AlGaAs crystalline coatings is discussed in \cref{subsec:crystalline_coatings}.
\end{itemize}

\begin{figure}[h]
	\centering
	\includegraphics[width=0.75\textwidth]{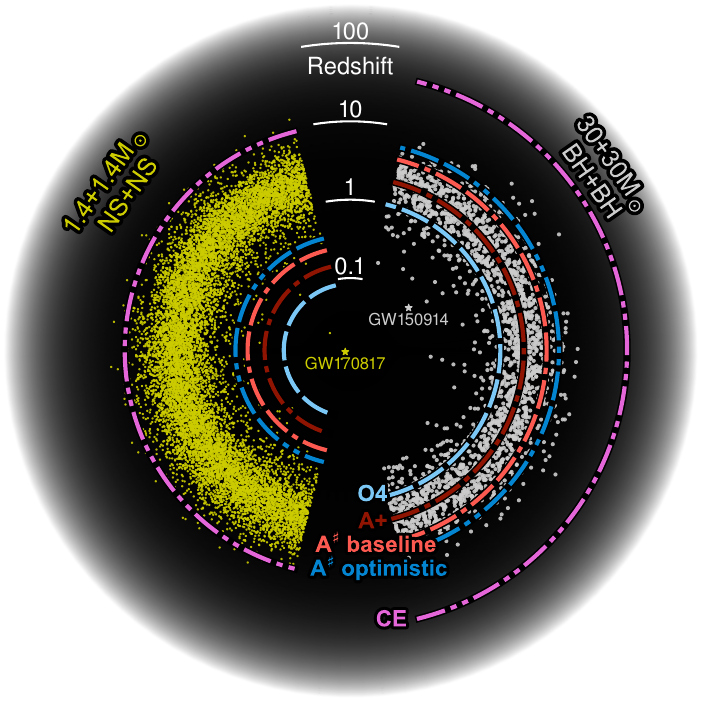}	
	\caption[Horizon redshifts with compact binary populations]
	{Horizon redshifts for representative compact binary coalescences and selected detector configurations shown in \cref{fig:postO5-overview}, together with the projected reach of the next-generation observatory Cosmic Explorer (``CE'') for comparison. The yellow and white dots indicate simulated populations of BNS and BBH mergers, respectively.}
	\label{fig:horizon-donut}
\end{figure}

\Cref{fig:horizon-donut} illustrates the horizon redshifts for representative compact binary coalescences ($1.4+1.4 M_\odot$ BNS and $30+30 M_\odot$ BBH systems) achievable by the selected detector configurations listed above, together with the performance of O4 and the projected reach of the Cosmic Explorer next-generation observatory for comparison (see \cref{subsubsec:horizons}).
The design of Cosmic Explorer is based on much longer arm lengths and incorporates nearly all of the advances described above.
%implementing these technologies within the current LIGO facilities enhances near-term astrophysical output and facilitates the transition to the next generation of observatories.

\begin{figure}[h]
	\centering
	\includegraphics[width=0.85\textwidth]{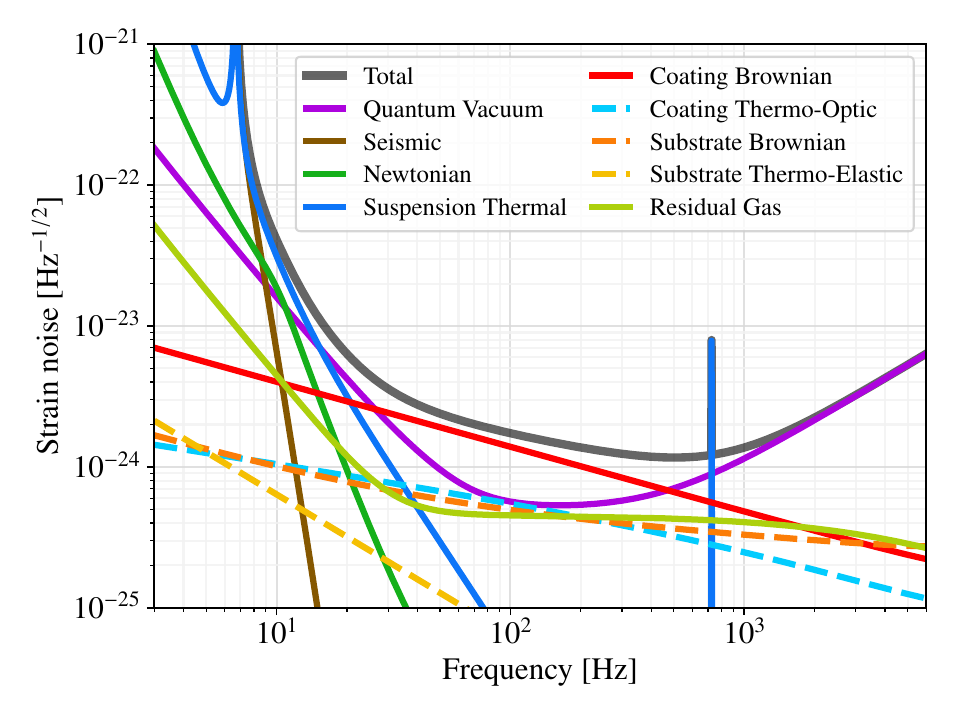}
	\caption{Noise budget for the \Asharp{} baseline target design sensitivity. The total coating thermal noise, the sum of both Brownian and thermo-optic noise, corresponds to a factor-of-two reduction from the Advanced LIGO level. For the noise budgets of the \Asharp{} optimistic and \Asharp{} wideband configurations, see \cref{fig:Asharp-wideband-nb,fig:Asharp-AlGaAs-nb}, respectively.}
	\label{fig:Asharp-nb}
\end{figure}

The noise budget for the \Asharp{} baseline design is shown in \cref{fig:Asharp-nb}; noise budgets for the other configurations are provided in \cref{sec:mfupgrades,sec:hfupgrades}.
\Cref{tab:main_params} list the main detector parameters of the A+ and \Asharp{} baseline configurations.
The projected strain noise is decomposed into its dominant contributions, highlighting the frequency regions in which seismic and Newtonian noise, suspension thermal noise, coating and substrate thermal noise, and quantum noise limit the detector performance. These limiting noise sources motivate the structure of the following discussion, where we describe the low-, mid-, and high-frequency upgrades in \cref{sec:lfupgrades,sec:mfupgrades,sec:hfupgrades}, respectively.

\begin{table}[h]
	\centering
   \renewcommand{\arraystretch}{1.2}
	\sisetup{table-alignment-mode=none}
	\makebox[\textwidth]{
		\begin{tabular}{r l S S}
			\hline
			\textbf{Parameter} & \textbf{Units} & \textbf{{A+}} & \textbf{\Asharp{} baseline}  \\
			\hline
			Arm cavity power & \unit{\kilo\W} &
            % hard coding this because A+ is now O5c and I don't want to rewrite the
            % entire infrastructure just to fix this one number
			550 & % \fpeval{1e-3 * \Val{Aplus.ArmPower_W}} &
			\fpeval{1e-3 * \Val{Asharp.ArmPower_W}} \\
			Laser wavelength & \unit{\nm} & {1064} & {1064} \\
                      	Test mass material & & {Silica} &  {Silica} \\
                      	Temperature & \unit{\kelvin} &
                          \Val{Aplus.Materials.Substrate.Temp} &
                          \Val{Asharp.Materials.Substrate.Temp} \\
			Mass of test mass & \unit{\kilo\gram} &
			\sisetup{round-mode=places, round-precision=0}\Val{Aplus.Suspension.Stage[0].Mass} &
            % hard coding this because the auto generated one gives 104; probably some minor parameter change since the report
            105 \\
			% \sisetup{round-mode=places, round-precision=0}\Val{Asharp.Suspension.Stage[0].Mass}  \\
			Size of test mass (diameter $\times$ thickness) & \unit{\cm} & 
			\fpeval{200 * \Val{Aplus.Materials.MassRadius}} $\times$ \fpeval{100 * \Val{Aplus.Materials.MassThickness}} & 
			\fpeval{200 * \Val{Asharp.Materials.MassRadius}} $\times$ \fpeval{100 * \Val{Asharp.Materials.MassThickness}} \\
			Test mass bounce frequency & \unit{\Hz} & 
			\sisetup{round-mode=places, round-precision=1}\Val{Aplus_sus.BounceMode_Hz} & 
			\sisetup{round-mode=places, round-precision=1}\fpeval{6.57}\\
			Test mass roll frequency & \unit{\Hz} & 
			\sisetup{round-mode=places, round-precision=1}\fpeval{sqrt(2) * \Val{Aplus_sus.BounceMode_Hz}} & 
			\sisetup{round-mode=places, round-precision=1}\fpeval{sqrt(2) * 6.57} \\
			Observed squeezing & \unit{\decibel} &
      			\sisetup{round-mode=places, round-precision=0}\fpeval{-\Val{Aplus.ObsSqueezeHigh_dB}} &
      			\sisetup{round-mode=places, round-precision=0}\fpeval{-\Val{Asharp.ObsSqueezeHigh_dB}}  \\
			Rayleigh wave suppression & \unit{\decibel} &
      			\sisetup{round-mode=places, round-precision=0}\fpeval{\Val{Aplus.RayleighSuppr_dB}} &
      			\sisetup{round-mode=places, round-precision=0}\fpeval{\Val{Asharp.RayleighSuppr_dB}}  \\
			Suspension thermal noise (\qty{15}{\Hz}) & Hz$^{-1/2}$ & $1.2 \times 10^{-23}$ & $7.5 \times 10^{-24}$ \\
			Coating thermal noise (\qty{100}{\Hz}) & Hz$^{-1/2}$ & $1.9 \times 10^{-24}$ & $1.4 \times 10^{-24}$\\
			\hline
		\end{tabular}
	}
	\caption{Major parameters of the A+ and \Asharp{} baseline detector configurations.}
	\label{tab:main_params}
\end{table}

\subsection{Low-frequency upgrades}
\label{sec:lfupgrades}

%Improvements to the low-frequency performance have a significant impact on the science output of the detector, as described in \cref{sec:science}.
Significant research efforts have led to continuous improvements in LIGO's sensitivity in the \qtyrange{10}{50}{\Hz} band and have further improved our understanding of the technical noises still present in this region~\cite{LIGOScientific:2014pky}.
Despite these improvements, the detectors have not achieved the design sensitivity for Advanced LIGO~\cite{Capote2025PRD} at low frequencies. \Asharp{} is designed to address known technical noises, with the intent to reach the limits imposed by thermal and quantum noise at frequencies as low as \qty{10}{\Hz}.

The low-frequency sensitivity improvements for \Asharp{} stem from the following changes to the interferometers: the test masses are increased in size and mass; the test-mass suspensions are redesigned to reduce thermal noise and control noise; the active seismic isolation platforms are upgraded with lower-noise sensors to reduce control noise; and ground-motion sensors are added to enable subtraction of Newtonian noise.
These upgrades target performance improvements from \qtyrange{10}{50}{\Hz}, and maintain the performance increase even with very high-power operation.

The upgrades targeting improved low-frequency performance are described in the following subsections, with \cref{subsubsec:sus-105kg,subsubsec:sus-fibers} discussing improvements that address fundamental noise sources, such as increased mass of the test mass and increased stress in the suspension fibers.
In \cref{subsubsec:sus-control}, we describe how changes to the suspension design improve its dynamics and controllability, leading to noise reductions both below and above \qty{10}{\Hz}.
New sensors for the suspension and seismic isolation systems are described in \cref{subsubsec:sus_sensors}. These sensors are key to reducing the test-mass motion below \qty{10}{\Hz}, and thus to reducing noise from the interferometer control signals.
Additional advanced sensors to improve seismic-isolation performance are described in \cref{subsubsec:seismic_isolation}.
With reduced technical noise around \qtyrange{10}{20}{\Hz}, the impact of Newtonian noise will, for the first time, not be negligible. Suppression of Newtonian noise is described in \cref{subsubsec:newtonian_noise}.

\subsubsection{Heavier test masses} 
\label{subsubsec:sus-105kg}

The Advanced LIGO suspension and isolation systems provide excellent passive isolation and control of thermal noise of the test masses. The \Asharp{} suspension upgrade builds on this extensive design expertise, incorporating updates from experience of operating Advanced LIGO. The most apparent upgrade is the increased mass and dimensions of the test mass. The mass of the fused-silica test mass is increased to \qty{105}{\kg} (from \qty{40}{\kg}), with a diameter of \qty{45}{\cm} (from \qty{34}{\cm}) and a thickness of \qty{30}{\cm} (from \qty{20}{\cm}). 

The increased mass, $m$, has several benefits to the strain noise: quantum radiation pressure noise is reduced (as $1/m$); suspension thermal noise is reduced (as $1/\sqrt{m}$); the impact of gas damping force noise is reduced (as $m^{-2/3}$ under certain assumptions); and the increased moment of inertia (by a factor of 4.6) counters radiation-pressure torques, helping to stabilize arm-cavity alignment. 
Increasing the test mass significantly beyond \qty{105}{\kg} would require increasing the payload capability of the seismic isolation system (a major endeavor), and would provide only minimal additional performance benefit.

All test-mass substrates will be made of Heraeus Suprasil fused silica, using type 312 for the end test masses and the lower-OH content, and thus lower optical absorption, type 3002 for the input test masses. Both types are available in sizes up to hundreds of kilograms. The Suprasil~3002 may require a compensating polish on the back side of the substrate to correct for bulk inhomogeneity. 

 %In general, at high enough power, this mode must be damped by the angular controls of the interferometer. To keep that angular control noise out of the detection band, this mode frequency well below \qty{10}{\Hz}. For \Asharp{} the Sidles-Sigg stiffness becomes comparable with the mechanical stiffness at a cavity power of about \qty{1.1}{\mega\W} for pitch. For yaw, the mechanical stiffness dominates up to the \qty{1.5}{\mega\W} design point.  

\subsubsection{Suspension fiber upgrade} 
\label{subsubsec:sus-fibers}

The test masses will be suspended from the penultimate mass of the suspension with four fused-silica fibers, to minimise mechanical dissipation, as in the current LIGO suspension~\cite{Cumming2012.CQG,Cumming2020.CQG}. The fiber characteristics (length, diameter, and stress) determine the suspension thermal noise and the frequencies of the modes that involve fiber motion, namely the highest vertical (or `bounce') and roll modes of the test mass and the internal (or `violin') modes of the fibers. The fiber length will be the same as in the current suspension, with the capacity to increase the length by approximately \qty{10}{\%} to aid lowering the roll mode of the suspension~\cite{G2600209}. The fiber will have a starting stock diameter of \qty{5}{\mm} to aid with the energy distribution through the suspension fiber. Along the main length of the fiber it will have a diameter of \qty{450}{\um}, with a thermoelastic nulling region on either end of \qty{1300}{\um} in diameter, with a length of \qtyrange{25}{30}{\mm} \cite{PhysRevB.65.174111}.

The suspension fibers for \Asharp{} will operate at a stress of at least \qty{1.6}{\giga\Pa}, twice that of the current LIGO fibers, with the potential to increase to \qty{2}{\giga\Pa}~\cite{G2600209}. 
Suspension thermal noise is only marginally the dominant strain noise contributor from \qty{10}{\Hz} to \qty{14}{\Hz} with this design (see \cref{fig:Asharp-nb}).
%\comments{Updated numbers from T2600209 are 1.5e-23 strain at 10 Hz, not dominant for latest advanced SUS TN modelling; we think figure 3 might need updated, we can supply the numbers if required}. 
The test-mass bounce and roll modes, at \qty{6.6}{\Hz} and \qty{9.3}{\Hz} respectively, are both below \qty{10}{\Hz}, and the fundamental violin-mode frequency is increased from \qty{500}{\Hz} to \qty{740}{\Hz}, reducing its impact in the mid-frequency detection band.

Research and development of fibers that operate at this increased stress of \qty{1.6}{\giga\Pa} is advanced~\cite{PhysRevApplied.17.024044}. Previous investigations of stress fatigue in fused silica suggest the expected lifespan of these fibers in air is over 300 years~\cite{Toland2020.PhD, Lee2019.PhD}. Ongoing research investigating high-stress operation of these fibers under vacuum suggests that the lifespan may increase by an order of magnitude~\cite{T2500387, G2600366, G2500474}, implying a significantly reduced risk of stress-fatigue failure over the operating lifetime of \Asharp{}.

% LS removed Table 1; kept numbers below for reference:
% Config | Bounce mode (Hz) | Roll mode (Hz) | Violin mode (Hz) | Early warning (min)
% A+ | 9.7 | 13.8 | 497 | 3
% 1.6 GPa fibers | 6.6 | 9.3 | 738 | 6

\subsubsection{Suspension Dynamics and Controllability} 
\label{subsubsec:sus-control}

\begin{figure}
	\centering
	\includegraphics[width=0.9\textwidth]{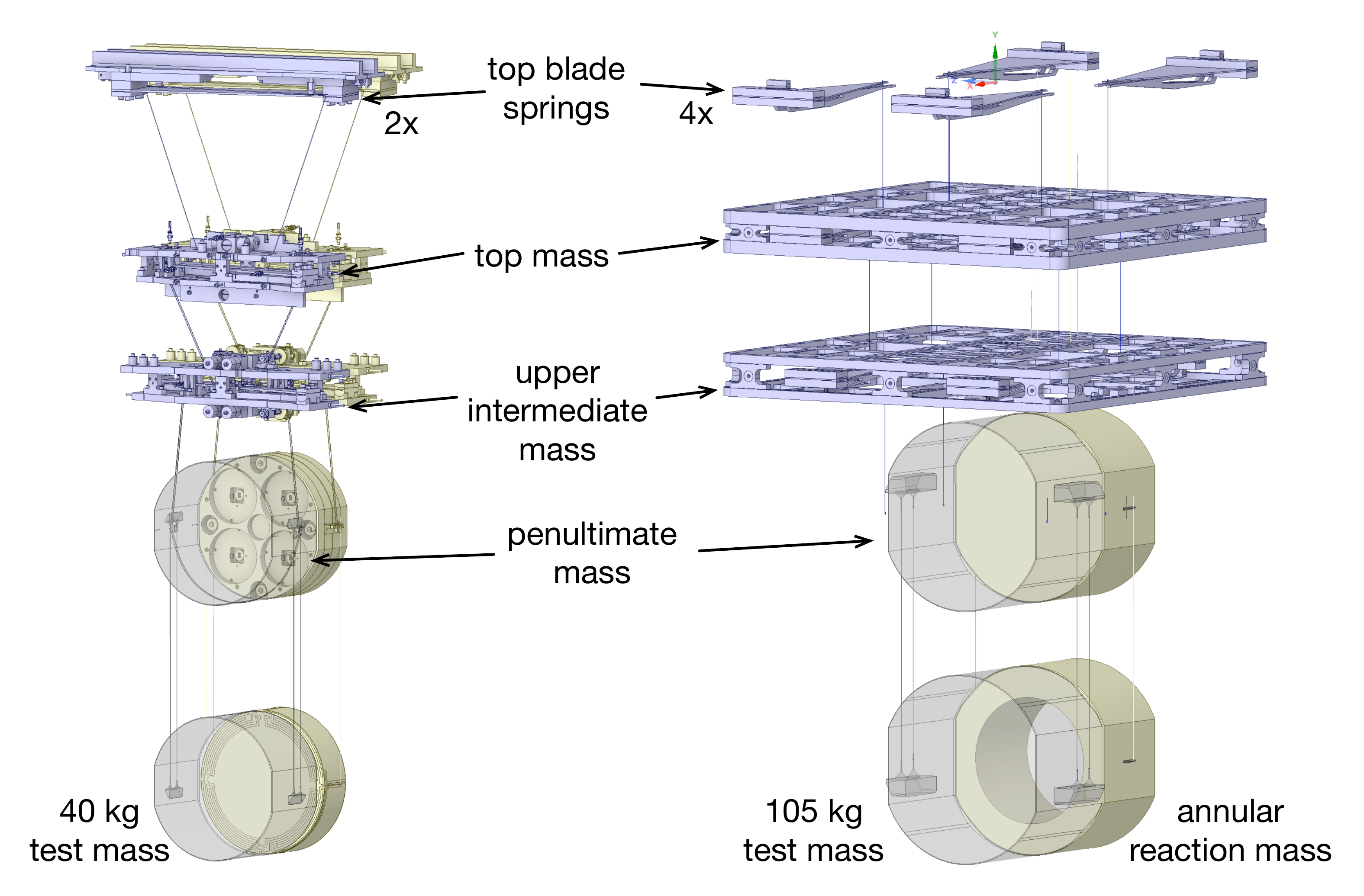}
	\caption[]
	{Advanced LIGO and A+ suspension (left) for the input test mass, and a conceptual drawing of the \Asharp{} suspension (right)~\cite{G2401599}. The overall designs are similar, but several changes are evident. The fused silica optic (lowest grey mass) is larger (increased from \qty{40}{\kg} to \qty{105}{\kg}). The two upper stages (grey) are heavier and much larger for higher moments of inertia. 
	To reduce cross-coupling, the suspension wires are vertical, and there are four blades, each supporting one wire, at each of the three levels, instead of two blades with one wire each at the top level and two blades with two wires each below.
	The separation of the upper stages is more uniform, which improves the isolation and reduces the frequency of several high-frequency modes. The reaction chain (yellow) has been reduced to two stages (from four). There is a \qty{34}{\cm}-diameter clear aperture in the lower reaction mass to minimize scattered light and reduce gas damping forces. The springs that support the reaction mass and the actuators for the reaction mass are not shown. Local sensing and actuation, caging, and baffles are also not shown.}
	\label{fig:2_sus}
\end{figure}

The four-stage suspension system has been redesigned to support the larger and heavier \Asharp{} test-mass mirrors. This redesign integrates modifications that address lessons learned during operation of Advanced LIGO. Two of the largest sources of low-frequency technical noise in Advanced LIGO are the local damping of the suspension pendulum modes and global feedback required to actively maintain alignment of the optical cavities. The main goal is to mitigate these and other technical noise contributions in the detection band by reducing test-mass motion below \qty{10}{\Hz}, particularly the test-mass angular motion \cite{G2402461}. A conceptual drawing of the new suspension is shown on the right in \cref{fig:2_sus}.

The total mass of the main suspended chain is increased to \qty{420}{\kg} (from \qty{120}{\kg}). Each of the four stages has approximately the same mass, in contrast to the current suspension, where the top two stages have half the mass of the lower two. This updated mass distribution keeps the suspension modes more closely spaced in frequency, which improves the passive isolation and reduces the required bandwidth of the local damping loops (thus reducing the infiltration of damping noise into the detector's strain measurement). 

The \Asharp{} suspension has several features that greatly reduce cross-couplings, especially from horizontal (longitudinal) motion of the suspension point and longitudal drives on the suspension to angular (pitch) motion of the test mass. All the wires in the \Asharp{} suspension are vertical, whereas many Advanced LIGO wires are mounted at angles, to give more flexibility in setting the mode frequencies at the expense of increased cross coupling.
The \Asharp{} suspension has four independent blades, each with a separate wire, supporting the three upper stages. The effective flexure points of the wires and fibers are aligned with the vertical center of mass for associated stages so that translations do not generate torques on the stages. 
Since the Advanced LIGO suspension has two blades at each level, with only two wires at the top and two wires per blade at the top mass and upper-intermediate mass, the flexure points must be offset from the center of mass to ensure stability, but this couples angular and translational motion into the test-mass motion.
Even so, having four blades/wires at the top directly couples low-frequency angular motion from the seismic isolation table to the suspension. The impact of this angular drive is relatively small and will be further reduced by installing new rotation sensors in the isolation system, as discussed further in \cref{subsubsec:seismic_isolation}.

%Each stage is suspended by 4 wires, so that the effective wire attachment point at each stage can be aligned with that stage’s center of mass, reducing the cross-coupling from horizontal to pitch motion. In addition, all of the suspension wires will be vertical, rather than angled as in Advanced LIGO, which also reduces nominal cross couplings between degrees of freedom. 

In \Asharp{}, the two upper stages are not only relatively heavier than in the current suspension, but also much larger, especially in the horizontal dimensions. 
With the increased moment of inertia of these stages, the frequencies of the highest angular modes of the suspension are reduced and their visibility in the top stage is increased, so that they can be damped. In addition, it makes the design more robust against cross-coupling from small errors in alignment of actuators and wire flexure points.  

 As in the Advanced LIGO suspension, a suspended reaction chain is included, from which low-noise control of the main chain is applied. Modeling has shown that a two-stage suspension for the reaction chain (rather than four stages as in the current suspension) is sufficient to provide low-noise feedback to the penultimate mass and test mass \cite{G2402448}. As shown in \cref{fig:2_sus}, the masses of the reaction chain hang adjacent to the penultimate mass and test mass of the main chain, with a nominal gap of \qty{5}{\mm} between adjacent masses. The bottom reaction-chain mass is an annulus rather than a solid cylinder, in order to further reduce force noise from residual gas in the gap~\cite{Dolesi2011}. In addition, the large, \qty{34}{\cm} diameter of the annulus will help minimize scattered light. 
 
%Two of the largest sources of low-frequency technical noise in Advanced LIGO are local damping of the suspension pendulum modes and global feedback required to actively maintain alignment of the optical cavities. The \Asharp{} suspension is designed to mitigate both noise sources. \comments{KLD: This para fits better in the first para of Sec 2.1.3?}

\paragraph{Local Damping}

All modes of the main chain, except for the highest-frequency bounce and roll modes of the test mass, are designed to be actively damped from the top stage. This is the same strategy that is used in Advanced LIGO.
The challenge is to adequately damp these modes, while preventing the noise from the local sensors from compromising the test-mass motion in the GW band. This challenge is eased in the \Asharp{} design by several factors. First, as noted, the mass distribution of the chain keeps the mode frequencies low and closely spaced, reducing the required bandwidth of the local damping. Second, in addition to the increased moment of inertia of the upper masses, their larger size means that the local displacement sensors will have a larger lever arm for the angular sensing, reducing the effective angular-sensing noise. Finally, lower-noise interferometric displacement sensors will be used for the local sensing, as described in \cref{subsubsec:sus_sensors}. In addition, with sufficiently low-noise sensors, they can also be used at the second stage of the suspension to provide more damping authority for some modes.

The bounce and roll modes of the test mass
%, at 
%\qty{6.6}{\Hz} and 
%\sisetup{round-mode=places, round-precision=1}\Val{1p6GPa.BounceMode_Hz}{\Hz} and
%\qty{9.3}{\Hz},
%\sisetup{round-mode=places, round-precision=1}{\fpeval{sqrt(2) * \Val{1p6GPa.BounceMode_Hz}}{\Hz},
%respectively, 
will be damped with passive, tuned-mass dampers mounted at the ends of the second-stage blade springs, similar to that of the current design.

\paragraph{Global Angular Sensing and Control}
Global interferometric alignment signals are used to maintain precise alignment of all interferometer optics using feedback controls~\cite{Barsotti_2010}. In the current Advanced LIGO design, the angular motion of the test masses is stabilized to a residual root-mean-square (RMS) alignment error of approximately \qty{1e-9}{rad}, with control bandwidths of up to a few hertz. Angular-sensing noise, though well filtered, still drives the test-mass motion in the \qtyrange{10}{20}{\Hz} region; this motion couples to the differential arm length at a level of order \qty{1e-3}{\m/rad}, creating strain noise that exceeds the target design level below \qty{20}{\Hz}. 

The \Asharp{} design mitigates this global alignment noise in two ways. First, with the improved suspension dynamics described above, lower-noise local sensors, and improved sensors in the seismic-isolation system described in the following subsections, the RMS angular motions of the test masses will be significantly reduced. This will translate to lower global control bandwidths, enabling more filtering of the angular-sensing noise at \qty{10}{\Hz} and above.

Second, the much greater moments of inertia of the larger test masses and the upper stages of the suspension will help mitigate radiation-pressure torques. 
As the optical power increases in the arm cavities, the photon pressure of the beam creates significant torques on the optic. 
The Sidles-Sigg effect~\cite{SidlesSigg2006PRA} changes the angular dynamics and creates virtual optical springs between the cavity optics. The `hard mode' stiffens the symmetric alignment mode and the `soft mode' softens the anti-symmetric mode of the cavity. For the \Asharp{} suspension, above about \qty{1.1}{\MW} of arm power, the stiffness of the hard mode will exceed the local mechanical stiffness of the suspension, such that the highest-frequency pitch mode is no longer visible at the top two stages; therefore, global interferometer sensing must be used to damp this mode. (In comparison, for the current \qty{40}{\kg} test-mass suspensions, this crossover occurs at about \qty{300}{\kW}.)  The suspension's yaw motion is relatively stiffer, and is larger than the optical stiffness even at \qty{1.5}{\MW}, so it can always be damped locally. 

For arm powers below \qty{1.1}{\MW}, the global angular-control bandwidth should only be needed up to about \qty{300}{\mHz}, making it much easier to remove angular-sensing noise at \qty{10}{\Hz} and above. 
At an arm power of \qty{1.5}{\MW}, global alignment control is required to damp the highest-frequency pitch mode, which will be around \qty{3}{\Hz}. Modeling suggests that this can be done without compromising the low-frequency noise in the detector~\cite{G2402461}. 
 
\subsubsection{Local Interferometric Sensors for the \Asharp{} Suspension}
\label{subsubsec:sus_sensors}
The sensors used for local damping in the current Advance LIGO suspension are so-called `shadow sensors'; they comprise an LED--photodiode pair mounted to the suspension support structure, and a flag on the suspended mass that partially blocks the LED light to provide a photodiode signal proportional to the position of the flag relative to the structure (BOSEM shadow sensors are commonly used in LIGO~\cite{Cooper2023.RSI}). These sensors have a noise floor of approximately \qty{1e-10}{\m\rtHz} above about \qty{5}{\Hz}, with increasing noise below \qty{1}{\Hz}. This level of sensor noise has been challenging to the interferometer performance in two ways: the sensor noise up to a frequency of \qty{10}{\Hz} limits how strongly and sufficiently the suspension modes can be damped; sufficiently filtering out the sensor noise above \qty{10}{\Hz} so that it is below the target strain noise is very challenging.

%In the \Asharp{} suspension,
%%the local sensors at the top two stages will instead be interferometric sensors, which in principle can have a much lower noise floor than the shadow sensors. 
%improved sensors will be used throughout, a DC sensitive beam deflection sensor (QOSEM~\cite{Roocke2026.GWADW, Roocke2026.Prep}) at the Top Mass, and \todo{interferometric sensors on the remaining masses}. The QOSEM achieves a noise floor of \qty{4}{\pm\rtHz} at \qty{1}{\Hz}, while using the same mounting and actuation scheme as the current shadow sensors. It is capable of DC readout, which is required for optic positioning and alignment. 

In the \Asharp{} suspension, improved sensors will be used throughout, with interferometric sensors on the main chain and DC sensitive beam deflection sensors (QOSEM~\cite{Roocke2026.GWADW, Roocke2026.Prep}) or BOSEM sensors on the reaction chain. The QOSEM achieves a noise floor of \qty{4}{\pm\rtHz} at \qty{1}{\Hz}, while using the same mounting and actuation scheme as the current shadow sensors. This performance and form factor make the QOSEM an ideal candidate for local sensing and actuation on the triple suspensions used for the mirrors in the two recycling cavities, including the main beamsplitter. 
Several compact interferometric displacement sensors have been developed by research groups in the GW community; these typically have sensing-noise levels of \qtyrange{0.1}{1}{\pm\rtHz}, down to frequencies of \qty{1}{\Hz} or lower. All of them use a single-frequency, low-noise, fiber-coupled laser source. One design uses a Mach--Zender interferometer and homodyne readout of two orthogonal quadratures using a polarization scheme~\cite{Cooper2018.CQG}. Two other designs are based on deep frequency modulation of the laser source, in conjunction with a macroscopic asymmetry between a reference arm and the sensed target~\cite{Gerberding:15, PhysRevApplied.12.034025, PhysRevApplied.18.034040}. 
The readouts of the QOSEM and interferometric sensors will be blended together to optimally damp the suspension modes.

For the \Asharp{} suspension, the local sensor noise floor must be \qty{5}{\pm\rtHz} or less at \qty{10}{\Hz}~\cite{G2502605}, which all of these displacement sensors are able to achieve. To help determine which sensors will be used for \Asharp{}, a testing program is planned to compare their sensitivity, alignment complexity, operational reliability, vacuum compatibility, and cost.

\subsubsection{Improved sensors for seismic isolation}
\label{subsubsec:seismic_isolation}

The \Asharp{} upgrade will continue to use the existing Advanced LIGO internal seismic isolation (ISI) system~\cite{MATICHARD2015273, MATICHARD2015287}, with additional advanced sensors to improve the isolation performance in important frequency bands. Low frequency isolation is key to reliable, low noise interferometer performance~\cite{LIGOScientific:2014pky, Yu2018.PRL}. To achieve the strain equivalent seismic noise level shown in \cref{fig:postO5-overview}, the ISIs that support the test masses (known as the BSC-ISI) must achieve motion levels at \qty{10}{\Hz} of \qty{5e-13}{\m\rtHz} in translation and \qty{5e-13}{\radian\rtHz} in rotation (or lower). Though there is some variation among the eight BSC-ISI platforms, they all essentially meet this requirement already. The \Asharp{} seismic upgrades are thus targeted at reducing the ISI platform motion below \qty{10}{\Hz}, in the so-called control band. Reductions in the platform motion in the control band translate to smaller RMS motion of the test masses, both in translational and angular motion. This in turn means that the interferometer global controls, which keep the optics at the operating point within tight tolerances, can operate with reduced bandwidths, thereby reducing the coupling of noise in the auxiliary channels to the strain channel in the detection band.

One type of upgrade is to equip the ISI systems with new or additional, lower-noise inertial sensors. Currently the ISIs use commercial inertial sensors (geophones and seismometers), chosen approximately two decades ago. Since then, several inertial sensors that have better performance and are inherently designed to be used in the LIGO vacuum chambers have been developed by labs within the LIGO Scientific Collaboration. 
Under the equivalence principle, a horizontal seismometer cannot distinguish between horizontal acceleration and tilt. Dedicated rotation sensors therefore add a new sensing dimension by directly measuring rotation.
One such sensor is the Cylindrical Rotation Sensor, or CRS~\cite{Ross_CRS}. This is a single-axis inertial rotation sensor, that can be used to measure, and thus to reduce, very low-frequency platform tilt, particularly at the microseism peak around \qty{160}{\mHz}.
Another type is a six-axis inertial sensor, based on a fused-silica mass suspended by a fused-silica fiber~\cite{Smetana_6D}. Known as the C-6D, it has very low eigenfrequencies for all three rotational degrees-of-freedom.
Another rugged portable multi-orientation rotation sensor (ALFRA) could provide access to rotation sensing around the vertical axis using crossed flexures~\cite{McCann2021.RSI}, with the potential to be inherent iso-elastic preventing seismic motion from being down-converted. 
It uses a multi-bounce laser lever-arm ("walk-off" sensor)~\cite{McCann2019.RSI}, while both the CRS and C-6D achieve their sensitivity partially through the use of compact interferometric displacement sensors, of the type described in \cref{subsubsec:sus_sensors}, to sense their proof mass. 

The Birmingham Inertial Sensor, or BIS, is a single-axis seismometer designed for use in the ISI platforms that support the optics of the recycling cavities (the HAM-ISI). Below \qty{100}{\Hz}, they are more than an order of magnitude more sensitive than the Geotech Instruments GS-13 seismometers that are currently used in those platforms~\cite{Amit_5Hz}. Another seismometer in development, designed to be very compact, is based on a mechanical resonator etched from monolithic fused silica, and read out interferometrically~\cite{Guzman_APL2023}.

The other type of upgrade for the seismic system is the Seismic Platform Interferometer (SPI).  The goal of the SPI is to stabilize the relative motion between seismic platforms. The separations between platforms are sensed with a large-working-range interferometer, using a phasemeter developed for the LISA Pathfinder~\cite{GHeinzel_2004}; the relative angular motions between platforms are sensed with two optical levers.  
Feedback loops then use these sensing signals to reduce the relative motions by an order of magnitude or more~\cite{koehlenbeck_study_2023}. 
The SPI will be used between the ISIs in the corner station, but not along the long arms, due to the increased difficulty of implementing the required telescopes. Thus SPIs will reduce the global controls required to stabilize the recycling cavities.

\subsubsection{Newtonian noise suppression.}
\label{subsubsec:newtonian_noise}

Newtonian gravitational noise will limit the low-frequency sensitivity of the \Asharp{} detector when all improvements discussed above in \cref{sec:lfupgrades} are implemented. The dominant source of Newtonian noise for the \Asharp{} detectors is Rayleigh seismic waves near the test masses~\cite{Driggers:2012ac}. Newtonian noise from other sources is expected to be subdominant and should not limit the \Asharp{} detectors. As indicated in \cref{fig:Asharp-nb}, the target is a relatively modest suppression of Newtonian noise from Rayleigh waves by a factor of two. This can be achieved by combining an array of seismometers within the corner station~\cite{PhysRevLett.121.221104} with additional ground-tilt sensors underneath the individual test masses. Their ground-motion readout will effectively predict the seismically induced Newtonian noise contribution to the test masses, so it can be addressed in online processing. Other sensing methods are also under development using distributed acoustic sensing~\cite{Ophardt2026.CQG,rading2025das} or direct measurement of local Newtonian noise~\cite{Chua2023.APL}. The model used for the Newtonian noise curve in \cref{fig:Asharp-nb} corresponds approximately to the \qty{50}~th percentile seismic noise level at the two LIGO sites, reduced by the planned sensing and processing.

\subsection{Mid-frequency upgrades}
\label{sec:mfupgrades}

Improving the sensitivity significantly at mid-frequencies will require new test-mass coatings with lower thermal noise. As noted at the beginning of this section, the LIGO A+ interferometers are expected to achieve a 30\% reduction in coating thermal noise compared to the current O4 level. If there were no further reduction for \Asharp{}, coating thermal noise would be about 3.5 times larger than quantum noise at \qty{100}{\Hz}, strongly limiting the interferometer sensitivity. Thus, there is a major opportunity to improve the sensitivity and increase the science impact of \Asharp{} with lower-thermal-noise coatings. The \Asharp{} test masses will of course use the best coatings available when the choice needs to be made; in this section, we consider the range of thermal-noise levels that could be expected.

\subsubsection{Baseline coatings.}
\label{subsec:aplus_coatings}

The metal-oxide coatings used in the current interferometers are composed of alternating layers of a low-index film (silica) and a high-index film (titania-doped tantala)~\cite{Granata_2020}; the doped tantala layers have much higher mechanical loss than the silica layers~\cite{StevenPenn_2003,Crooks_2004,Harry_2007}, and are therefore the dominant contribution to the total coating thermal noise. This noise results from the thermally activated atomic reconfigurations in the materials and is fundamentally linked to their internal friction, or mechanical loss~\cite{Levin_PhysRevD.57.659}.
Mechanical loss is characterized by the inverse quality factor, $Q^{-1}$, where $Q$ is the quality factor, and can be measured for individual layers.
Note that the noise goes as  $\sqrt{Q^{-1}}$.
In 2019, germanium oxide was identified as a promising material for the high-index layers~\cite{sciadv.abh1117}, also combined with titania doping to raise the index from 1.6 to around 1.9. Mechanical-loss measurements of this titania-germania mixture (mechanical loss of $9 \times 10^{-5}$, four times lower than for titania-doped tantala), with appropriate post-deposition annealing, pointed to coating designs based on titania-germania and silica layers (mechanical loss of $1.3 \times 10^{-5}$) that would exhibit a 50\% reduction in coating thermal noise~\cite{tigeo2_2021prl}. Subsequent development of high-reflection coatings has to date achieved a 40\% measured reduction in thermal noise in small samples~\cite{Davenport2024}. The target design sensitivity for A+ is thus now based on a 40\% reduction in coating thermal noise for the end test masses. The A+ input test-mass coatings, however, still use titania-doped tantala, although with modifications to the deposition parameters a 12\% reduction in their thermal noise is possible.

For the \Asharp{} baseline coatings, we assume that the 50\% reduction for titania-germania coatings suggested in Ref.~\cite{tigeo2_2021prl} can be realized with further development and applied to both input and end test masses. Titania mixed with silica is another high-index material that could potentially reach this target, or even slightly lower~\cite{ti-silica_2023}. Thus, the \Asharp{} baseline design incorporates this 50\% reduction in coating thermal noise. 
%Significant further reduction with titania-germania or other amorphous metal-oxide materials seems unlikely on the \Asharp{} timescale. 
Unfortunately, there is no clearly defined pathway to obtain a significant further reduction with amorphous metal-oxide materials.
There are, however, efforts to develop amorphous semiconductors as high-index materials, with the potential for lower mechanical loss. Amorphous silicon nitride (\textit{a}-SiN$_x$) and amorphous silicon (\textit{a}-Si) are the leading examples, although both materials currently exhibit unacceptably high levels of optical absorption~\cite{PhysRevD.103.042001,PhysRevLett.131.256902,Wallace_2024,PhysRevD.111.042003,5n1c-tjhq}. Multi-material coatings, which combine layers having low mechanical loss but high optical loss close to the substrate, with layers having the inverse characteristics close to the surface, are another approach that could yield lower-noise amorphous coatings~\cite{PhysRevD.91.042002,PhysRevD.91.042001,PhysRevLett.125.011102,76xz-q8k1, 8mhb-mmtz}.

Recent advances indicate that the incorporation of small amounts of hydrogen may substantially improve the prospects of \textit{a}-Si as a coating material. Hydrogen concentrations of only a few atomic percent have been shown to dramatically suppress optical absorption while simultaneously reducing the mechanical loss through dangling-bond passivation and structural relaxation without compromising the refractive index, which remains above 3.7 over the \qtyrange{1064}{2000}{\nm} wavelength range~\cite{PhysRevLett.131.256902,5n1c-tjhq}. Furthermore, annealing at temperatures up to 600 °C, just below the crystallization temperature of \textit{a}-Si, could further reduce its optical absorption and mechanical loss, while remaining compatible with the thermal treatments that may be required for low-index coating materials. These advances establish hydrogenated amorphous silicon (\textit{a}-Si:H) as a promising candidate for future low-noise high-index coatings.

\subsubsection{Optimistic coatings.}
\label{subsec:crystalline_coatings}

\begin{figure}[h]
	\centering
	\includegraphics[width=0.85\textwidth]{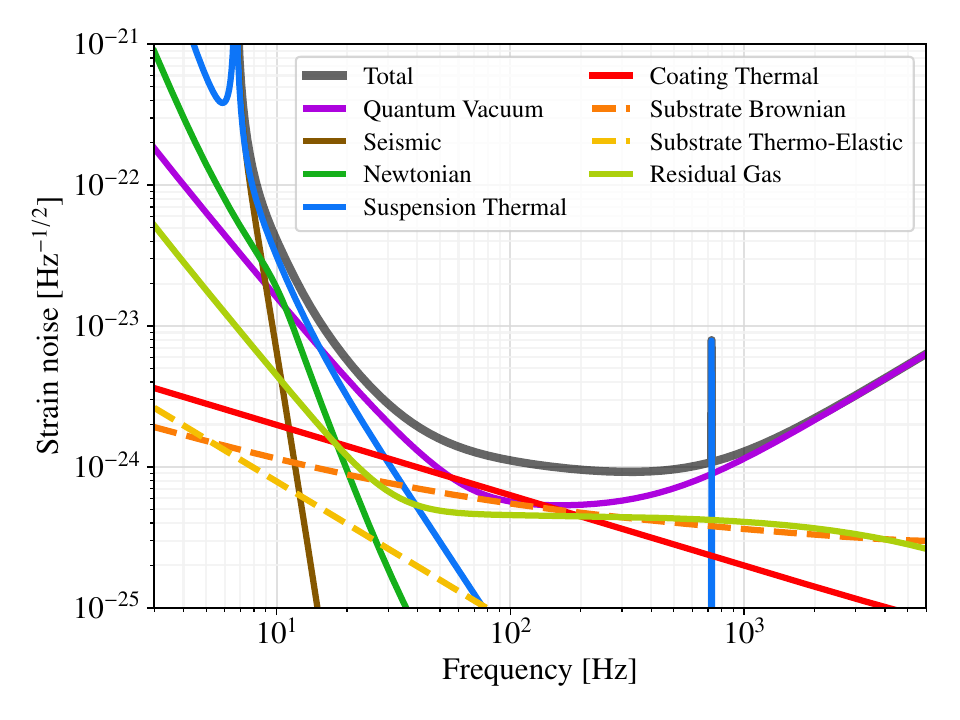}
	\caption{\Asharp{} optimistic noise budget. The total coating thermal noise, the sum of both Brownian and thermo-optic noise, corresponds to a factor-of-four reduction from the Advanced LIGO level.}
	\label{fig:Asharp-AlGaAs-nb}
\end{figure}

To evaluate the science impact of better coatings, we consider a reduction in coating thermal noise by approximately a factor of two relative to the baseline case. At this level, coating thermal noise is roughly comparable to substrate thermal noise and quantum noise around \qty{100}{\Hz} (see \cref{fig:Asharp-AlGaAs-nb}), so further reductions would have diminishing impact on the overall sensitivity.

One coating technology that could potentially reach this lower-noise target is crystalline coatings made from GaAs/AlGaAs Bragg stacks. 
Such mirror coatings offer very low Brownian noise because of the low mechanical loss of the crystalline material, which, at $2.5 \times 10^{-5}$ at room temperature, is more than ten times lower than that of titania-doped tantala~\cite{AlGaAs_noise_2013}.

There are several challenges to implementing GaAs/AlGaAs-based coatings in a GW interferometer. These are active areas of research and are reviewed in Refs.~\cite{Cole_APL2023,Cole2026.OPTICA}. One challenge is to scale up the crystal-growth and substrate-transfer processes from reference cavity sizes (\qty{25.4}{\mm} diameter) to the optic size required for multi-kilometer scale interferometers, and substantial progress has been made on this over the past several years. Recent demonstrations of \qty{200}{\mm} diameter GaAs/AlGaAs coatings optimized for thermo-optic noise cancellation~\cite{chalermsongsak} at \qty{1064}{\nm} have validated the full process flow, including large-area wafer-scale epitaxy, direct bonding, and substrate removal~\cite{AlGaAs_SPIE, AlGaAs_PRR2026}.
	
A remaining challenge is scaling the coating diameter to \qty{300}{\mm} to accommodate the large beam size. This requires either the development of custom large-diameter GaAs substrates (\qty{300}{\mm}) or the adoption of heteroepitaxial growth approaches, such as GaAs/AlGaAs on Ge, which offers a lower-cost and more scalable alternative. The latter approach has been established for vertical-cavity surface-emitting laser (VCSEL) manufacturing~\cite{johnson}; however, test mass coatings impose far more stringent constraints on optical absorption (sub-ppm, requiring extremely low background doping) and scatter (sub-ppm, requiring angstrom-level surface quality and interface control). The use of offcut Ge substrates—commercially available from vendors such as Umicore—is expected to suppress antiphase domain formation and improve crystalline quality, though rigorous process optimization remains necessary.

In parallel with epitaxial materials development, scaling of the substrate-transfer process requires advances in tooling and fabrication infrastructure. Notably, new bonding systems capable of handling substrates up to \qty{450}{\mm} in diameter and masses of up to \qty{105}{\kg} are now under development through the Canadian GRAIN/CFI initiative~\cite{GRAIN_link}. These systems will enable direct bonding of large crystalline multilayers onto superpolished silica (or silicon) substrates. These developments set a clear path for realizing large-area, low Brownian-noise crystalline coatings suitable for next-generation gravitational-wave observatories.

 %There are, however, open questions regarding other noise mechanisms that might prevent these coatings from reaching the Brownian-noise limit, in particular birefringence noise~\cite{AlGaAs_noise_2026} and charge-carrier generation-recombination (GR) noise~\cite{wu2025birefringence}. These are also under active investigation, with recent results showing the potential to modify the GR transfer function in order to reach the system Brownian noise floor (for example with external LED illumination to saturate this process~\cite{ma2026laserstabilizedroomtemperature}). Although not readily compatible with large-scale interferometer implementations, this has enabled record performance in room temperature \cite{ma2026laserstabilizedroomtemperature} and also cryogenic reference cavities~\cite{wu2025birefringence}.

There are, however, open questions regarding other noise mechanisms that might prevent these coatings from reaching the Brownian-noise limit, in particular birefringence noise~\cite{AlGaAs_noise_2026} and charge-carrier generation-recombination noise~\cite{wu2025birefringence}; both mechanisms remain under active investigation.\footnote{While in some applications, birefringence noise can be mitigated by averaging over the two polarization eigenmodes.}
Recent results have shown the potential to modify the generation-recombination transfer function in order to reach the system Brownian noise floor, for example, through external LED illumination that saturates the process~\cite{ma2026laserstabilizedroomtemperature}. Although not readily compatible with large-scale interferometer implementations, this approach has enabled record performance in both room-temperature~\cite{ma2026laserstabilizedroomtemperature} and cryogenic reference cavities~\cite{wu2025birefringence}.

Some other changes would need to be made to the interferometer if AlGaAs coatings were used. The coatings would be grown on \qty{30}{\cm}-diameter wafers, and the usable optical aperture would be somewhat (up to \qty{1}{\cm}) smaller. This is not quite large enough for the current end-test-mass beam size of
\qty[round-mode=places, round-precision=1]
{\fpeval{200 * \Val{Aplus.wETM_m}}}{\cm}
diameter ($1/e^2$), so the test-mass curvatures would be changed to produce slightly smaller beams.
With GaAs/AlGaAs, the coating mechanical loss is low enough that the Brownian-noise target could be achieved even with somewhat smaller beam sizes, and the concomitant more stable cavity geometry would be a benefit in terms of reduced sensitivity to misalignments.
In this case, the curvature of some of the recycling-cavity optics would need to change to match the new arm-cavity mode.

In addition, with GaAs/AlGaAs coatings the Arm Length Stabilization (ALS) system~\cite{Mullavey_ALS}, used in the interferometer lock-acquisition process, would need to be modified. The Advanced LIGO ALS system uses \qty{532}{\nm} light, derived by frequency-doubling the main laser, to initially control the long arm cavities. However, since AlGaAs coatings are opaque to green light, a different ALS wavelength would be required. Options being investigated include the use of a \qty{2128}{\nm} beam generated via wavelength doubling from the main laser~\cite{Tanioka:24}, and the use of a \qty{1596}{\nm} auxiliary laser that is tied to the main laser by phase locking its third harmonic (at \qty{532}{\nm}) to the second harmonic of the main \qty{1064}{\nm} beam~\cite{Liu2026ALS}. 

\paragraph{Metamaterial coatings}
A different technology that has recently been explored for its low-noise potential is the meta-mirror concept~\cite{Dickmann:2023}. This approach uses a single subwavelength layer of periodic nanostructures to produce guided-mode resonances with engineered reflective properties~\cite{Wang:1993}. Such a metasurface could be used on its own, or in combination with a few traditional Bragg layers, to produce high reflectivity~\cite{PhysRevD.98.082003}. The thermal noise of this type of mirror would be significantly lower than that of a traditional quarter-wave Bragg stack because the amount of coating material would be reduced by more than an order of magnitude~\cite{Kroker:2017}. Much of this research has been directed at producing small mirrors operating at telecommunications-band wavelengths (where silicon can be used), and significant challenges remain in producing large metasurfaces for \qty{1064}{\nm} operation~\cite{Atikian:2022}.

\subsection{High-frequency upgrades}
\label{sec:hfupgrades}

\begin{figure}
  \centering
  \includegraphics[width=\textwidth]{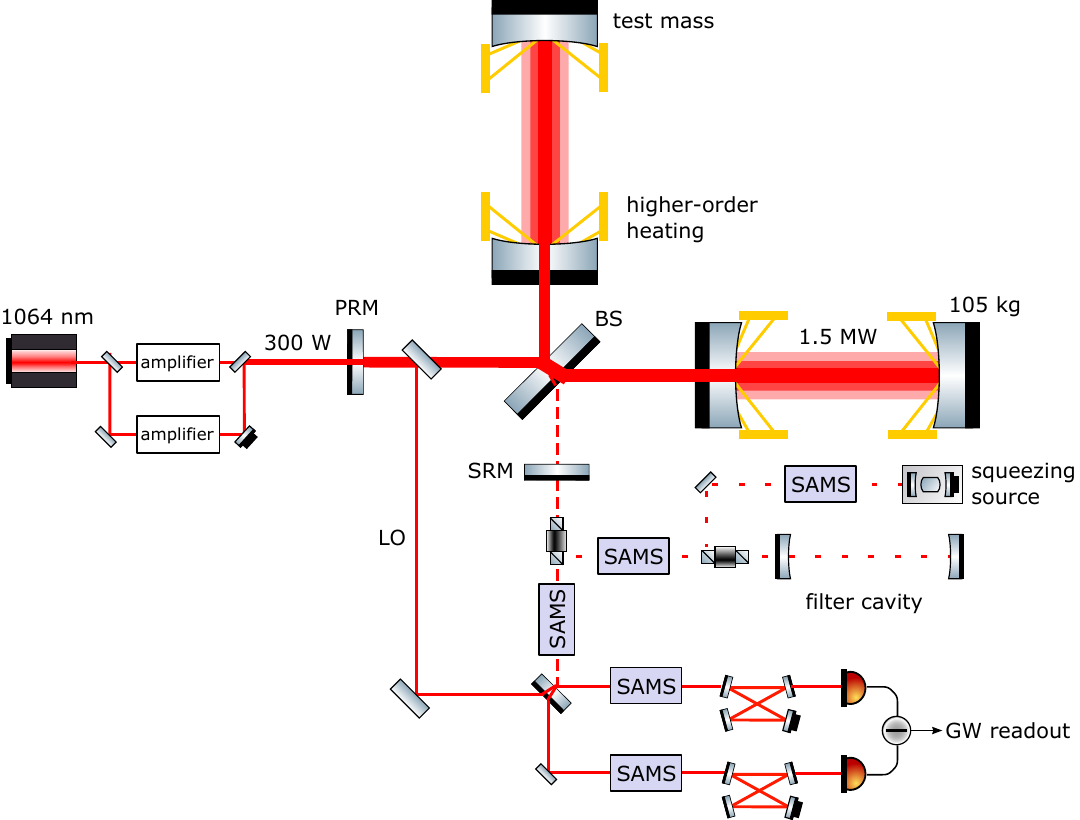}
\caption{Conceptual optical layout of the \Asharp{} upgrades, building on A+~\cite{AplusASD}, as discussed in \cref{sec:hfupgrades}. The design includes \qty{105}{\kg} test masses, increased arm-cavity power of \qty{1.5}{\MW}, an observed squeezing level of \qty{10}{\dB}, advanced test-mass thermal-deformation control and spatial modematching actuation. PRM: power-recycling mirror, BS: interferometer beamsplitter, LO: balanced homodyne local oscillator field, SAMS: suspended adaptive mode-matching stage - see~\cref{sub3sec:spatial_mm}, SRM: signal-recycling mirror, forming the signal extraction cavity (referred to in~\cref{subsec:wideband}) with the Michelson.}
  \label{fig:Asharp-layout}
\end{figure}

Further increases in high-frequency sensitivity can be achieved through a combination of higher operating optical power and an increase in effective squeezed vacuum enhancement level~\cite{MartynovPRD2019,Ganapathy2021}. The operating optical-power goal for \Asharp{} is to have \qty{1.5}{\MW} resonantly stored in each arm cavity, about four times the level achieved during O4~\cite{Capote2025PRD} and about three times the A+ design value~\cite{AplusASD}. 
The goal for squeezed light enhancement is set at \qty{10}{\dB} of broadband, effective squeezing, matching the target specified for the Cosmic Explorer detector design~\cite{Evans2021} and improving on the \qty{6}{\dB} realized during O4~\cite{Capote2025PRD} and the \qty{7}{\dB} assumed for A+~\cite{AplusASD}. Reaching these \Asharp{} baseline goals will require advances in optical and thermal technologies, which will also improve the interferometer's operational stability and sensitivity. These technologies are discussed in the following subsections and conceptually shown in \cref{fig:Asharp-layout}.

We conclude this section with a discussion of the signal-recycling configuration, comparing the standard broadband and wideband signal-recycling operations. This affects high-frequency sensitivity, trading increased performance above ${\sim}\qty{1}{\kHz}$ against decreased performance at low and mid frequencies.

\subsubsection{Increased optical power}
\label{subsec:optical_power}

Increasing the power circulating in each arm cavity to \qty{1.5}{\MW} will present a familiar but increasingly stringent set of high-power challenges: laser-power generation, angular instabilities, parametric instabilities, and thermal distortion.

% STARTING TEXT Increasing the power circulating in the arm cavities to \qty{1.5}{\MW} in each arm will present the usual suite of high power challenges: Siddles-Sigg angular instabilities; parametric instabilities; thermal distortions and their compensation. 

\paragraph{Increased laser power generation} 
\label{sub3sec:laser_power}

To achieve \qty{1.5}{\MW} of circulating power in the arm cavities, the input laser would need to operate at \qtyrange{300}{350}{\W}. Advanced LIGO used systems with sequential free-space amplifiers \cite{Bode2020OE}, which achieved an operating power of \qtyrange{125}{140}{\W} for O4. However, at this power, a higher-order mode content of nearly \qty{14}{\%} was observed; subsequent free-space amplification would be expected to further increase higher-order mode content. Solutions for reaching the required \Asharp{} laser power include fiber-based amplification and  coherent combination of separately amplified lasers \cite{Wellmann2021OE}.

%STARTING TEXT The laser power required to achieve \qty{1.5}{\MW} in the arms, assuming a power recycling gain of 50, is \qty{210}{\W} at the power recycling mirror input, which would require a source laser power of \qtyrange{300}{350}{\W}.  The current design approach of using sequential free-space amplifiers has achieved \qty{195}{\W}, though with a higher-order mode content of nearly \qty{14}{\%} that probably represents the limit of this approach (see \href{https://dcc.ligo.org/LIGO-G2000893}{LIGO-G2000893}). Higher power is possible with a fiber-amplifier, and/or coherent combining of separately-amplified laser beams (see \href{https://dcc.ligo.org/LIGO-G2100481}{LIGO-G2100481}).

\paragraph{Angular instability impacts}
\label{sub3sec:angular_instability}

In suspended optical cavities, angular motions of the mirrors cause the optical beams in these cavities to become offset from the optical axis. This leads to radiation pressure exerting a torque on the mirrors, inducing angular instabilities~\cite{SidlesSigg2006PRA, Hirose2010AO}. Increasing the mass and moment of inertia of the test masses will help control angular instabilities, as described in \cref{subsubsec:sus-control}. Given the expected increase in \Asharp{} test-mass dimensions, the moment of inertia will increase by a factor of 4.6. This increase in moment of inertia will compensate for the factor-of-two increase in optical torque relative to A+ due to the higher target circulating power in the arm cavities~\cite{AplusASD}.

% STARTING TEXT Increasing the mass and moment of inertia of the test masses will help control Siddles-Sigg angular instabilites. If the test mass dimensions are simply scaled by $(100/40)^{1/3}$, the moment of inertia will increase by a factor of 4.6; this will more than compensate for the factor of 2 increase in optical torque due to the power increase (compared to A+ design power)\,--\,at \qty{1.5}{\MW} the opto-mechanical angular dynamics will be similar to those of Advanced LIGO at \qty{330}{\kW} arm power.

\paragraph{Parametric instability mitigation}
\label{sub3sec:parametric_instability}
Parametric instabilities arise from the transfer of energy between the optical mode of the interferometer and a mechanical mode of an optic. This is due to the optomechanical interaction from radiation pressure, which leads to an unstable spatial modulation of the optical field by the excited mechanical mode~\cite{Braginsky2001PRA, Evans2010PRA, EvansEtAl2015PRL}. For \Asharp{}, parametric instabilities will have a greater impact, both from the higher circulating optical power and from the increased test-mass size, which increases the internal mechanical-mode density. Therefore, improved mitigation through targeted choice of test-mass radius of curvature and prioritized placement of refined passive acoustic mode dampers~\cite{Biscans2019PRD}, as well as advanced active independent optical and electrostatic damping techniques~\cite{BlairEtAl2017PRL,Bossilkov2024PRD}, will be deployed.

% STARTING TEXT The parametric instability (PI) situation will be worse both because of the factor of 2 increase in arm power, and because the larger test masses will have a higher density in frequency of internal modes of vibration. 
% The design and number of acoustic mode dampers will likely need to change in accordance. Active damping of PIs may be required, using the electro-static test mass actuators or possibly a radiation pressure actuator using an external laser beam. In Advanced LIGO, parametric instability mitigation was an add-on, designed after the O1 run. For \Asharp{}, accommodations for PI mitigation will be designed in at the beginning (e.g., the shape of the test mass could be chosen to minimize PI risk, and to mount acoustic mode dampers in more ideal locations).

\paragraph{Internal and Surface Wavefront Thermal Compensation}
\label{sub3sec:thermal_distortion}
The higher circulating optical power in each arm cavity will induce faster-onset and larger-magnitude thermal distortion effects. Without compensation, optical coating distortions and coating point-absorber impacts can limit the maximum stored power by up to \qty{50}{\%}~\cite{Brooks2021AO}, making the \Asharp{} target unobtainable. The need for further reductions in coating absorption is discussed above in \cref{sec:mfupgrades}, and for the remainder of the section, it is assumed that thermally-induced wavefront distortions are not dominated by point absorbers. Thermoelastic distortions of the test-mass surfaces and thermorefractive lensing in the input test-mass substrates induce spatial mode mismatch. This degrades the quantum-noise-limited strain sensitivity and effective squeezing improvement, as discussed further in \cref{sub3sec:spatial_mm}, reduces the buildup of signals used to control the interferometer, and increases technical-noise coupling.

Achieving the necessary distortion mitigation will require recently developed technologies. This will first involve improved sensing of the changing optical spatial modes within the interferometer cavities~\cite{MaganaSandoval19PRD, GoodwinJones23OE, GoodwinJones24Optica}. A combination of advances in the current test-mass ring heaters, which radially counteract the effects of thermal distortions~\cite{Brooks16AO}, new thermal-radiation shields around the test-mass barrels, and advances in laser-heating compensation systems will be used to partially correct the residual substrate lensing~\cite{Brooks16AO}. In addition, Central Heater for Transient Attenuation (CHETA) systems~\cite{JaberianHamedan_2018CHETA}, involving the pre-heating of the test masses to tune and maintain the test-mass ``hot'' operating state, will stabilize the test-mass surface distortions and substrate thermal lens, thus minimizing thermal transient distortions. Finally, new Front Surface Type Irradiator (FROSTI) systems~\cite{Tao2025PRL,Rosauer2025OPTICA}, which produce blackbody radiation shaped and imaged into complex spatial patterns onto the test-mass front surface, will enable direct compensation to the high-power-facing optical surface.

\subsubsection{Improved squeezed-light quantum noise reduction}
\label{subsec:quantum_noise}

Achieving the \Asharp{} target of \qty{10}{\dB} frequency-dependent broadband squeezing enhancement will require several improvements over current squeezing implementations~\cite{Evans2021}.

\paragraph{Squeezed light generation}
\label{sub3sec:squeezed_source}
Current squeezed-light sources used in GW detectors generate frequency-independent squeezed states from optical parametric oscillators (OPOs)~\cite{Oelker16Optica}, which are then reflected from \qty{300}{\m} optical filter cavities, producing a frequency-dependent rotation of the squeezing angle. This frequency-dependent squeezing is then injected into the dark port of the interferometers~\cite{Ganapathy23PRX, Capote2025PRD}. Current squeezed-light systems are capable of generating the levels of squeezing necessary to reach the \qty{10}{\dB} \Asharp{} target, with operational improvements needed for long-term robust performance.

% the squeezer crystal degrades over time necessitating the crystal being moved to recover this performance. Work is ongoing to reduce this degradation.

% Before detailing these improvements, the \Asharp{} baseline quantum noise budget is shown in \cref{fig:quantum_noise}, and the corresponding set of target parameters are summarized in \cref{tab:squeezing_requirements}.

%STARTING TEXT Approaching the \Asharp{} target of \qty{10}{\dB} broadband squeezing will require several improvements over the current detectors~\cite{Evans2021}. We consider as a baseline design using the current \qty{300}{\m} filter cavity in addition to increasing the arm power to \qty{1.5}{\mega\W}, increasing the generated squeezing, and improving losses and technical noises of all kinds.The quantum noise budget for this baseline is shown in \cref{fig:quantum_noise} and a set of parameters that would meet this target is summarized in \cref{tab:squeezing_requirements}. As a rule of thumb, since squeezing is limited to$10\log_{10}(\Lambda + 2\theta_\text{rms})$ decibels, the \qty{10}{\dB} goal is achieved with total effective losses of$\Lambda\sim\qty{8}{\%}$ and total effective phase noise of $\theta_\text{rms}\sim\qty{10}{\milli\radian}$.

\paragraph{Improved optical losses}
\label{sub3sec:optical_losses}
The optical detection loss budget for \Asharp{} is given in \cref{tab:squeezing_requirements}. The dominant optical losses are currently from the filter cavity, readout chain, and the injection path~\cite{Evans2021}. The effective losses in the OPO currently total to \qty{2}{\%} in LIGO O4 run, and only incremental improvements are needed to reach the \Asharp{} goal of \qty{1}{\%}. Four optical passes through Faraday isolators are currently required for squeezed-state injection, with an additional pass required for readout. Since backscatter noise is reduced through the use of BHD and with reduced optic motion, one Faraday isolator for the injection can be removed. Only minor improvements are needed to reach the goal of \qty{2.5}{\%} loss from the Faraday isolators, with the current low-loss isolators achieving \qtyrange{0.5}{1}{\%} loss per pass~\cite{Genin2018}. The installation of new low-loss output mode cleaners (OMCs) and minor improvements to the output steering optics are expected to reduce the readout loss from \qty{4.5}{\%} to \qty{3.5}{\%}. Readout photodiode quantum efficiency is expected to remain close to \qty{99}{\%} as in O4.

\paragraph{Improved spatial modematching}
\label{sub3sec:spatial_mm}
The mismatch between the spatial modes of the various optical cavities making up the detector is responsible for complex frequency-dependent squeezed-state degradations due, in part, to the coherent nature in which the higher-order modes excited by the mismatch interfere with themselves and with the fundamental mode~\cite{QRSS,DnA,DnD}. Suspended adaptive mode-matching stages (SAMS) are deformable mirrors used as telescopes to spatially match the OPO, OMC, filter cavity, and interferometer to each other~\cite{Srivastava2021,Cao2020}. The external modematching between these cavities will be improved through further optimization of these telescopes. The development of SAMS with reduced higher order aberrations may be needed to improve the beam quality. The mismatch between the optical cavities internal to the interferometer is mainly due to the thermal aberrations generated by power absorbed in the test-mass coatings, and will be controlled by the significant improvements to the thermal actuators described in \cref{sub3sec:thermal_distortion}.

\begin{table}
	\centering
	 \setlength{\tabcolsep}{10pt}
	\renewcommand{\arraystretch}{1.2}
	\begin{tabular}{r l r r}
		\hline
		Parameter & Units & {A+} & {\Asharp} \\
		\hline
		Arm power & \unit{\kilo\W} &
        % hard coding this because A+ is now O5c and I don't want to rewrite the
        % entire infrastructure just to fix this one number
        550 & % \fpeval{1e-3 * \Val{Aplus.ArmPower_W}} &
		\fpeval{1e-3 * \Val{Asharp.ArmPower_W}} \\
		Source squeezing level & \unit{\decibel} &
		\Val{Aplus.Squeezer.AmplitudedB} &
		\Val{Asharp.Squeezer.AmplitudedB} \\
		Injection loss & \unit{\%} &
		\fpeval{100 * \Val{Aplus.Squeezer.InjectionLoss}} &
		\fpeval{100 * \Val{Asharp.Squeezer.InjectionLoss}} \\
		Readout loss & \unit{\%} &
		\fpeval{100 * (1 - \Val{Aplus.Optics.PhotoDetectorEfficiency})} &
		\fpeval{100 * (1 - \Val{Asharp.Optics.PhotoDetectorEfficiency})} \\
		Roundtrip arm loss & \unit{\ppm} &
		\fpeval{2 * 1e6 * \Val{Aplus.Optics.Loss}} &
		\fpeval{2 * 1e6 * \Val{Asharp.Optics.Loss}} \\
		Roundtrip SEC loss & \unit{\ppm} &
		\fpeval{1e6 * \Val{Aplus.Optics.BSLoss}} &
		\fpeval{1e6 * \Val{Asharp.Optics.BSLoss}} \\
        Roundtrip FC loss & \unit{\ppm} &
		% \fpeval{1e6 * \Val{Aplus.Squeezer.FilterCavity.Lrt}} &
        100 &  % hard coding for the same reason
		\fpeval{1e6 * \Val{Asharp.Squeezer.FilterCavity.Lrt}} \\
		RMS phase noise & \unit{\milli\radian} &
		\fpeval{1e3 * \Val{Aplus.Squeezer.SQZAngleRMS}} &
		\fpeval{1e3 * \Val{Asharp.Squeezer.SQZAngleRMS}} \\
		\hline
	\end{tabular}
	\caption{Parameters used for quantum-noise projections for A+ and \Asharp{}. SEC: signal extraction cavity; FC: filter cavity. The source squeezing level is the idealized, lossless squeezing generated before encountering any losses. The injection loss is the total of OPO effective loss and the propagation-path loss between the OPO and the interferometer output Faraday isolator. The readout loss is the total interferometer output-path loss, including output mode cleaner and photodetector losses. All loss is the total equivalent loss along a path including both optical loss and the coherently summed mode mismatch loss.}
	\label{tab:squeezing_requirements}
\end{table}

\subsubsection{Interferometer sensitivity tuning}
\label{subsec:wideband}

\begin{figure}[h]
	\centering
	\includegraphics[width=0.85\textwidth]{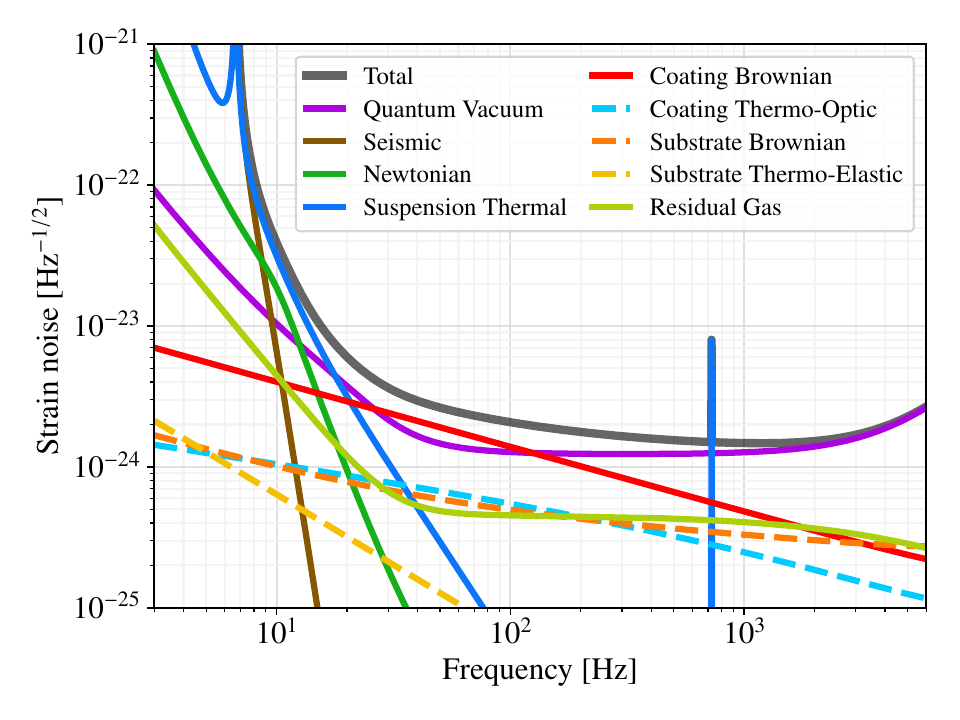}
	\caption{\Asharp{} wideband noise budget. Compared to the \Asharp{} baseline design, the signal recycling mirror (SRM) transmission is reduced to 5\% (from 32.5\%), and the filter cavity finesse and detuning is re-optimized.}
	\label{fig:Asharp-wideband-nb}
\end{figure}

The high-frequency sensitivity can also be improved by increasing the signal extraction mirror reflectivity in order to broaden the bandwidth of the interferometer~\cite{Ganapathy2021}. This \Asharp{} wideband configuration is shown by the purple trace in \cref{fig:postO5-overview} (see noise budget in \cref{fig:Asharp-wideband-nb}). In reference~\cite{Ganapathy2021} this configuration is compared with a narrowband configuration in which the signal extraction cavity (SEC) is also detuned from resonance. Although such additional detuning can lead to higher sensitivity within a narrow window in the high-frequency region, the nominal configuration is preferred because its noise reduction over a broader range of frequencies provides the best overall astrophysical performance, while also benefiting from simpler control requirements.

Relative to the \Asharp{} baseline configuration, there are several trade-offs associated with using the wideband configuration. First, as seen in \cref{fig:postO5-overview}, the mid-frequency sensitivity is significantly reduced, leading to a decrease in astrophysical signal detection range. Second, the filter-cavity finesse must be increased in order to match the lower-frequency crossover of the quantum noise~\cite{Ganapathy2021}, which makes the detector more susceptible to filter-cavity noise~\cite{QRSS,DnD}. Third, the detector is more susceptible to mode mismatch and optical loss in the SEC, since this loss is not broadened along with the other quantum-noise contributions~\cite{Miao2019PRX,DnA}.

Changing from the \Asharp{} baseline configuration to the wideband configuration, or vice versa, would require physically switching optics with different reflectivities in the SEC and filter cavity. This physical switching would result in significant detector downtime. One consideration could thus be to designate one detector for wideband sensitivity, while the other maximizes astrophysical detection reach.

% STARTING TEXT The high frequency sensitivity can also be improved by altering the frequency response of the interferometer. Increasing the high frequency sensitivity in this way does in general give up sensitivity at mid-frequencies. Therefore, one of these options is probably most attractive as an upgrade to a single interferometer in a network where there are at least two (three?) other interferometers providing maximum detection range.

% STARTING TEXT The \Asharp{} wideband configuration is shown in \cref{fig:postO5-overview} (purple). The simplest way to do this is to widen the bandwidth by increasing the level of signal extraction (using a higher reflectivity signal extraction mirror). If injecting squeezing using a filter cavity, its finesse also needs to be increased to manage quantum radiation pressure noise. In Ref.~\cite{Ganapathy2021} this simple wideband configuration is compared to a narrowband configuration where the signal extraction is detuned in addition to being increased. Although such detuning can lead to lower strain noise over a narrow band of frequencies, Ref.~\cite{Ganapathy2021} concludes that the wideband configuration is preferred given its relative simplicity and broader noise reduction. The losses in the SEC must be reasonably low for this to be effective since this loss is not broadened along with the other noises; see \cref{fig:Asharp-wideband-quantum-nb}.

\section{Astrophysical prospects}
\label{sec:science}
Since the first detections by the current GW detector network \cite{LIGOScientific:2014pky,VIRGO:2014yos,KAGRA:2020tym}, successive improvements in instrumental sensitivity have driven rapid growth in the observed populations of compact binary mergers \cite{GWTC-4,GWTC-5}.
%The detector upgrades described in \cref{sec:upgrades} enhance the science reach of \Asharp{} in different ways across the observing band. 
The \Asharp{} upgrade continues this progression by improving sensitivity across the observing band, with different frequency regions contributing to different science goals.
At low frequencies, reduced technical noise and Newtonian noise lead to an increase in the duration over which compact-binary signals remain in band, and improve the sensitivity to higher-mass and higher-redshift systems whose signals enter the detector band at low frequencies.
At mid-frequencies, reduced coating thermal noise increases the accumulated signal-to-noise ratio (SNR) for a broad range of compact-binary sources, and therefore has a direct impact on detection rates, parameter estimation, and population inference.  
At high frequencies, increased arm-cavity power, improved squeezing, and possible changes to the signal-recycling configuration improve the sensitivity to tidal effects, kHz ringdown signals from lower-mass systems, and prospects of detecting continuous waves from millisecond pulsars.

The science projections presented below are intended to quantify these connections for the detector configurations introduced in \cref{sec:upgrades}.  They do not constitute a complete set of metrics for optimizing the detector design, but are instead selected examples illustrating how the broadband sensitivity improvements of \Asharp{} translate into astrophysical measurements across different source classes and frequency bands. 
%At the same time, \Asharp{} remains an upgrade within the existing LIGO facilities, and some science targets will remain limited by sensitivity, arm length, or infrastructure until next-generation observatories become available.

We base the compact-binary projections on the current understanding of the astrophysical source population inferred after the first part of the fourth observing run (O4a)~\cite{GWTC-4}.  
We first consider source classes that have already been observed, using current empirical constraints on their rates, parameter distributions, and representative source properties to assess the science outcomes expected for the \Asharp{} baseline design and for the alternative wideband and optimistic configurations considered in this work.
We then discuss source classes and searches for which \Asharp{} improves the discovery potential, including continuous waves, 
%bursts, 
stochastic backgrounds, and probes of fundamental physics.

In \cref{subsec:rates}, we compare the mass-dependent horizon redshifts, present the plausible range of detectable event catalog sizes, and assess the early-warning and localization capabilities for BNSs.
In \cref{subsec:BH}, we investigate the sensitivity of each configuration to BBH population inference, higher-order modes, intermediate-mass BH binaries, and BH ringdown signals.
\Cref{subsec:ns} evaluates the prospects for NS science, including constraints on the cold nuclear equation of state, sensitivity to post-merger signals and the hot equation of state, and continuous-wave (CW) signals.
%\todo{In \cref{subsec:burst}, we discuss burst searches, using tests of the accretion-disk-instability model as a representative science case.}
We also consider the prospects for detecting stochastic GW backgrounds from compact binary mergers in \cref{subsec:background}, and for probing Beyond-Standard-Model particles and dark matter in \cref{subsec:dm}.

%%%%%%%%%%%%%%%%%%%%%%%%%%%%%%%%%
\subsection{Compact binary detection prospects}
\label{subsec:rates}
%%%%%%%%%%%%%%%%%%%%%%%%%%%%%%%%%

Building on the population-inference results and merger rates reported in GWTC-4~\cite{GWTC-4}, we quantify how the improved sensitivity of the \Asharp{} configurations translates into observational outcomes: (i) increased reach in redshift and surveyed spacetime volume (\cref{subsubsec:horizons}), (ii) the resulting growth in the size of detected event catalogs  (\cref{subsubsec:catalog}) , (iii) enhanced early-warning capabilities for BNS systems (\cref{subsubsec:early}), and (iv) improved BNS sky localization (\cref{subsubsec:bns_loc}). 

\subsubsection{Detection horizons and ranges}
\label{subsubsec:horizons}

\begin{figure}[h!]
	\centering
	\includegraphics[width=0.8\textwidth]{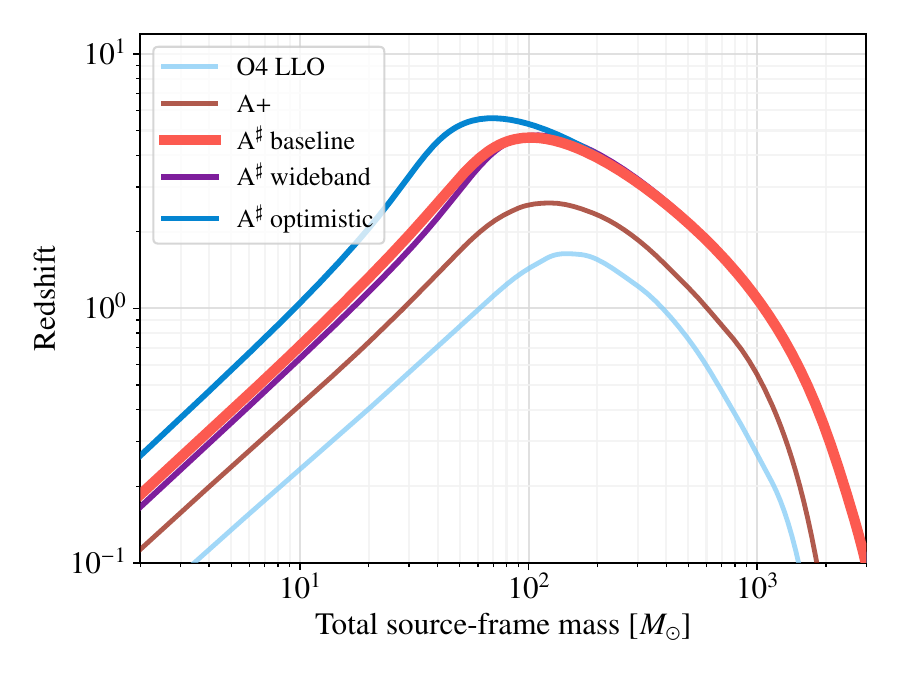}
	\caption[Detectability of equal mass black hole binaries for A+ and post-O5 detectors]
	{Horizon redshifts corresponding to an SNR of 8 for face-on, optimally located, non-spinning, equal-mass binaries, for the detector configurations shown in \cref{fig:postO5-overview}.}
	\label{fig:horizons}
\end{figure}

To assess the relative sensitivity of the detector configurations to compact binary mergers, we first consider the horizon, defined as the redshift at which a face-on, optimally located (overhead), non-spinning, equal-mass binary produces a SNR equal to a detection threshold of $\rho=8$ in a single detector. This metric provides a convenient upper bound on the detector reach. We compute the horizon redshift as a function of source-frame total mass, with the results shown in \cref{fig:horizons}.\footnote{The horizon-redshift calculation assumes a flat $\Lambda$CDM cosmology with $H_0 = \qty{67.9}{\kilo\m\per\second\per\mega\pc}$ and $\Omega_m = 0.3065$~\cite{Planck:2015fie}.} The mass dependence of the horizon reflects the interplay between the detector noise curve and the characteristic frequency evolution of the binary waveform. 

While horizon distances characterize the furthest detectable sources, the effects of cosmological volume evolution and the non-uniform detector antenna response make it non-trivial to translate horizon improvements into changes in detection rates. 
We therefore compute the sky-averaged and inclination/orientation-averaged spacetime (comoving) volume $V$ surveyed per unit detector time.
For a population with a constant comoving source-frame merger rate density, multiplying $V$ by that rate density provides the expected detection rate~\cite{Chen_2021}. 
We then also report estimates of the range, $R$, defined via $(4\pi/3)R^3 = V$.
\Cref{tab:astro_metrics} lists the ranges for fiducial BNS ($1.4{+}1.4\,M_\odot$) and BBH ($30{+}30\,M_\odot$) systems, as well as the maximum redshift at which an equal-mass, non-spinning binary of \emph{any} total mass is detectable.

\begin{table}[h]
  \centering
  \sisetup{table-alignment-mode=none, table-number-alignment=center}
  \setlength{\tabcolsep}{10pt}
  \renewcommand{\arraystretch}{1.2}
  \begin{tabular}{
      r
      S[round-mode=figures, round-precision=2, exponent-mode=fixed, fixed-exponent=0]
      S[round-mode=figures, round-precision=2, exponent-mode=fixed, fixed-exponent=0]
      S[round-mode=places, round-precision=1]
      S[round-mode=places, round-precision=1]
      S[round-mode=places, round-precision=1]
      S[round-mode=places, round-precision=1]
  }
  \hline
   & \multicolumn{2}{c}{Range [\unit{\mega\pc}]} & &  \\
    Configuration & BNS & BBH & $z_\text{max}$ \\
  \hline
  O4 &
      \Val{O4LLO.bns_range_Mpc} &
      \Val{O4LLO.bbh_range_Mpc} &
      \Val{O4LLO.max_redshift} \\
  A+ &
      \Val{O5c.bns_range_Mpc} &
      \Val{O5c.bbh_range_Mpc} &
      \Val{O5c.max_redshift}  \\
  \Asharp{} baseline &
      \Val{Asharp.bns_range_Mpc} &
      \Val{Asharp.bbh_range_Mpc} &
      \Val{Asharp.max_redshift}  \\
  \Asharp{} wideband &
      \Val{Asharp_wideband.bns_range_Mpc} &
      \Val{Asharp_wideband.bbh_range_Mpc} &
      \Val{Asharp_wideband.max_redshift} \\
  \Asharp{} optimistic &
      \Val{Asharp_AlGaAs.bns_range_Mpc} &
      \Val{Asharp_AlGaAs.bbh_range_Mpc} &
      \Val{Asharp_AlGaAs.max_redshift} \\
  \hline
  \end{tabular}
  \caption{Astrophysical performance of the detector configurations shown in \cref{fig:postO5-overview} for equal-mass, non-spinning compact binaries. 
  The BNS and BBH range estimates correspond to $1.4{+}1.4\,M_\odot$ and $30{+}30\,M_\odot$ systems, respectively. 
  The maximum detectable redshift $z_\text{max}$ corresponds to the peak in \cref{fig:horizons}.}
  \label{tab:astro_metrics}
\end{table}

Overall, \Asharp{} substantially extends the compact-binary reach relative to A+. 
For the fiducial systems considered here, the \Asharp{} baseline configuration increases the BNS range from \qtyrange{280}{440}{\mega\pc} and the BBH range from \qtyrange{2.2}{3.0}{\giga\pc}. 
It also increases the maximum horizon redshift for equal-mass binaries from 2.6 to 4.7, with further gains possible for the optimistic configuration with lower coating noise.
These gains translate directly into larger compact-binary samples extending to higher redshift, improving the prospects for population studies, rare-event discoveries, and precision measurements of individual mergers.

%%%%%%%%%%%%%%%%%%%%%%%%%%%%%%%%%
\subsubsection{Catalog size}
\label{subsubsec:catalog}

The latest population analyses based on observations up to O4a constrain the local astrophysical merger rates of BNS, NSBH, and BBH systems to \num{\ValInt{O4LLO.BNS.merger_rate_density.mid}(\ValInt{O4LLO.BNS.merger_rate_density.hi}:\ValInt{O4LLO.BNS.merger_rate_density.lo})}, \num{\ValInt{O4LLO.NSBH.merger_rate_density.mid}(\ValInt{O4LLO.NSBH.merger_rate_density.hi}:\ValInt{O4LLO.NSBH.merger_rate_density.lo})}, and \qty{\ValInt{O4LLO.BBH.merger_rate_density.mid}(\ValInt{O4LLO.BBH.merger_rate_density.hi}:\ValInt{O4LLO.BBH.merger_rate_density.lo})}{\per\giga\pc\cubed\per\yr}, respectively~\cite{GWTC-4_pop}, where the quoted values correspond to the median and central \qty{90}{\%} credible intervals.
These rates differ from those adopted in earlier observing-scenario studies \cite{evans2023,Branchesi:2023mws}, which were based primarily on O3-era population constraints and therefore reflected a different event catalog and the rate uncertainties available at that time.
The updated O4a population analysis changes both the median rates and their credible intervals, especially for source classes with small numbers of confident detections, namely BNS and NSBH events.
%and these changes propagate directly into the expected catalog sizes quoted below.

%By assuming an astrophysical source distribution and detector duty cycle, we estimate
In \cref{tab:obs_scn}, we present estimates of the number of binaries of each class detectable per year with each detector configuration.
Following an approach similar to that of \cite{KAGRA:2013rdx}, we adopt a simplified astrophysical population model: BBHs are assumed to be equal-mass systems with component masses drawn from a power-law distribution with index $-3$; NSs are assumed to have masses of \qty{1.4}{\solarmass}; and merger rates are taken to evolve proportionally to the Madau--Dickinson star formation rate as given in \cite{Madau:2014bja}.
In this simplified model, we neglect any time delay between binary formation and merger.
For all configurations, we consider a LIGO Hanford--LIGO Livingston network with single-observatory duty cycles of \qty{80}{\%}, and adopt a network SNR threshold of $8$ for detection.\footnote{The adopted SNR threshold is relatively optimistic for BNS and NSBH searches, and should be regarded as one source of systematic uncertainty in the rate forecasts.}

\begin{table}[h]
  \centering
  \begingroup
	\let\OriginalVal\Val
	\renewcommand{\Val}[1]{%
		\expandafter\num\expandafter[
		round-mode=figures,
		round-precision=2
		]{\OriginalVal{#1}}%
	}
  \setlength{\tabcolsep}{10pt}
  \renewcommand{\arraystretch}{1.2}
  \begin{tabular}{rccc}
  \hline
  & \multicolumn{3}{c}{Annual Detections} \\
  Configuration & BNS & NSBH &  BBH \\
  \hline
    A+ &
    $\Val{O5c.BNS.rate.mid}_{-\Val{O5c.BNS.rate.lo}}^{+\Val{O5c.BNS.rate.hi}}$ &
    $\Val{O5c.NSBH.rate.mid}_{-\Val{O5c.NSBH.rate.lo}}^{+\Val{O5c.NSBH.rate.hi}}$ &
    $\Val{O5c.BBH.rate.mid}_{-\Val{O5c.BBH.rate.lo}}^{+\Val{O5c.BBH.rate.hi}}$ \\
    \Asharp{} baseline & $\Val{Asharp.BNS.rate.mid}_{-\Val{Asharp.BNS.rate.lo}}^{+\Val{Asharp.BNS.rate.hi}}$ & $\Val{Asharp.NSBH.rate.mid}_{-\Val{Asharp.NSBH.rate.lo}}^{+\Val{Asharp.NSBH.rate.hi}}$ & $\Val{Asharp.BBH.rate.mid}_{-\Val{Asharp.BBH.rate.lo}}^{+\Val{Asharp.BBH.rate.hi}}$ \\
     \Asharp{} wideband & $\Val{Asharp_wideband.BNS.rate.mid}_{-\Val{Asharp_wideband.BNS.rate.lo}}^{+\Val{Asharp_wideband.BNS.rate.hi}}$ & $\Val{Asharp_wideband.NSBH.rate.mid}_{-\Val{Asharp_wideband.NSBH.rate.lo}}^{+\Val{Asharp_wideband.NSBH.rate.hi}}$ &  $\Val{Asharp_wideband.BBH.rate.mid}_{-\Val{Asharp_wideband.BBH.rate.lo}}^{+\Val{Asharp_wideband.BBH.rate.hi}}$ \\
    \Asharp{} optimistic &
    $\Val{Asharp_AlGaAs.BNS.rate.mid}_{-\Val{Asharp_AlGaAs.BNS.rate.lo}}^{+\Val{Asharp_AlGaAs.BNS.rate.hi}}$ &
    $\Val{Asharp_AlGaAs.NSBH.rate.mid}_{-\Val{Asharp_AlGaAs.NSBH.rate.lo}}^{+\Val{Asharp_AlGaAs.NSBH.rate.hi}}$ &
    $\Val{Asharp_AlGaAs.BBH.rate.mid}_{-\Val{Asharp_AlGaAs.BBH.rate.lo}}^{+\Val{Asharp_AlGaAs.BBH.rate.hi}}$ \\
  \hline
  \end{tabular}
  \caption{Plausible ranges of the expected number of detections in a calendar-year observing run for each binary class, based on the central \qty{90}{\%} credible intervals of the astrophysical merger rates reported in GWTC-4~\cite{GWTC-4}.}
  \label{tab:obs_scn}
  \endgroup
\end{table}

%\Cref{tab:obs_scn} shows the estimated number of annual detection for each class, based on the current median and central \qty{90}{\%} credible intervals of their astrophysical merger rates. 
Relative to A+, the \Asharp{} baseline configuration increases the expected annual detections by factors of several across all compact-binary classes, yielding catalogs with tens to hundreds of BNS and NSBH events and several thousands of BBH events per year. 
Catalogs of this size would enable detailed studies of the underlying astrophysical populations, including the inference of mass, spin, and redshift distributions.

With catalogs containing several $10^3$ BBH events, we expect to constrain the redshift-dependent Hubble expansion rate, $H(z)$, at the few-percent level through measurements of the luminosity-distance--redshift relation, provided that a characteristic mass scale is imprinted in the BBH mass distribution, e.g., through the pair-instability supernova process~\cite{Farr:2019twy}. 
While standard-siren measurements with BNS mergers that have identified electromagnetic counterparts could ultimately provide tighter constraints, the number of such events remains uncertain and will represent only a subset of the total BNS detection rate. For dark-siren analyses, the achievable precision depends sensitively on source localization and the completeness of galaxy catalogs, both of which remain difficult to predict reliably at present.

%%%%%%%%%%
\subsubsection{Early warning for binary neutron stars}
\label{subsubsec:early}

%As the sensitivity and bandwidth of the detectors improve, BNS signals can be observed sufficiently early in their inspiral to enable approximate localization \emph{prior to} merger. This could allow electromagnetic observatories to observe the localization region before the source becomes electromagnetically bright. \cref{tab:astro_metrics} shows, for each configuration, the time prior to merger at which a $1.4{+}1.4\,M_\odot$ system at $z=\num{\Val{Asharp.z_early}}$ accumulates sufficient SNR to exceed a detection threshold of 8.

Early warning of BNS mergers is one of the most direct science drivers for improved low-frequency sensitivity. Because BNS signals remain in band for a much longer time before merger than BBH signals, lowering the detector noise at tens of hertz allows the SNR to accumulate earlier in the inspiral. If a detection threshold can be reached before merger, a low-latency alert can be issued while the binary is still evolving, allowing electromagnetic facilities to begin observations before, or close to, the onset of prompt emission, early afterglow, or kilonova emission. 
This is particularly important for observations of short gamma-ray bursts and early kilonova emission, as well as for wide-field facilities that require time to tile a localization region
~\cite{Gehrels:2004Swift,Meegan:2009GBM,SoaresSantos:2017lru,Bellm:2018sui,Thomas:2020rubin}.

We quantify this capability by computing, for each detector configuration, the time before merger at which the cumulative SNR in a given detector reaches a threshold of 8 for a $1.4{+}1.4\,M_\odot$ BNS at $z=\num{\Val{Asharp.z_early}}$ (${\sim}\qty{130}{\mega\pc}$).\footnote{This distance approximately corresponds to the expected distance to the nearest BNS merger in one year, based on the merger-rate density inferred from O4a observations~\cite{GWTC-4_pop}.}
This metric is especially sensitive to the low-frequency noise level. In this metric, the improvement provided by \Asharp{} is substantial: the warning time is only about \qty{30}{s} for the representative O4 sensitivity and \qty{1.6}{min} for A+, but increases to approximately \sisetup{range-phrase = {--}}\qtyrange{6}{7}{\min} for the considered \Asharp{} configurations. The similar performance across the three \Asharp{} configurations indicates that this observable is driven primarily by the low-frequency design rather than by the mid- or high-frequency improvements.

These warning times should be interpreted as an idealized single-source metric: the practical utility of an alert also depends on the detector network, duty cycle, sky localization, low-latency data quality, and alert-generation pipelines.  However, the comparison demonstrates that the low-frequency upgrades in \Asharp{} can move BNS detections from sub-minute or minute-scale warning to several minutes of warning for nearby events of the type considered here. This improves the prospects for coordinated multi-messenger observations and provides a direct example of how low-frequency technical improvements translate into multi-messenger time-domain astronomy.

%\begin{table}[h]
%	\centering
%	\sisetup{table-alignment-mode=none, table-number-alignment=center}
%	\begin{tabular}{
%			r
%			S[round-mode=places, round-precision=1]
%		}
%		\hline
%		Configuration & \multicolumn{1}{c}{Early-warning time [\unit{\minute}]}  \\
%		\hline
%		O4 &
%		\fpeval{\Val{O4LLO.early_warning_s} / 60} \\
%		A+ (O5c) &
%		\fpeval{\Val{O5c.early_warning_s} / 60} \\
%		\Asharp{} &
%		\fpeval{\Val{Asharp.early_warning_s} / 60} \\
%		\Asharp{} wideband &
%		\fpeval{\Val{Asharp_wideband.early_warning_s} / 60} \\
%		\Asharp{} optimistic &
%		\fpeval{\Val{Asharp_AlGaAs.early_warning_s} / 60} \\
%		\hline
%	\end{tabular}
%	\caption{Early-warning time for BNS signals for different detector configurations. 
%		The early-warning time is defined as the time before merger at which the cumulative SNR reaches 8 for a $1.4{+}1.4\,M_\odot$ system at $z=\num{\Val{Asharp.z_early}}$. 
%		This metric is sensitive to low-frequency noise, as illustrated by the comparison between the \Asharp{} wideband and baseline \Asharp{} configurations: the latter has slightly higher low-frequency noise despite its larger BNS range.}
%	\label{tab:early_warning}
%\end{table}

%%%%%%%%%%%%%%%%%%%%%%%%%%%%%%%%%%%%

\subsubsection{Localization of binary neutron stars}
\label{subsubsec:bns_loc}

To assess the impact of the \Asharp{} sensitivity improvements on BNS localization, we simulate a one-year population of compact binary mergers based on the population models inferred from GWTC-4.
We use the \textsc{FullPop-4.0} mass model~\cite{GWTC-4_pop}, adopt the Madau--Dickinson star-formation model~\cite{Madau:2014bja} convolved with an inverse-time-delay model for redshift, and use a local merger-rate density of \qty{130}{\per\giga\pc\cubed\per\yr} for the full population~\cite{GWTC-4_pop}.
Sources are simulated up to a redshift of $z=0.5$. 
From these sources, we identify binaries with both component masses below \qty{2.06}{\solarmass} as BNS mergers and localize them using timing-based triangulation~\cite{Fairhurst:2017mvj, Fairhurst:2023idl}. 

\begin{figure}[h]
    \centering
    \includegraphics[width=0.49\linewidth]{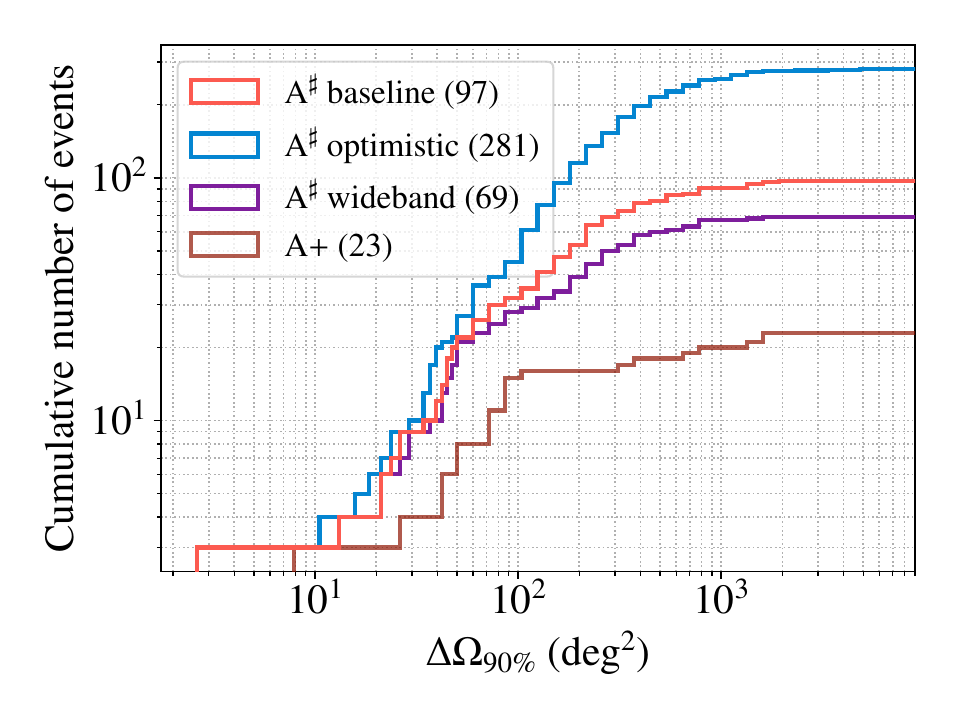}
    \includegraphics[width=0.49\linewidth]{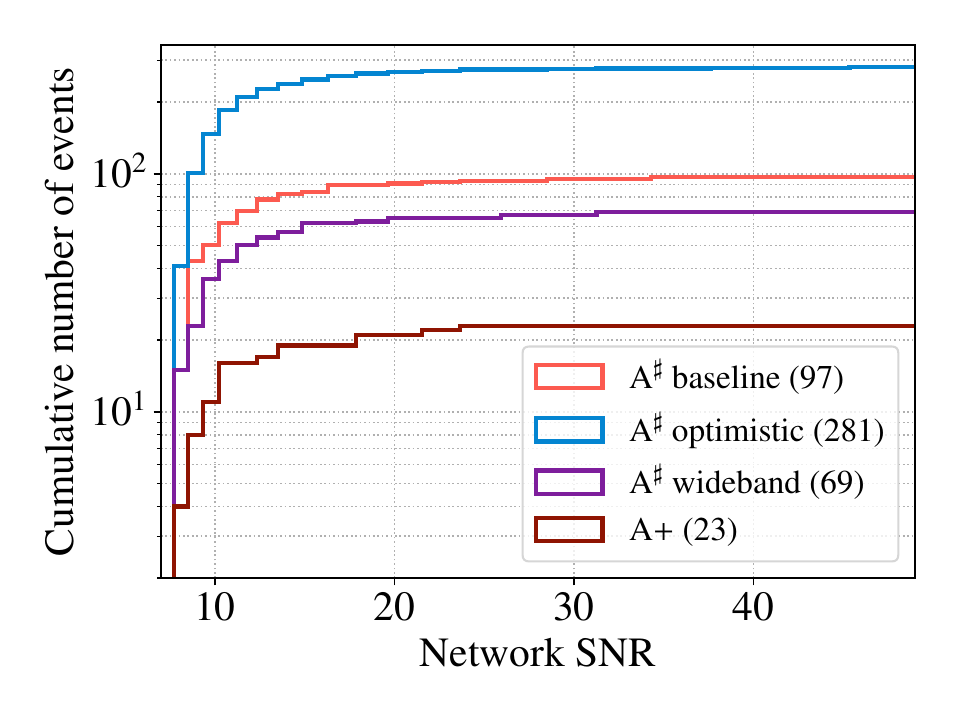}
    \caption{Cumulative numbers of BNS events as a function of localization area (left) and network SNR (right) detected by the \Asharp{} baseline, \Asharp{} wideband, \Asharp{} optimistic, and A+ networks. The simulated population extends to $z=0.5$ and follows the Madau--Dickinson redshift model with an inverse-time-delay function. 
    The numbers in parentheses indicate the number of detected events for each network.
    }
    \label{fig:bns_loc:loc}
\end{figure}

For our analysis, we use a three-detector network consisting of LIGO Hanford, LIGO Livingston, and Virgo.
For the \Asharp{} networks, the two LIGO detectors are modeled to operate at the \Asharp{} baseline, \Asharp{} wideband, or \Asharp{} optimistic sensitivities shown in \cref{fig:postO5-overview}; for the A+ network, they are assumed to operate at the target A+ sensitivity~\cite{AplusASD}. 
In all cases, Virgo is assumed to operate at its O5 target sensitivity~\cite{VirgoPSDs,acernese2026advancedvirgopluso5}.\footnote{We use \textit{O5Stage1HighSensPSD.txt} in this work.} 
We consider an event detected if it has an SNR above 3 in at least two of the three detectors and a network SNR of at least 8. Out of 2910 simulated BNS events, the largest number is detected by the \Asharp{} optimistic network.

\Cref{fig:bns_loc:loc} shows the cumulative number of events as a function of localization area (left) and network SNR (right) for the different detector networks. The numbers in parentheses indicate the number of detected events for each network. 
%For the A+ network, we find that, on average, ${\sim}\qty{13}{\%}$ of detected events are localized to better than \qty{20}{\deg\squared}. However, because the overall detection efficiency of A+ is relatively small, corresponding to ${\sim}\qty{0.7}{\%}$ of the simulated population, which yields only $\mathcal{O}(3)$ events localized to ${<}\qty{20}{\deg\squared}$ per year. 
%For the \Asharp{} networks, the detection efficiencies increase to  \sisetup{range-phrase = {--}}\qtyrange{2.4}{9.6}{\%}, with \sisetup{range-phrase = {--}}\qtyrange{2}{6}{\%} of detected events localized to ${<}\qty{20}{\deg\squared}$, corresponding to $\mathcal{O}(4$--$6)$ such events per year. 
For the A+ network, we find that $\mathcal{O}(14)$ events per year are localized to better than \qty{100}{\deg\squared}. The higher detection rates of the \Asharp{} networks increase this to $\mathcal{O}(26$--$43)$ events per year.
For reference, GW170817 was localized to ${\sim}\qty{28}{\deg\squared}$~\cite{LIGOScientific:2017vwq}, and no subsequent BNS event in O1--O4a has been localized to a comparable area. Localization to better than \qty{100}{\deg\squared}, combined with the improved early-warning capability of \Asharp{}, would enable more rapid electromagnetic follow-up and strengthen the prospects for multi-messenger observations~\cite{thomas2020vera}.

%%%%%%%%%%%%%%%%%%%%%%%%%%%%%%%%%
\subsection{Binary black hole science}
\label{subsec:BH}
%%%%%%%%%%%%%%%%%%%%%%%%%%%%%%%%%

Building on the compact-binary reach discussed in \cref{subsec:rates}, we assess how the improved sensitivity of the \Asharp{} configurations enhances BBH science beyond increased detection numbers: (i) population inference across cosmic time (\cref{subsubsec:pop_inf_bbh}), (ii) measurements of higher-order modes (\cref{subsubsec:HM}), (iii) observations of intermediate-mass BH binaries (\cref{subsubsec:imbh}), and (iv) pure-ringdown measurements and BH spectroscopy (\cref{subsubsec:ringdown}).

\subsubsection{Population inference for stellar-mass binary black holes}
\label{subsubsec:pop_inf_bbh}

With the \Asharp{} network, it becomes possible to identify the peak of the BBH merger-rate evolution and constrain both its redshift and the overall shape of the redshift distribution (cf. the simplified illustration in \cref{fig:horizon-donut}). This is difficult with current GW observations, because existing detector networks do not provide enough high-SNR detections beyond $z \sim 1.2$. The improved reach of \Asharp{} therefore opens the possibility of using BBH observations to complement electromagnetic constraints from galaxy surveys and further our understanding of the cosmic star-formation history.

\begin{figure}[ht]
    \centering
    \includegraphics[width=0.6\linewidth]{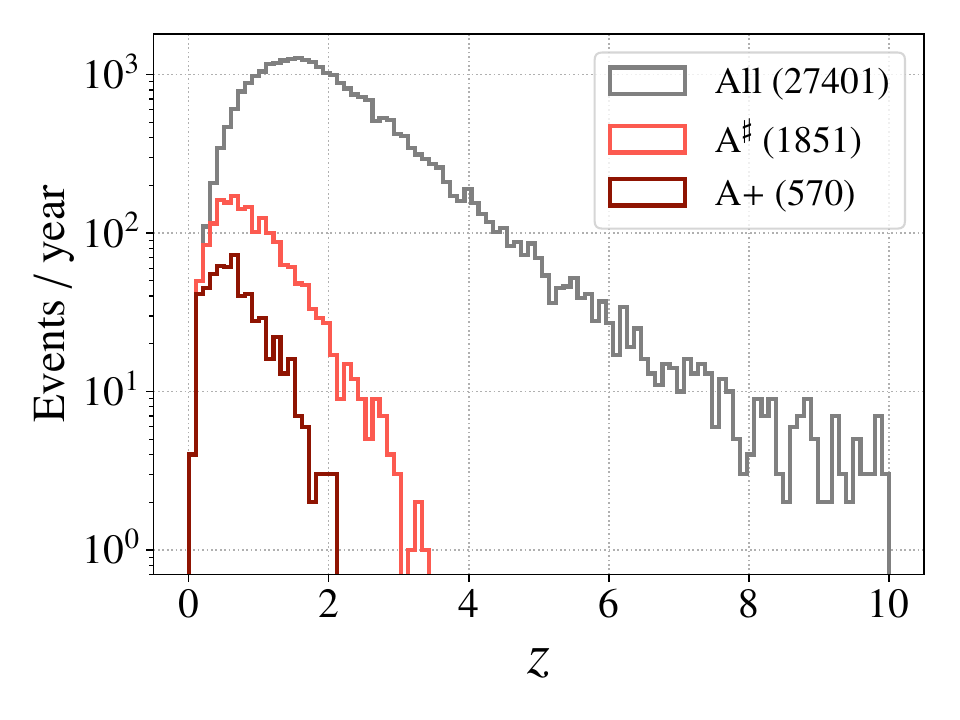}
    \caption{Input population used for population inference. The grey histogram shows all simulated BBH systems for one year of observing time up to redshift $z=10$. The brown and red histograms show the detected events for the A+ and \Asharp{} baseline networks, respectively, using a network optimal SNR threshold of $\rho_{\rm opt} \geq 10$. The numbers in parentheses indicate the total number of events in the full simulated population and in the detected samples for each network.}   
    \label{fig:pop_inference:input}
\end{figure}

\begin{figure}[ht]
\centering
\includegraphics[width=0.65\linewidth]{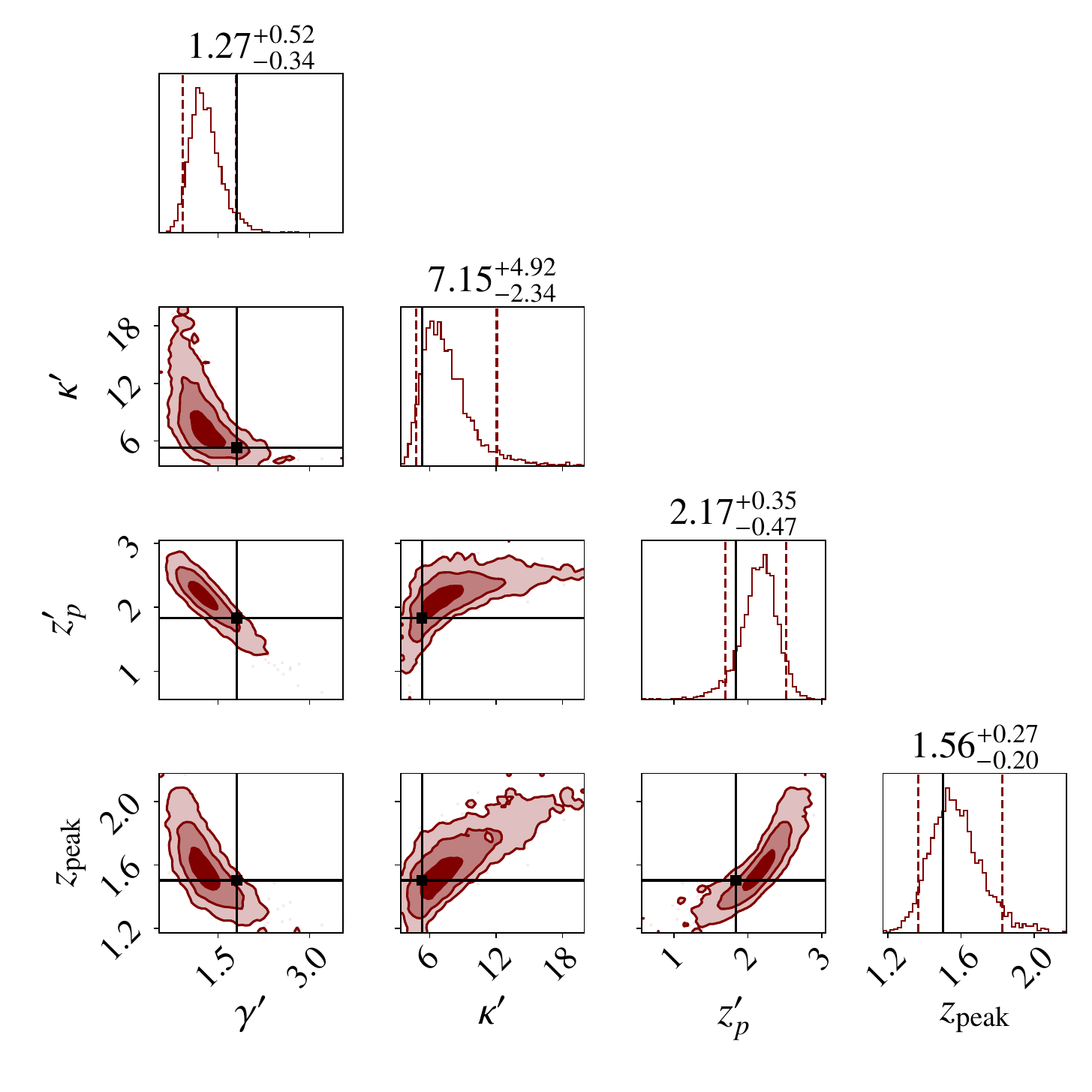}
\includegraphics[width=0.58\linewidth]{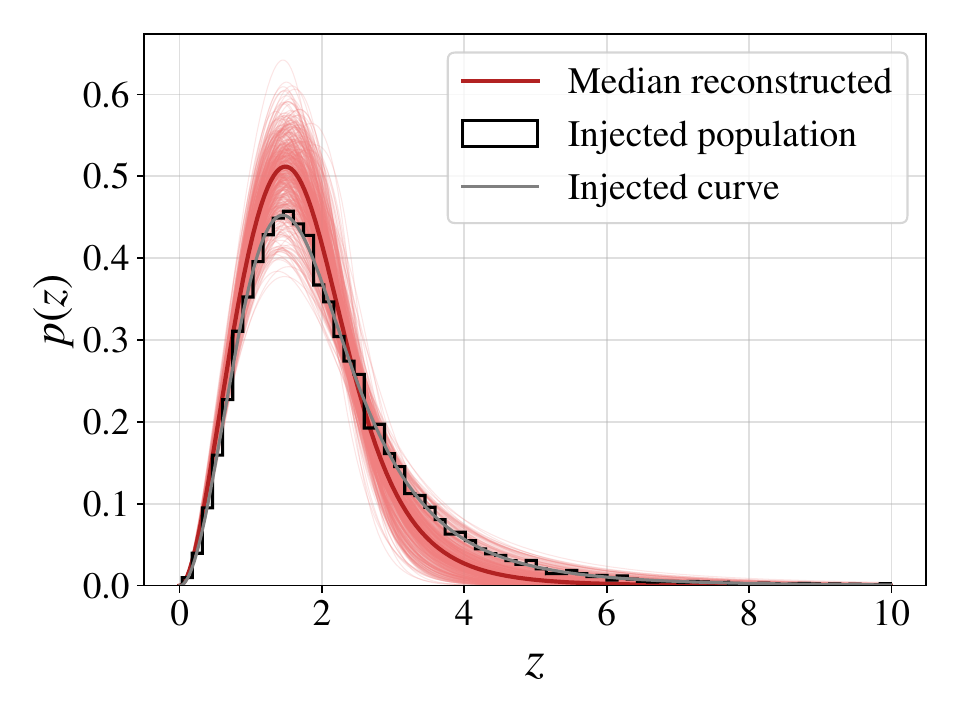}
\caption{Population-inference results for the \Asharp{} baseline network. Top: posterior distributions for the redshift-distribution hyperparameters. Black lines indicate the true injected values. The intervals quoted above the one-dimensional posteriors give the 90\% credible intervals, which are also marked by dashed vertical lines. Bottom: reconstructed redshift distribution $p(z)$. The dark-red curve shows the reconstruction obtained using the median values of the posterior samples for $(\gamma', \kappa', z_p')$, while the thin red curves show reconstructions from 500 random posterior samples to illustrate the uncertainty. For reference, the black histogram shows the input population from \cref{fig:pop_inference:input}, and the grey curve shows the injected functional form of $p(z)$.}
\label{fig:pop_inference:corner}
\end{figure}

We simulate a one-year BBH population using the Madau--Dickinson star-formation model~\cite{Madau:2014bja}, convolved with an inverse time-delay distribution, and adopt a local merger-rate density of \qty{19}{\per\giga\pc\cubed\per\yr}~\cite{GWTC-4_pop}. 
As in \cref{subsubsec:bns_loc}, we consider a network consisting of the LIGO Hanford, LIGO Livingston, and Virgo detectors, with the LIGO detectors operating at either \Asharp{} baseline or A+ sensitivity and Virgo operating at its O5 target sensitivity~\cite{VirgoPSDs,acernese2026advancedvirgopluso5}.

The Madau--Dickinson star-formation rate can be written as
\begin{equation}
	\psi(z \mid \gamma, \kappa, z_p) = 0.015 \frac{(1+z)^{\gamma}}{1+\left[(1+z)/(1+z_p)\right]^\kappa},
	\label{eq:madau-dickinson}
\end{equation}
with hyperparameters $\gamma = 2.7$, $\kappa=5.6$, and $z_p=1.9$. Convolving this model with an inverse-time-delay distribution~\cite{Fishbach:2018edt} yields a BBH merger-rate distribution that can be described by the same functional form but with modified hyperparameters $(\gamma',\kappa',z_p')$~\cite{Vitale:2018yhm}. We simulate a population with a merger-rate peak at $z_{\rm peak}=1.5$. The choice of time-delay model can shift the peak to lower or higher redshift; however, the population-inference model used here does not assume a specific time-delay distribution. Ref.~\cite{Divyajyoti:2026gds} showed that \Asharp{} can constrain the peak location well, provided it lies within $z<2$.

For this simulated population, we adopt a matched-filter network SNR threshold of 10 and perform a Fisher-matrix analysis of the detected events with the \texttt{GWFish} package~\cite{Dupletsa:2022scg}. \Cref{fig:pop_inference:input} shows the input population together with the detected events for each network. We then construct event-level redshift posteriors from the resulting errors and covariance matrices, and perform hierarchical Bayesian inference with the \texttt{gwpopulation} package~\cite{Talbot:2024yqw} to infer the hyperparameters describing the redshift distribution. A detailed description of the procedure can be found in Ref.~\cite{Divyajyoti:2026gds}. Since the A+ network does not yield sufficient high-SNR events to constrain the merger-rate peak, we restrict the population-inference study to the \Asharp{} baseline network.

The top panel of \cref{fig:pop_inference:corner} shows the corner plot for the three hyperparameters ($\gamma', \kappa', z_p'$), along with the posterior on the derived parameter $z_{\rm peak}$. The \Asharp{} baseline network is able to constrain the position of the redshift peak well, while also providing constraints on the other hyperparameters that determine the shape of the distribution. In the bottom panel of \cref{fig:pop_inference:corner}, we show the reconstructed $p(z)$ distribution obtained from the posterior samples of ($\gamma', \kappa', z_p'$). The dark red curve denotes the reconstruction using the median posterior values, while the thin red curves correspond to 500 random posterior samples. For reference, we also show the histogram of the input population from \cref{fig:pop_inference:input} in black, together with the true functional form of $p(z)$ in grey. These results demonstrate the potential of \Asharp{} for population inference at redshifts where current GW detector networks have limited constraining power.

%%%%%%%%%%%%%%%%%%%%%%%
\subsubsection{Higher-order modes}
\label{subsubsec:HM}
%%%%%%%%%%%%%%%%%%%%%%%
The GW signal emitted by merging BHs is typically dominated by the quadrupole emission but the excitation of additional modes can arise from relativistic effects such as spin-induced precession and asymmetries in the component masses or spins. 
While higher-order modes are often subdominant, they can influence the observed signal by altering its amplitude and phase.  
Their relative strength depends on the mass and spin parameters of the BHs, as well as the orientation of the binary relative to the detector.
They are of particular interest as they help to break the distance-inclination degeneracy~\cite{Cutler:1994ys} and improve parameter estimates. Moreover, the detection of individual higher-order modes during either the inspiral or the ringdown phase of the remnant BH enables precision tests of BH dynamics and the no-hair theorem~\cite{VanDenBroeck:2006ar}. 

The total GW signal can be decomposed into a basis of spin-weighted spherical harmonics ${}_{-2}Y_{\ell m}$ with spin weight $s=-2$, 
\begin{equation}
    h = h_+ - i h_\times = \sum_{\ell \geq 2} \sum_{m=-\ell}^{\ell} h_{\ell m} \, {}_{-2}Y_{\ell m},
\end{equation}
where the coefficients $h_{\ell m}$ denote the multiples (modes) of the gravitational radiation field. The quadrupolar modes are the $\ell=|m|=2$ modes; higher-order modes such as the $(2,|1|)$ and $(3,|3|)$ modes are of particular interest for asymmetric binaries, while the $(4,|4|)$ modes are expected to be important for more equal mass, nonspinning binaries. 

% Analysis
To assess the detectability of higher-order modes by different detector configurations, we generate a mock population of BBHs following the O4a population analysis: We draw the BH masses from the fiducial \textsc{broken power law + 2 peaks} mass model, the BH spin magnitudes from a Gaussian truncated on the interval $[0,1]$ and the spin tilt angles from a mixture model consisting of an isotropic and a truncated Gaussian component~\cite{GWTC-4_pop}. 
Current GW detectors can measure the merger rate $R(z)$ up to a redshift of $z \sim 1.5$. As the horizon of \Asharp{} is significantly larger than that of current detectors, the redshift evolution of the merger rate $R(z)$ beyond $z \sim 1.5$ needs to be considered. 
Therefore, we distribute the binaries in redshift $z$ according to the phenomenological merger rate model of Ref.~\cite{Callister:2020arv} that consists of a power law model at low redshift, then transitions to follow the star formation rate~\cite{Madau:2014bja} peaking at $z_p$ (see eq.~\eqref{eq:madau-dickinson}) before decaying to zero at high redshift as star formation is suppressed in the early Universe. 
%
%Therefore, we distribute the binaries in redshift $z$ according to the phenomenological rate model of Ref.~\cite{Callister:2020arv} that consists of a power law with index $\alpha$ at low redshift, then transitions to follow the star formation rate~\cite{Madau:2014bja} peaking at $z_p$ before decaying to zero, controlled by a power-law with index $\beta$, at high redshift as star formation is suppressed in the early Universe:
%\begin{equation}
%R(z) = \Big(1 + (1 + z_p)^{-\alpha -\beta}\Big) \times \frac{R_0 (1+z)^\alpha}{1+ \Big(\frac{1+z}{1+z_p}\Big)^{\alpha+\beta}},
%\end{equation}
%where $R_0$ is the local merger rate, i.e. $R(z=0)$.
For the simulated BBH population we choose the following merger rate model parameters: $\gamma = 3.2$ in accordance with the median measured value after O4a~\cite{GWTC-4_pop}, $z_p=2$, $\kappa=6.7$ and a redshift range of $z \in [0,6]$ consistent with the expected horizon for the most sensitive \Asharp{} configuration. We note that current GW observations do not place meaningful constraints on $z_p$ and $\kappa$, and, for simplicity, we do not consider a time delay between binary formation and merger in this analysis.
All other extrinsic binary parameters are drawn from their standard prior distributions~\cite{LIGOScientific:2018mvr}.
%the coalescence time is fixed and set to zero.
Assuming a median local BBH merger rate density of $19\, {\rm Gpc^{-3}\, yr^{-1}}$, considering one calendar year of observing time and including a Poisson error to account for statistical fluctuations, our mock population consists of approximately $10^5$ BBHs.  
%$N=126,247$ binaries. 

\begin{figure}
    \centering
    \includegraphics[width=\textwidth]{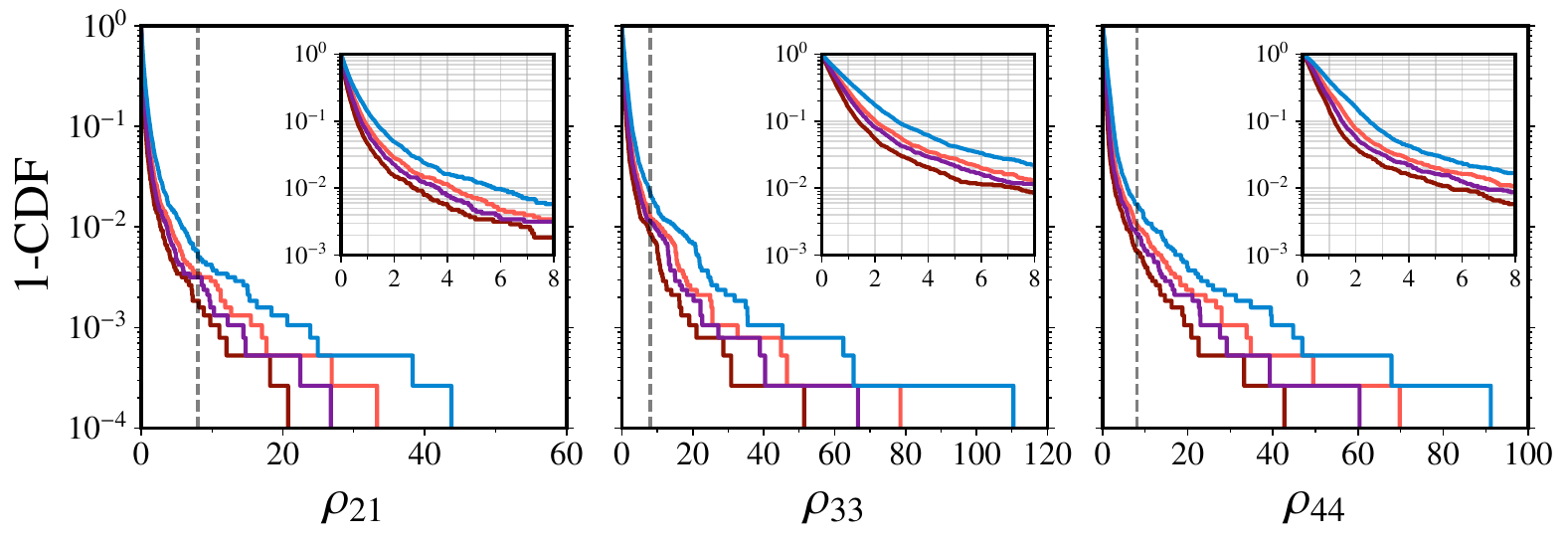}
    \caption[Expected distribution of SNR in the higher-order modes of BBH events]
     {Expected distribution of the SNR in the higher-order modes $(\ell, |m|) = (2, |1|), (3,|3|), (4,|4|)$ for the fraction of events detected with a network SNR $ > 8$  in the least sensitive network configuration drawn from the simulated BBH population. 
     The vertical axis shows the complementary cumulative distribution (CDF is the cumulative distribution function), i.e., the fraction of detected events with higher-order mode SNR greater than the value on the horizontal axis.
     Results are shown for four LIGO sensitivity configurations: A+ (brown), \Asharp{} baseline (red), \Asharp{} wideband (purple), and \Asharp{} optimistic (blue). The inset shows a zoomed-in view of the region with $\rho_{\ell m} \leq 8$. The vertical dashed line marks a higher-order mode SNR of $8$.
     }
    \label{fig:HM}
\end{figure}

We consider the three-detector network consisting of LIGO Hanford, LIGO Livingston and Virgo with four different sensitivity configurations for the two LIGO detectors and a fixed O5 target sensitivity for Virgo~\cite{VirgoPSDs,acernese2026advancedvirgopluso5}.  
Using the complete inspiral--merger--ringdown model \texttt{IMRPhenomXPHM}~\cite{Pratten:2020ceb,Garcia-Quiros:2020qpx,Pratten:2020fqn} to simulate the signals, and imposing an optimal network SNR threshold of 8 on the full strain signal, we identify the binaries detectable in the least sensitive network, with the LIGO detectors operating at A+ sensitivity. This yields 3,790 BBHs above the detection threshold.
In addition to the total network SNR, we also compute the SNR, $\rho_{\ell m}$, contained in the $\{(\ell, |m|) = (2,|1|), (3,|3|), (4,|4|)\}$-modes for this common set of detected binaries. The results are summarized in~\cref{fig:HM}:
We find comparable performance between the A+ and \Asharp{} wideband configurations, while \Asharp{} optimistic provides a significant enhancement in the fraction of events with detectable higher-order modes exceeding an SNR of $\rho_{\ell m} = 8$. 
The \Asharp{} baseline configuration improves notably on A+, while the \Asharp{} optimistic provides an even greater improvement.
%but falls short of \Asharp{} optimistic, lying approximately halfway between the two.
%
For a given SNR, \Asharp{} optimistic yields at least a factor-of-a-few increase in the fraction of events with detectable higher-order modes relative to the A+ or \Asharp{} configurations without major improvements to the coating thermal noise. 
The sensitivity enhancement of the \Asharp{} optimistic configuration in the mid-frequency range has a significant impact on the detectability of higher-order modes. 

In addition to the conditional fractions shown in~\cref{fig:HM}, \cref{tab:HOMs} presents the expected number of annual detections with a higher-order mode SNR exceeding $8$ assuming the fiducial population described above. Unlike the conditional fractions, these absolute yields account for the larger binary black hole population accessible by the \Asharp{} configurations. We estimate that the baseline \Asharp{} configuration would increase the annual number of higher-order-mode detections by a factor of approximately $1.5$--$1.8$ relative to A+, while the \Asharp{} optimistic configuration could yield up to three times as many detections.

\begin{table}[t]
  \centering
  \setlength{\tabcolsep}{10pt}
  \renewcommand{\arraystretch}{1.2}
  \begin{tabular}{rccc}
  \hline
  & \multicolumn{3}{c}{Annual Detections of HOMs} \\
  Configuration & $\rho_{21} > 8$ &  $\rho_{33} > 8$  & $\rho_{44} > 8$  \\
  \hline 
  A+ & 7 & 32 & 22 \\
  \Asharp{} baseline & 13 & 49 & 40 \\
  \Asharp{} wideband & 12 & 43 & 33 \\
   \Asharp{} optimistic & 21 & 83 & 64\\
  \hline
   \end{tabular}
  \caption{Expected number of BBH detections per year with a higher-order mode SNR exceeding $8$ for the fiducial BBH population model. These absolute yields complement the conditional fractions shown in~\cref{fig:HM} by accounting for the number of binaries detectable by the more sensitive detector configurations. A local merger-rate density of $19\, \mathrm{Gpc}^{-3}\,\mathrm{yr}^{-1}$ is assumed.}
  \label{tab:HOMs}
\end{table}

Finally, we note that the BBH population, especially at  high redshift, is highly uncertain and hence the absolute numbers depend on assumptions about the population and the evolution of the merger rate with redshift. However, our results indicate that for the various \Asharp{} configurations up to a few percent of BBHs will have higher-order modes that are loud enough to be detected individually. This will not only improve estimates of the source parameters by breaking degeneracies, but also enhance the prospects for detecting the GW memory effect~\cite{Christodoulou:1991cr, Lasky:2016knh, Boersma:2020gxx, Mitman:2026zfg}.
 
 %%%%%%%%%%%%%%%%%%%%%%%%%%%%%%%%%
 \subsubsection{Intermediate-mass black holes}
 \label{subsubsec:imbh}
 %%%%%%%%%%%%%%%%%%%%%%%%%%%%%%%%%
 
% \begin{figure}[h]
%    \centering
%    \includegraphics[scale=0.5]{Science/waveform.pdf}
%    \caption[]
%    {Characteristic strain $h_c$ of a GW231123-like signal at a luminosity distance of $d_L\sim \qty{2.4}{\giga\pc}$ ($z\sim 0.4$; black) and at $d_L \sim \qty{12.1}{\giga\pc}$ ($z\sim 1.6$; orange), and the characteristic noise for A+ and the different \Asharp{} configurations. The grey shaded area highlights the low frequency range between 10 and \qty{20}{\Hz}.}
%    \label{fig:strain}
%\end{figure}
 
During O4a, the two LIGO detectors observed the GW signal from the merger of two BHs,  GW231123\_135430 (hereafter referred to
as GW231123)~\cite{LIGOScientific:2025rsn}, with source-frame masses $m_1\sim \qty{137}{\solarmass}$ and $m_2 \sim \qty{101}{\solarmass}$ at a distance of $d_L \sim \qty{2200}{\mega\pc}$ or $z \sim 0.4$. In addition, the BHs were also found to have large spin magnitudes and hints of spin-induced orbital precession~\cite{Apostolatos:1994mx} were identified.
Its high component masses are indicative of back hole formation beyond standard stellar collapse~\cite{Woosley:2021xba}, and its high final mass ($M\sim \qty{222}{\solarmass}$) suggests the formation of intermediate mass black holes (IMBHs) via GW-driven mergers (see, e.g., Ref.~\cite{Gerosa:2021mno} and references therein).

The key advantage of \Asharp{} for such heavy systems is its improved sensitivity below \qty{20}{\Hz}, which brings more of the inspiral of high-mass BBH mergers into band. This can make otherwise inaccessible effects, such as spin-induced orbital precession, measurable and thereby provide new information about IMBH formation and dynamical assembly in dense environments, including active galactic nuclei (AGN).

\begin{figure}[t]
    \centering
    \includegraphics[scale=0.6]{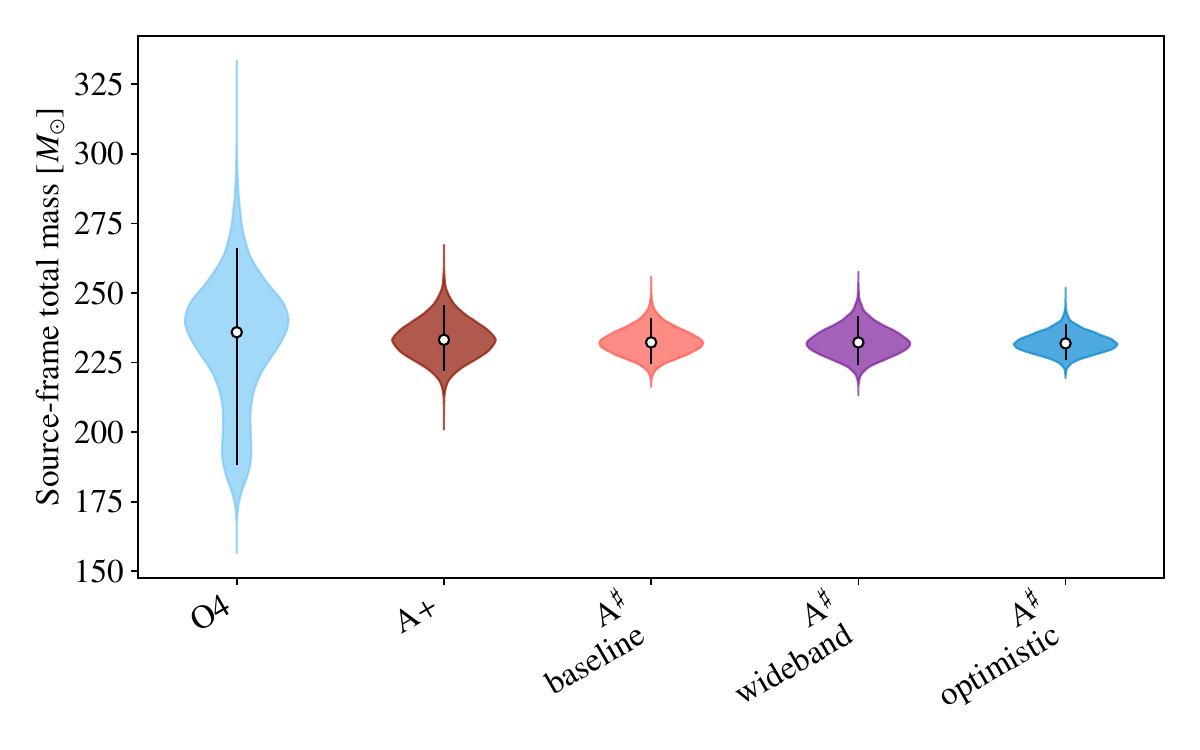}
    \caption[]
    {One-dimensional posterior probability distribution of the source-frame total mass of GW231123 (light blue), as well as for a GW231123-like zero-noise simulation at a redshift of $z=0.4$ in A+ and the three different \Asharp{} configurations. The white dots mark the median posterior values, while the vertical lines indicate the 90\% credible interval. If such a signal were to be observed in an \Asharp{} network, the parameter uncertainty is estimated to decrease by approximately $67\%$ in comparison to the O4 measurement.}
    \label{fig:imbh_mass}
\end{figure}

GW231123 was observed with a network SNR of $\sim$21; a network with A+ sensitivity would approximately double the SNR, while an \Asharp{} network would yield SNRs between $\sim$70 and $\sim$100, depending on the configuration. 
%
%The characteristic strain of a GW231123-like signal, $h_c = 2f|\tilde{h}(f)|$, projected onto the LIGO-Livingston detector and the characteristic noise, $h_n = \sqrt{f S_n(f)}$, for different detector sensitivities are shown in~\cref{fig:strain}.
%
%For the current measurement, the uncertainties on the source parameters of GW231123 are relatively large, but a more sensitive detector network will allow us to place tighter constraints. 
Under the linear signal approximation for loud signals, the measurement uncertainty of a given parameter is expected to scale inversely with the SNR~\cite{Cutler:1994ys}. 
Therefore, if such a system was observed by an \Asharp{} detector network, the parameter uncertainties would be approximately one third of the current measurement, enabling tighter constraints on the binary's astrophysical origin. As an example of this improvement we show in~\cref{fig:imbh_mass} the one-dimensional posterior probability distributions of the source-frame total mass inferred for a GW231123-like signal assuming different \Asharp{} detector configurations in comparison to A+ and the actual O4 measurement for GW231123~\cite{ligo_scientific_collaboration_2025_17437902}. 
We note that the O4 measurement includes detector noise which broadens the distributions.

\begin{figure}[h]
    \centering
	\includegraphics[width=0.48\textwidth]{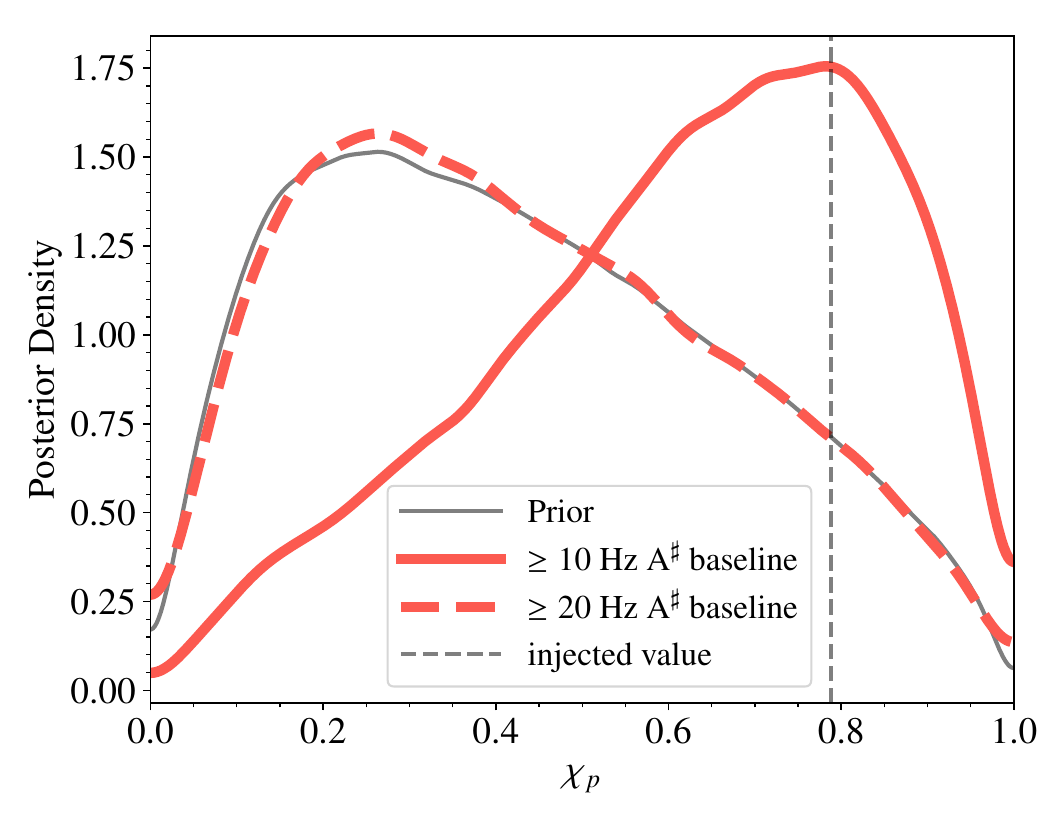}
	\includegraphics[width=0.48\textwidth]{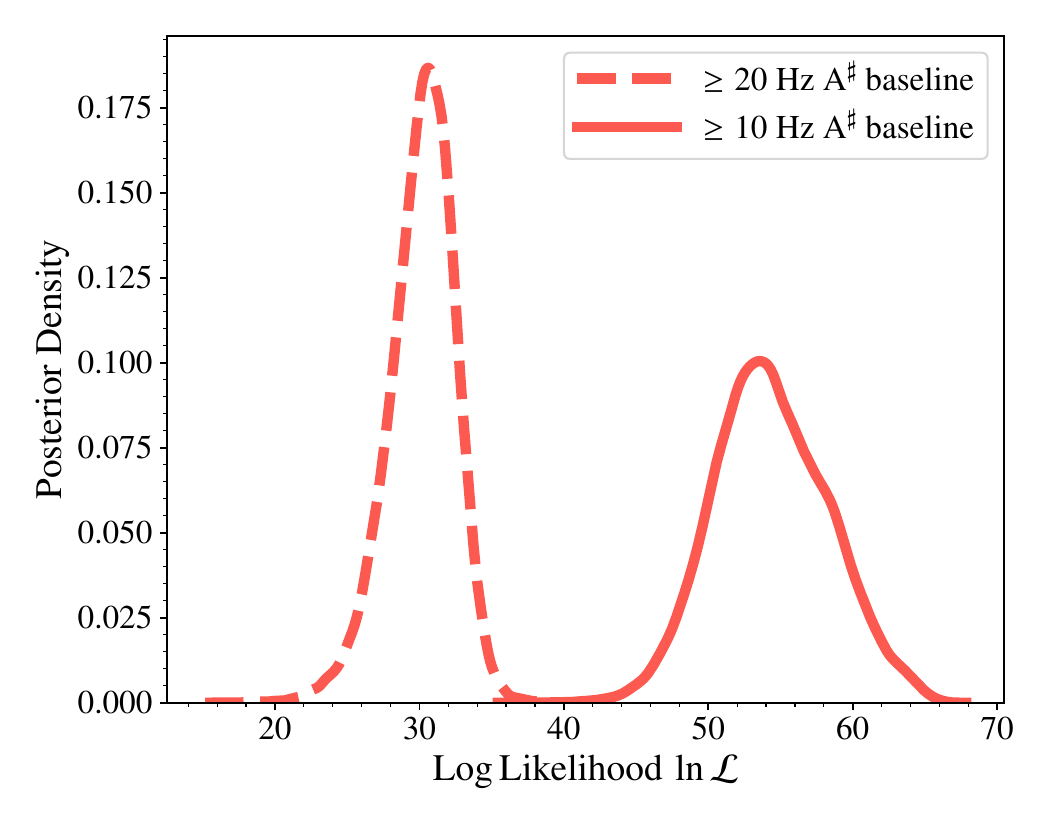}
    \caption[]
	{Left: One-dimensional posterior distribution of the effective precession spin parameter $\chi_p$ for a GW231123-like BBH at a redshift of $z=1.6$. The orbital precession present in the system cannot be inferred when only frequencies $\geq 20$ Hz are considered (red dashed) and the prior is recovered (grey solid line). The low-frequency improvements, however, enable an informative measurement of $\chi_p$ (red solid). The vertical dashed line indicates the injected value of $\chi_p \simeq 0.788$. Right: The log-likelihood for both analyses. When including frequencies from $10$ to $20$ Hz, the maximum log-likelihood increases by $\sim$26.}
    \label{fig:chi_p}
\end{figure}

%In addition to the overall sensitivity improvement, the various \Asharp{} configurations also enable access to frequencies below $20$ Hz, which is not achievable with current detectors or the projected A+ sensitivity. This low-frequency improvement provides an opportunity not only to observe even higher mass BBH mergers in band, but also to observe such high-mass mergers at higher redshifts. 
To concretely demonstrate the low-frequency advantage of \Asharp{}, we consider a representative IMBH merger, again similar to GW231123, located at a redshift of $z=1.6$. A three-detector network consisting of LIGO Hanford and LIGO Livingston at A+ sensitivity and Virgo at its O5 target sensitivity~\cite{VirgoPSDs,acernese2026advancedvirgopluso5} would miss this signal with a network SNR of only ${\sim}$6 from \qty{10}{\Hz}. With \Asharp{} baseline sensitivity for the two LIGO detectors, the same signal becomes detectable with a network SNR of ${\sim}$12, also from \qty{10}{\Hz}. Crucially, even a moderate SNR gain arising from the improved sensitivity in the \sisetup{range-phrase = {--}}\qtyrange{10}{20}{\Hz} band allows us to probe the binary evolution at lower frequencies than are accessible with A+. This substantially improves the measurability of binary parameters, particularly the spin of the primary black hole, as illustrated in \cref{fig:chi_p}.
Following a Bayesian analysis with the numerical-relativity surrogate model \texttt{NRSur7dq4}~\cite{Varma:2019csw}, \cref{fig:chi_p} compares the inferred one-dimensional posterior distribution of the effective precession spin $\chi_p$~\cite{Schmidt:2014iyl} (left) and the log-likelihood (right) for \Asharp{} baseline assuming \qty{20}{\Hz} and \qty{10}{\Hz} as the lower cutoff frequency, respectively. This highlights the impact of the low-frequency improvement in \Asharp{}, which, in this case, enables an informative measurement of spin-induced orbital precession, while the measurement from \qty{20}{\Hz} recovers the prior. 
The observation of significant spin precession in an intermediate mass BBH at a redshift of $z = 1.6$ would provide compelling evidence for hierarchical or dynamically assembly operating efficiently at early cosmic times~\cite{Rodriguez:2016kxx, Gerosa:2017kvu}. The combination of large mass, misaligned spins, and high redshift disfavours isolated stellar binary evolution and directly probes the formation and growth of intermediate mass black holes in dense environments, including AGN-assisted formation and hierarchical mergers~\cite{Miller:2001ez, Mckernan:2017ssq}. 
This example showcases the impact of the improved low-frequency sensitivity of the \Asharp{} design in enabling access to massive BBHs at significantly higher redshifts than can be observed today.

  %%%%%%%%%%%%%%%%%%%%%%%%%%%%%%%%%
 \subsubsection{Black hole ringdown}
 \label{subsubsec:ringdown}

The post-merger ringdown of the remnant BH formed in a merger is described, in general relativity, by a superposition of quasi-normal modes (QNMs)~\cite{teukolsky1973,berti2025} determined only by the remnant mass and spin~\cite{Carter:1971zc,Robinson:1975bv,Chrusciel:2012jk}. Measurements of these modes form the basis of BH spectroscopy~\cite{berti2025} and provide a direct probe of the final compact object, enabling tests of the no-hair theorem~\cite{Carter:1971zc,Robinson:1975bv,Chrusciel:2012jk} and of the consistency of the inspiral, merger, and ringdown portions of the signal. Ringdown analyses therefore form an important part of the GW strong-field science case, complementing tests based on the full inspiral-merger-ringdown (IMR) waveform and providing an independent probe of general relativity in the highly dynamical, strong-field regime.

Current ringdown measurements are limited by the available post-merger SNR. For most detected BBHs, GW emission is dominated by the fundamental $(\ell,m,n)=(2,2,0)$ mode, with informative studies of additional mode content, such as overtones or higher-order angular modes, possible only for the loudest events~\cite{GW250114:2025oiz,GW250114com:2025wao,LIGOScientific:2026wpt}. Pure-ringdown analyses, in which the remnant properties are inferred using only data after the merger, are therefore restricted to a small subset of high-SNR systems. For current detectors, these limitations are more pronounced for lower-mass binaries: their ringdown SNRs are lower, and their ringdown frequencies are shifted to the kHz band, where detector sensitivity is reduced.

The improved sensitivity of \Asharp{} yields higher ringdown SNRs, making it possible to measure the dominant QNM more accurately, improve the prospects for resolving subdominant higher-order modes and overtones, and extend ringdown studies to a broader range of BBH systems. This broadens the accessible source population for BH spectroscopy and increases the number of systems for which independent ringdown-based measurements of the remnant mass and spin can be compared with IMR predictions, enabling more stringent tests of general relativity. With larger post-merger SNRs, it will also be possible to carry out more detailed studies of the ringdown mode content, including tests of mode consistency, measurements of mode excitation across different classes of systems, such as highly precessing BBHs, and population-level analyses that combine information across events.

\begin{figure}[h]
	\centering
	\includegraphics[scale=0.7]{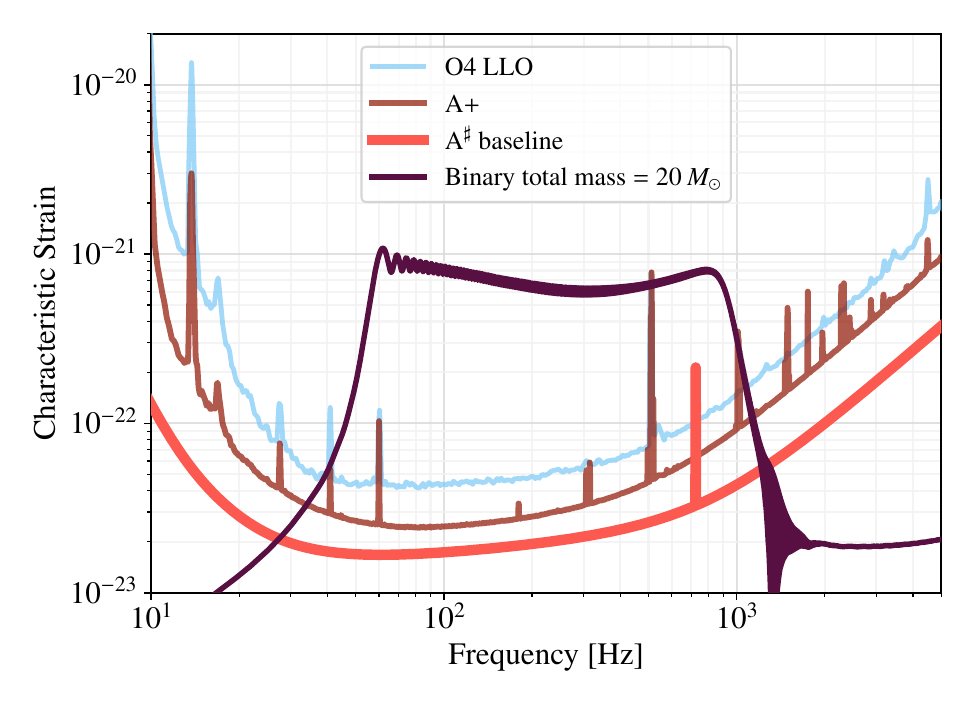}
	\caption{
		%Example high-frequency ringdown signal compared with detector sensitivities.
		Strain amplitude spectral density of a numerical-relativity waveform for a comparable-mass BBH system, \texttt{SXS:BBH:0305}, rescaled to a total binary mass of \qty{20}{\solarmass}, for which the ringdown signal lies in the high-frequency regime, and placed at a luminosity distance of \qty{400}{\mega\pc}.
		The detector curves show representative O4, A+, and \Asharp{} baseline amplitude spectral densities.
		For this low-mass system, the dominant $\ell=m=2$ ringdown mode lies close to \qty{900}{\Hz}, where the improved high-frequency sensitivity of \Asharp{} substantially increases the post-merger SNR.
		The low-frequency sweep reflects the finite duration of the numerical-relativity waveform, while the high-frequency end is affected by numerical noise.}
	\label{fig:ringdown_asd}
\end{figure}

As an illustrative example, we consider a relatively low-mass BBH system and demonstrate that the improved high-frequency sensitivity of \Asharp{} enables BH spectroscopy in the kHz regime. We use the numerical-relativity waveform for a comparable-mass system, \texttt{SXS:BBH:0305}~\cite{Lovelace_2016}, rescaled to a total binary mass of \qty{20}{\solarmass}, and place the system at a luminosity distance of \qty{400}{\mega\pc}, approximately the distance within which several mergers of $\sim 10{+}10\,M_\odot$ BHs are expected each year. For this system, the dominant $\ell=m=2$ ringdown mode lies close to \qty{900}{\Hz}, placing the post-merger signal in the frequency region where \Asharp{} provides a substantial gain. \Cref{fig:ringdown_asd} compares the strain amplitude spectral density of the example waveform with representative O4, A+, and \Asharp{} baseline noise curves. The signal is not accessible with an O4-level detector, similar to the challenge posed by relatively light events such as GW240925\_005809 in O4~\cite{astro-calib}, and remains too weak for meaningful pure-ringdown inference with A+. 
%In contrast, \Asharp{} baseline yields a ringdown SNR of 12.3 when the analysis is started at $8\,M_{\rm f}$ after merger, compared with 3.8 for O4 and 6.0 for A+ (time is in units of the final remnant BH mass in the detector frame, $M_{\rm f}$).
In contrast, \Asharp{} baseline yields a ringdown SNR of 12.3, compared with 3.8 for O4 and 6.0 for A+.\footnote{These SNRs assume the analysis is started at $8\,M_{\rm f}$ after merger, where time is in units of the remnant BH mass in the detector frame, $M_{\rm f}\sim0.1$ ms.}

\begin{figure}[h]
	\centering
	\includegraphics[scale=0.5]{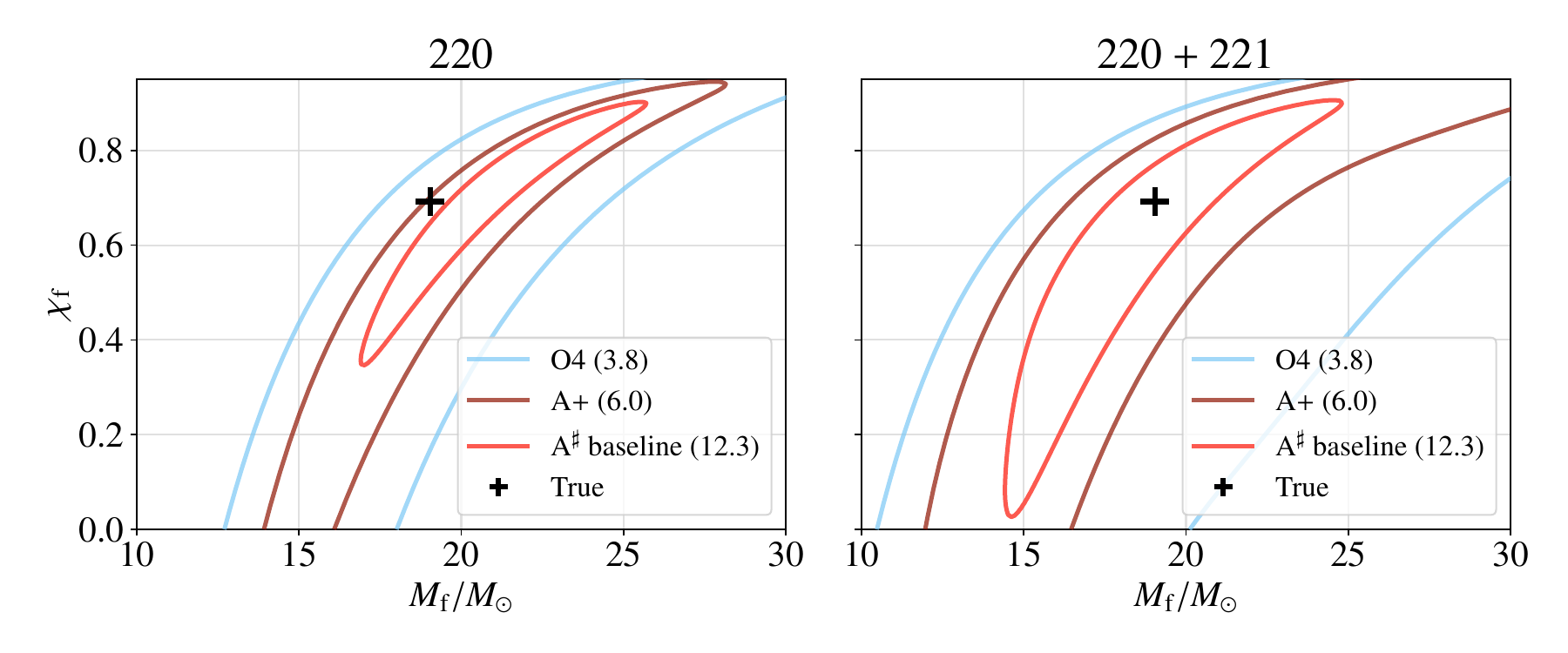}	
	\caption{
	Pure-ringdown constraints (90\% credible regions) on the remnant BH mass and spin for the example low-mass BBH system, using simulated data starting at $8\,M_{\rm f}$.
	The left panel uses a one-mode QNM model with the fundamental $(\ell,m,n)=(2,2,0)$ mode, while the right panel includes the first overtone, with $n=0,1$.
	The black plus marker indicates the injected true values.
	The numbers in parentheses indicate the ringdown SNR from $8\,M_{\rm f}$.
	The O4 and A+ configurations are largely uninformative, whereas \Asharp{} baseline yields informative posteriors; for \Asharp{}, the two-mode model recovers the injected parameters while the one-mode model is biased, supporting the identification of the first overtone.}
\label{fig:ringdown_posterior}
\end{figure}

\Cref{fig:ringdown_posterior} shows the corresponding remnant mass and spin constraints from a pure-ringdown analysis. We analyze the two-detector simulated data using the QNM rational filter~\cite{Ma:2022wpv,Ma:2023cwe,Ma:2023vvr,Lu:2025mwp}, comparing a one-mode model containing only the fundamental $\ell=m=2$ mode with a two-mode model that also includes its first overtone. (See alternative analysis methods in, e.g., Refs.~\cite{Isi:2021iql,Carullo:2019flw,bustillo2021,Ghosh:2021mrv,Finch:2022ynt,Wang:2023ljx,Correia:2023bfn,Capano:2021etf,Chandra:2025ipu}.) The two-mode model is preferred over the one-mode model by the Bayes factor and recovers remnant parameters consistent with the injected values, whereas the one-mode model gives biased constraints. The figure shows the results for both models, illustrating the importance of including the first overtone $(\ell,m,n)=(2,2,1)$ in the fit. Among the detector configurations considered, only \Asharp{} yields informative mass and spin posteriors; by contrast, the O4 and A+ configurations produce broad, largely prior-dominated posteriors. This example highlights the role of \Asharp{} in extending ringdown science into the kHz regime, enabling BH spectroscopy for lighter systems and strengthening IMR consistency tests.
Such low-mass systems are also valuable for model-specific tests of modified gravity: for example, in dynamical Chern--Simons gravity, projected ringdown constraints on the coupling scale approximately linearly with the remnant mass for fixed fractional QNM-frequency uncertainty, so lower-mass remnants can yield more stringent bounds~\cite{Chung2025}.

\subsection{Neutron star science}
\label{subsec:ns}

NS sources provide probes of dense matter and long-duration GW emission.
In this subsection, we quantify the impact of the \Asharp{} configurations on three representative science cases: constraints on the cold nuclear equation of state from BNS inspirals (\cref{subsubsec:eos}), prospects for observing high-frequency BNS post-merger signals that probe the hot equation of state~(\cref{subsubsec:pm}), and searches for persistent, quasi-monochromatic signals from spinning NSs~(\cref{subsubsec:cw}).

%%%%%%%%%%%%%%%%%%%%%%%%%%%%%%%%%
\subsubsection{Cold nuclear equation of state}
\label{subsubsec:eos}
%%%%%%%%%%%%%%%%%%%%%%%%%%%%%%%%%
\begin{figure}[h]
    \centering
    \includegraphics[scale=0.6]{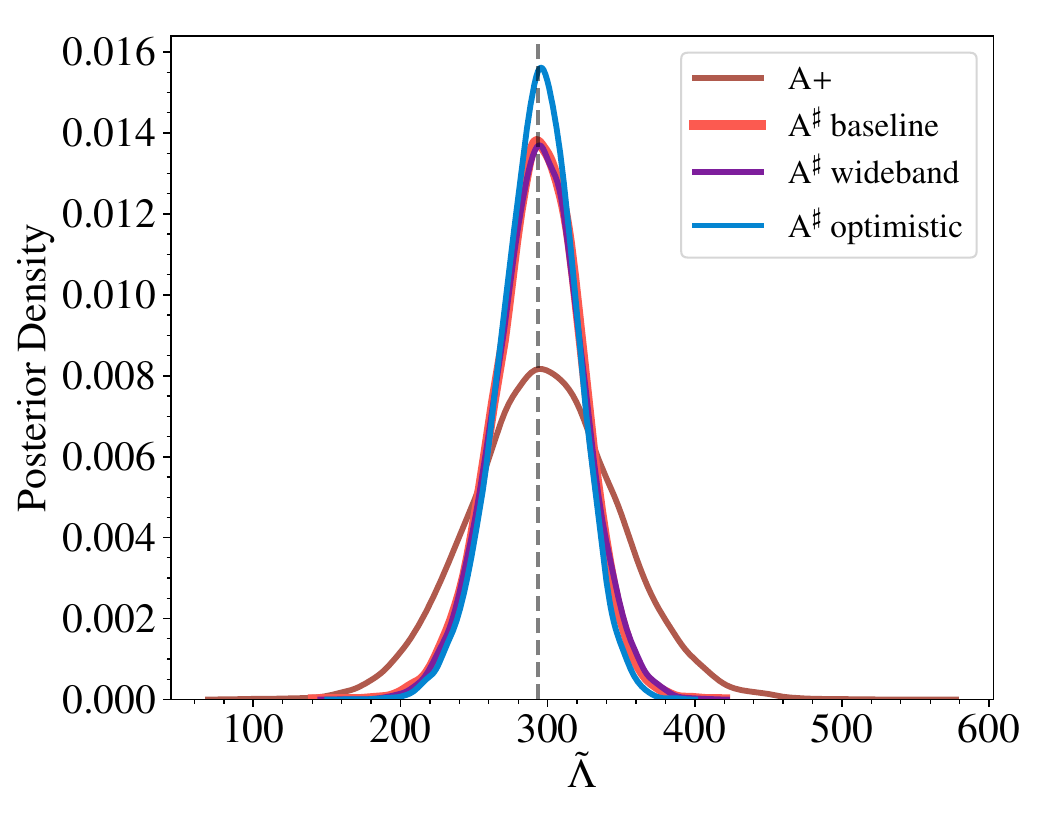}
    \caption[Tidal deformability parameter estimation]
    {One-dimensional posterior probability density of the binary tidal deformability $\tilde{\Lambda}$ for a simulated GW170817-like BNS event at a distance of \qty{100}{\mega\pc} shown for the different LIGO sensitivity configurations. The dashed vertical line indicates the simulated value of $\tilde{\Lambda}$.}
    \label{fig:LamT}
\end{figure}

\begin{figure}[h]
    \centering
    \includegraphics[width=0.48\textwidth]{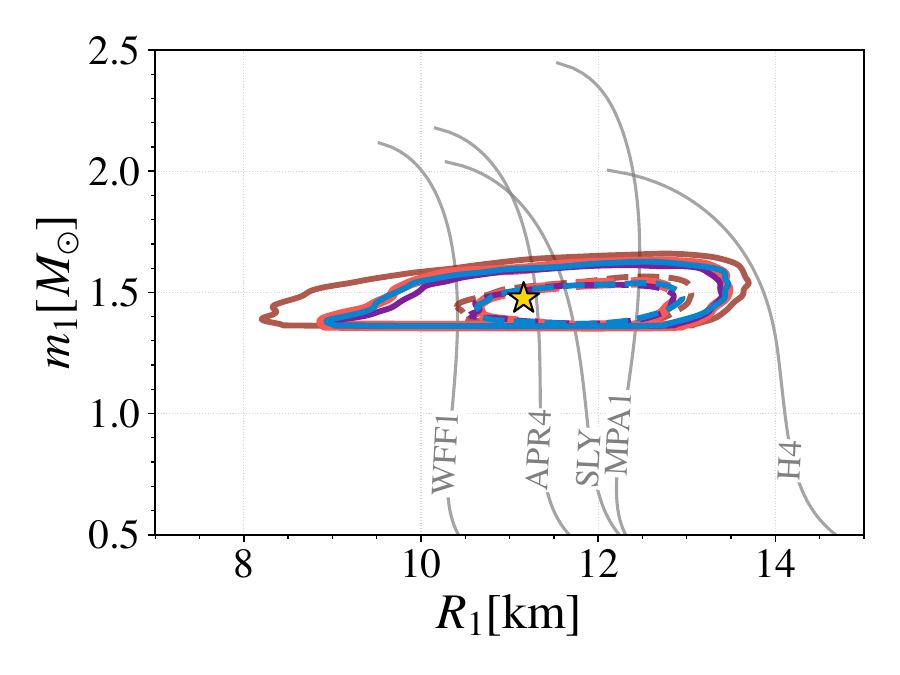}
    \includegraphics[width=0.48\textwidth]{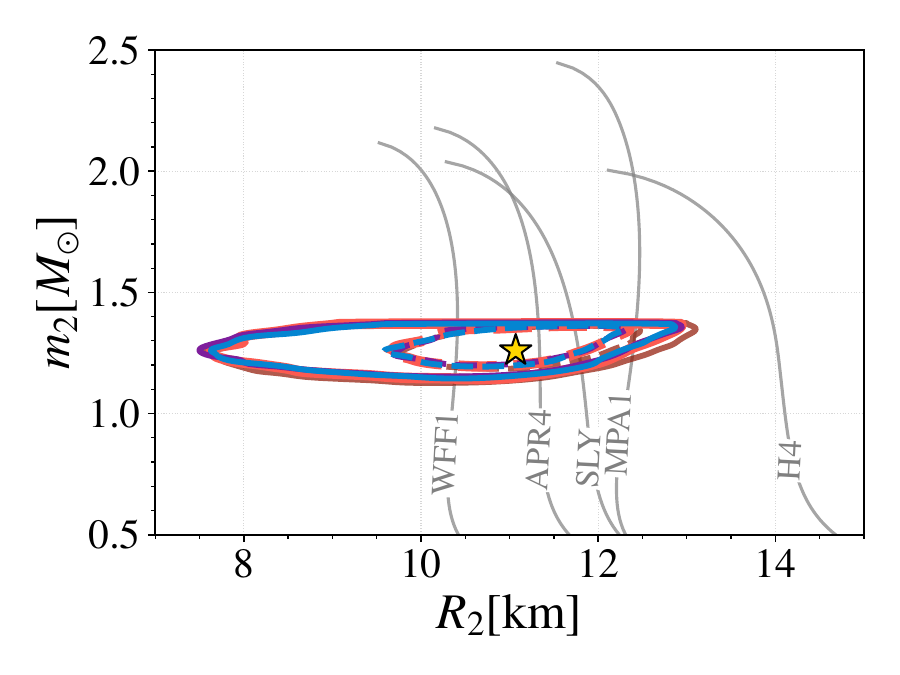}
    \caption[Mass-radius curves for a simulated GW170817-like BNS]
    {Mass-radius constraints for a simulated GW170817-like BNS event at a distance of \qty{100}{\mega\pc}. Universal relations are used to map the tidal parameters to the NS radius $R$. Note that we do not enforce that both stars obey the same EOS. The LIGO sensitivity configurations shown are A+ (brown), \Asharp{} baseline (red), \Asharp{} wideband (purple), and \Asharp{} optimistic (blue). The solid and dashed contours show the 90\% and 50\% credible intervals, respectively; the grey curves indicate predictions for a set of EOS, and the stars mark the injected parameters of the simulated signal. }
    \label{fig:MRcontours}
\end{figure}

GWs from BNS systems provide a unique opportunity to probe the nuclear equation-of-state (EOS) of cold, ultradense matter through the characteristic imprint the tidal interaction between the two stars leaves on the GW. 
The waveform emitted by a BNS differs from that of a BBH predominantly due to the deformation of each star in the gravitational field of its companion. At leading post-Newtonian order, this induced tidal deformation is encapsulated by the (dimensionless) binary tidal deformability parameter $\tilde{\Lambda}$~\cite{Flanagan:2007ix}. 

The observation of the BNS merger GW170817~\cite{LIGOScientific:2017vwq} yielded the first EOS constraints from GWs, with complementary information obtained from pulsar observations~\cite{Raaijmakers:2019qny} and terrestrial experiments~\cite{Reed:2021nqk}. 
To demonstrate the improved capability of \Asharp{} to constrain the properties of nuclear matter, we simulate a nonspinning GW170817-like BNS with source-frame component masses $m_1=\qty{1.475}{\solarmass}$ and $m_2=\qty{1.26}{\solarmass}$, located at a luminosity distance of \qty{100}{\mega\pc}, approximately the expected distance to the nearest BNS merger in one year based on the merger-rate density inferred from O4a observations~\cite{GWTC-4_pop}, and assuming the soft EOS APR4~\cite{Akmal:1998cf}.
The signal is analyzed in the LIGO-Hanford, LIGO-Livingston and Virgo three-detector network with four different sensitivities for the two LIGO detectors and a fixed O5 target sensitivity for Virgo~\cite{VirgoPSDs,acernese2026advancedvirgopluso5}.
We perform full Bayesian inference with \textsc{Bilby}~\cite{Ashton:2018jfp, Romero-Shaw:2020owr}, assuming zero noise, a maximal dimensionless NS spin of $0.05$ and employing the tidal waveform model \texttt{IMRPhenomXAS\textunderscore{}NRTidalv3}~\cite{Pratten:2020fqn,Abac:2023ujg}, to obtain posterior probability densities for the binary parameters.
The resulting posterior for the binary tidal deformability $\tilde{\Lambda}$ is shown in~\cref{fig:LamT}. 
In particular, we find that the constraints, as characterized by the \qty{90}{\%} credible interval, improve by approximately 40\% between A+ and various \Asharp{} configurations. 
We further observe comparable performance between \Asharp{} wideband and optimistic configurations. Since tidal effects enter the waveform predominantly at GW frequencies $\gtrsim \qty{400}{\Hz}$, the enhanced sensitivity at $f \gtrsim \qty{1000}{\Hz}$ for \Asharp{} wideband may trade-off the better performance of \Asharp{} optimistic at frequencies below \qty{1000}{\Hz}. 

Using quasi-universal relations~\cite{Yagi:2016bkt}, we map the tidal deformability parameter $\tilde{\Lambda}$ to a constraint on the NS radius $R$ via the compactness $C \equiv m/R$, where $m$ is the NS mass, and compare the resulting mass-radius constraints with predictions from representative EOS. \Cref{fig:MRcontours} shows the results for the four sensitivity configurations. 
In contrast to the case of the $\tilde{\Lambda}$ constraints, we find only marginal improvement in the inferred mass-radius constraints between A+ and the \Asharp{} sensitivities due to the uncertainty in the conversion from tidal deformability to compactness.
The uncertainty on the radius of the primary neutron star is approximately $32\%$ for \Asharp{}, constituting an improvement of around $8\%$ over A+. Percent-level constraints of the neutron star radius, and hence the EOS, will be achievable by the proposed next-generation of GW detectors~\cite{Chatziioannou:2021tdi}.

%%%%%%%%%%%%%%%%%%%%%%%%%%%%%%%%%%
\subsubsection{Post-merger signals and the hot equation of state}
\label{subsubsec:pm}
%%%%%%%%%%%%%%%%%%%%%%%%%%%%%%%%%%

The post-merger GW signal from a BNS merger carries information about a different thermodynamic regime from that probed during the inspiral. During the inspiral, the NSs are effectively cold, and tidal effects mainly constrain the cold equation of state through the stellar compactness and tidal deformability.  
After merger, however, the remnant can reach temperatures of order \qty{e11}{\kelvin} and, if it does not promptly collapse to a BH, may survive as a massive, differentially rotating NS.  The resulting GW emission is expected to lie predominantly in the $\sim$1--\qty{4}{\kHz} band and can persist for up to tens to hundreds of milliseconds.  Its spectral features, in particular the dominant post-merger peak frequency, are sensitive to the finite-temperature equation of state, the remnant compactness, and possible phase transitions in supranuclear-density matter. Post-merger observations therefore provide a unique probe of the hot equation of state, distinct from the cold-matter constraints obtained from the inspiral~\cite{Takami2015,Bauswein:2019qcd,Ackley:2020atn}.

%We use the time-domain BNS post-merger waveforms for the SLy, ALF2, and LS220 equations of state from the CoRe database~\cite{Dietrich2018} to assess the prospects for studying post-merger physics. The post-merger signal is defined as the portion of the waveform following the point at which the merger–ringdown emission has damped to a minimum. The post-merger SNR, $\rho_\text{pm}$, reported in \cref{tab:astro_metrics}, corresponds to the SNR computed over this post-merger portion of the signal.

We assess the post-merger sensitivity using time-domain BNS waveforms from the CoRe database~\cite{Dietrich2018}, considering the SLy, ALF2, and LS220 equations of state.  The post-merger signal is defined as the portion of the waveform after the coalescence, and we compute the corresponding post-merger SNR, $\rho_\text{pm}$, for an optimally oriented source at a luminosity distance of \qty{\Val{Asharp.Dl_pm_Mpc}}{\mega\parsec}.  Across the \num{\Val{Asharp.num_pm_sources}} simulated sources, the loudest 10\% of events reach $\rho_\text{pm}\sim 3$--4 for the \Asharp{} configurations, while the largest value among all simulations is $\rho_\text{pm}<6$ and is obtained for the \Asharp{} wideband configuration.  This represents a clear improvement over A+, for which the corresponding post-merger SNRs are of order unity, but the absolute SNRs remain too small for robust post-merger detection except possibly for the most favourable nearby events.

These results show that the improved high-frequency sensitivity of \Asharp{} increases the post-merger SNR relative to A+, with the wideband configuration giving the largest gain.  However, the absolute SNRs remain modest even for optimally oriented sources at \qty{\Val{Asharp.Dl_pm_Mpc}}{\mega\parsec}.  The \Asharp{} baseline configuration is therefore unlikely to enable regular detections or detailed measurements of the post-merger spectrum.  The wideband configuration could improve the prospects for exceptionally nearby events, and may provide useful constraints or upper limits in favourable cases, but robust measurements of post-merger spectral features and the hot equation of state will likely require the substantially improved high-frequency sensitivity of next-generation observatories~\cite{evans2021CEHS}.

%\begin{table}[h]
%	\centering
%	\sisetup{table-alignment-mode=none, table-number-alignment=center}
%	\begin{tabular}{
%			r
%			S[round-mode=places, round-precision=1]
%			S[round-mode=places, round-precision=1]
%		}
%		\hline
%		& \multicolumn{2}{c}{Post-Merger}  \\
%		Configuration & $\rho_\text{pm}^{(10)}$ & $\rho_\text{pm}^{(\text{max})}$ \\
%		\hline
%		O4 &
%		\Val{O4LLO.pm_snr_90} &
%		\Val{O4LLO.pm_snr_max} \\
%		A+ (O5c) &
%		\Val{O5c.pm_snr_90} &
%		\Val{O5c.pm_snr_max} \\
%		\Asharp{} &
%		\Val{Asharp.pm_snr_90} &
%		\Val{Asharp.pm_snr_max} \\
%		\Asharp{} wideband &
%		\Val{Asharp_wideband.pm_snr_90} &
%		\Val{Asharp_wideband.pm_snr_max} \\
%		\Asharp{} optimistic &
%		\Val{Asharp_AlGaAs.pm_snr_90} &
%		\Val{Asharp_AlGaAs.pm_snr_max} \\
%		\hline
%	\end{tabular}
%	\caption{Prospect for BNS post-merger signal with different detector configurations.
%		The post-merger SNR $\rho_\text{pm}$ is computed for an optimally oriented BNS at a luminosity distance of \qty{\Val{Asharp.Dl_pm_Mpc}}{\mega\parsec}. 
%		Three NS equations of state (SLy, ALF2, and LS220) are used to generate \num{\Val{Asharp.num_pm_sources}} sources; $\rho_\text{pm}^{(10)}$ denotes the SNR of the loudest 10\% of events, and $\rho_\text{pm}^{(\mathrm{max})}$ that of the loudest source.}
%	\label{tab:pm}
%\end{table}

%%%%%%%%%%%%%%%%%%%%%%%%%%%%%%%%%%
\subsubsection{Continuous waves}
\label{subsubsec:cw}
%%%%%%%%%%%%%%%%%%%%%%%%%%%%%%%%%%

Spinning, non-axisymmetric NSs can emit continuous gravitational waves (CWs) over timescales comparable to or longer than an observing run, in contrast to transient GW signals from compact binary mergers~\cite{Riles2022}. To date, no CW signal has been detected, though increasingly stringent upper limits have been placed by searches targeting known pulsars~\cite{O4a_known_pulsar,O4ab_narrow_band}, directed searches toward the Galactic center~\cite{2022PhRvD.106d2003A}, young supernova remnants~\cite{O4a_SNR}, low-mass X-ray binaries such as Scorpius X-1~\cite{Scox1_o4a}, and all-sky searches~\cite{O4a_all_sky,O4a_all_sky_binary,McGloughlin_2026,McGloughlin_2026_hi}. 

The typical amplitude of CW signals is uncertain, as it depends on poorly constrained quantities, in particular the star's ellipticity, which characterizes the degree of non-axisymmetry.
For a non-accreting, non-axisymmetric NS rotating about one of its principal axes of inertia, the signal amplitude is given by
\begin{equation}
	\label{eq:hexpected}
	h_0 = \frac{4\pi^2G}{c^4} \frac{\epsilon I_{zz}  f^2}{d}  \approx         %         1.06\times
	10^{-26}\left(\frac{\epsilon}{10^{-6}}\right) 
	\times\left(\frac{I_{zz}}{\qty{e38}{\kg.\m^2}}\right)\left(\frac{f}{\qty{100}{\Hz}}\right)^2
	\left(\frac{\qty{1}{\kilo\pc}}{d}\right), 
\end{equation}
where $G$ is Newton's gravitational constant, $c$ is the speed of light, $d$ is the source distance, $f$ is the GW frequency (twice the spin frequency for a NS rotating around one of its principal axes of inertia), $\epsilon = (I_{xx}-I_{yy})/I_{zz}$ is the ellipticity of the star, $I_{zz}$ is the moment of inertia about the rotation axis ($z$-axis), and $I_{xx}$ and $I_{yy}$ are the moments of inertia about the other two principal axes.

The scope of CW searches is broad, extending to more exotic scenarios such as ultralight beyond-Standard-Model particles (see \cref{subsec:dm}). In this section, we focus on two representative classes to illustrate the A+ and \Asharp{} detection prospects: targeted searches for known pulsars and all-sky searches for isolated NSs.
Targeted searches for known pulsars employ fully coherent matched filtering, which achieves optimal sensitivity under the assumption of Gaussian noise, but requires precise knowledge of the source parameters, typically obtained from electromagnetic observations.
In contrast, all-sky searches probe a broad parameter space for sources without electromagnetic counterparts and therefore rely on less sensitive methods to ensure computational feasibility.

For targeted searches, we use the observed pulsars in the ATNF catalog\footnote{Catalogue Version 2.7.0, https://www.atnf.csiro.au/research/pulsar/psrcat/} and summarize the results in \cref{tab:pulsar_eps}. The table lists, for each detector configuration, the number of potentially detectable known pulsars (at \qty{95}{\%} confidence level) under three assumptions for their ellipticity. We assume a fully coherent matched-filter search in the frequency domain~\cite{2010CQGra..27s4016A}, a total observation time of $T_\mathrm{obs}=\qty{3}{years}$ with a single detector, \footnote{For fully coherent analyses, this is equivalent to three detectors of equal sensitivity observing for one year.} and a duty cycle of \qty{80}{\%}. 

\begin{table}[h!]
	\centering
	\setlength{\tabcolsep}{10pt}
	\renewcommand{\arraystretch}{1.2}
	\begin{tabular}{cccc}
		\hline
		Configuration & $n_1~(\frac{\epsilon_\mathrm{min}}{10^{-10}},~\frac{\epsilon_\mathrm{median}}{10^{-5}})$ & $n_2~(\frac{\epsilon_\mathrm{min}}{10^{-10}},~\frac{\epsilon_\mathrm{median}}{10^{-9}})$   & $n_3~(\frac{\epsilon_\mathrm{min}}{10^{-10}},~\frac{\epsilon_\mathrm{median}}{10^{-10}})$ \\ \hline
		A+ & 119 ($13,~6.3$) & 30 ($13,~7.0$) &  0\\
		\Asharp{} baseline & 208 ($6.1,~3.5$) & 65 ($6.1,~5.1$) & 5 ($6.1,~7.3$)\\
		\Asharp{} wideband & 189  ($5.9,~4.5$) & 46 ($5.9,~4.7$) &  3 ($5.9,~7.1$) \\
		\Asharp{} optimistic & 238  ($5.8,~2.1$) & 95 ($5.8,~7.1$) & 8 ($5.8,~8.6$) \\
		\hline  
	\end{tabular}
	\caption{
		Expected number of detectable known pulsars under three assumptions for the ellipticity:
		$\epsilon=\epsilon_\mathrm{sd}$ (case $n_1$),
		$\epsilon=\mathrm{min}(\epsilon_\mathrm{sd},~10^{-6})$ (case $n_2$), and
		$\epsilon=\mathrm{min}(\epsilon_\mathrm{sd},~10^{-9})$ (case $n_3$).
		A total observation time of $T_\mathrm{obs}=\qty{3}{years}$ with a single detector and a duty cycle of 80$\%$ is assumed.
		For each case, the minimum and median ellipticities of the detectable signals are given in parentheses. 
	}
	\label{tab:pulsar_eps}
\end{table}

Column 2 assumes that pulsars emit at their spin-down limit, i.e., that the entire loss of rotational energy is attributed to GW emission, yielding an ellipticity $\epsilon_{\rm sd}$ that depends on the moment of inertia, spin frequency, and frequency derivative.
Column 3 assumes that each pulsar has an ellipticity $\epsilon = \mathrm{min}(\epsilon_\mathrm{sd}, 10^{-6})$, where $10^{-6}$ is of the order of the maximum theoretical ellipticity expected for a NS with a standard equation of state (see, e.g., Ref.~\cite{2022arXiv220903222M}).
Column 4 assumes $\epsilon = \mathrm{min}(\epsilon_\mathrm{sd},~10^{-9})$, a value considered plausible for millisecond pulsars~\cite{2018ApJ...863L..40W}.
These results indicate a substantial improvement in detection prospects relative to the current constraints~\cite{O4a_known_pulsar}, particularly for the \Asharp{} optimistic. Comparable results would be obtained with other matched-filter-based algorithms~\cite{Riles2022}; also see relevant studies in Refs.~\cite{Owen:2025ata,Owen:2025ecd}.

\begin{figure}[htb]
	\centering
	\includegraphics[width=0.7\columnwidth]{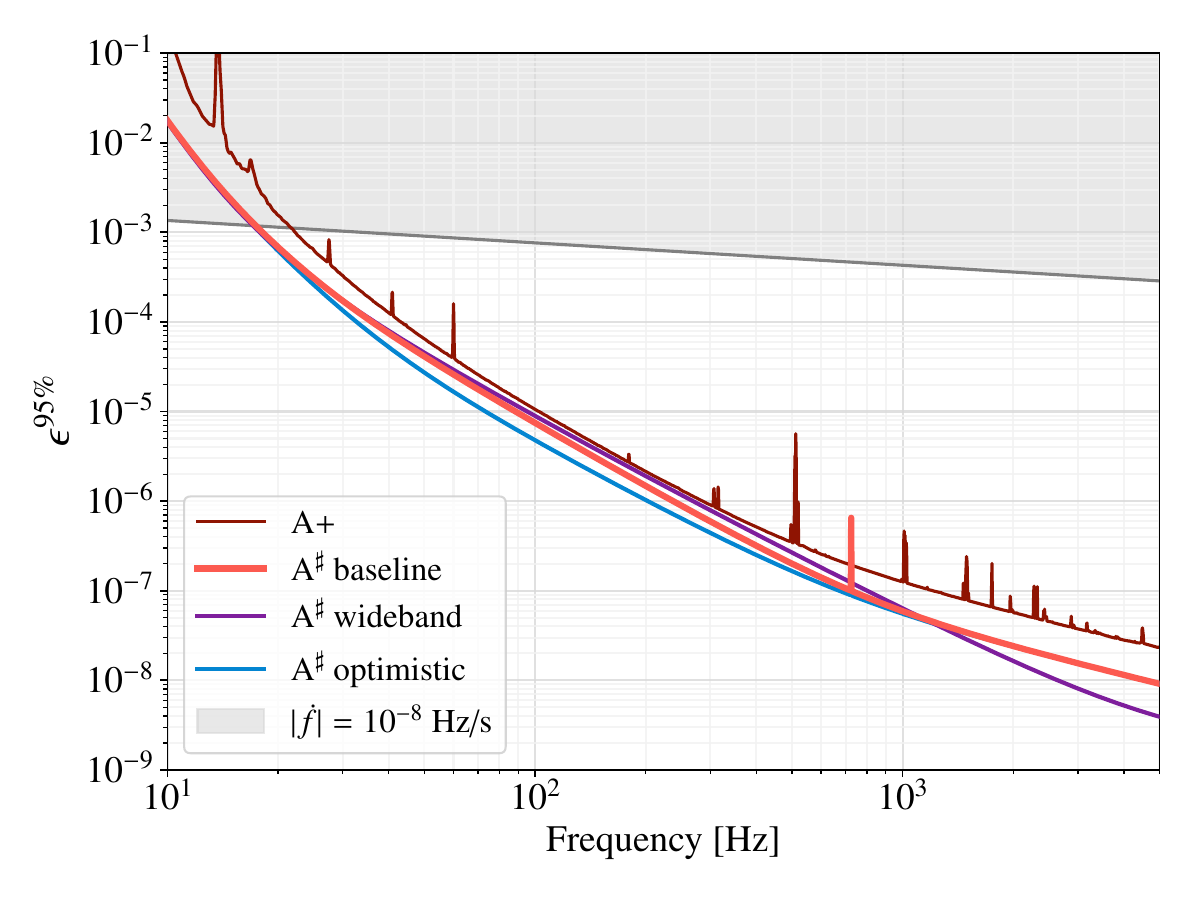}
	\caption[Sensitivity reach to all-sky searches of continuous waves]{
		Sensitivity reach for all-sky CW searches (95\% confidence level).
		The solid curves show the minimum detectable ellipticity as a function of frequency for the different detector configurations, assuming a source distance of \qty{5}{\kilo\pc}, an observation time $T_{\mathrm{obs}}=\qty{1}{year}$ with a single detector, a duty cycle of $\qty{80}{\%}$, and an FFT segment duration $T_{\rm FFT}=\qty{1}{day}$.
		The shaded region indicates spin-down rates $|\dot{f}| > 10^{-8}\,\mathrm{Hz\,s^{-1}}$, which are typically inaccessible to the search due to high computing cost.
	}
	\label{fig:allsky}
\end{figure}

For all-sky searches, \cref{fig:allsky} shows, for the different detector configurations, the minimum detectable ellipticity as a function of signal frequency, assuming that the source's rotational evolution is dominated by GW emission and a fixed source distance of \qty{5}{\kilo\pc}.
These results are obtained using the \textsc{FrequencyHough} pipeline~\cite{2014PhRvD..90d2002A}, a semi-coherent method routinely employed in current CW searches. Assuming Gaussian noise, the 95\% confidence-level sensitivity to the minimum detectable strain amplitude can be approximated as~\cite{peakmap_sensi,O4a_allsky}
\begin{equation}
	h_{\min,95\%}\approx
	\frac{4.71}{N^{1/4}}\sqrt{\frac{S_n(f)}{T_{\rm FFT}}}\sqrt{CR_{\rm thr}+1.6449},
	\label{eq:hul}
\end{equation}
where $N=T_{\rm obs}/T_{\rm FFT}$ is the ratio of the total observation time to the fast Fourier transform (FFT) segment duration (with 50\% overlap between segments, yielding $2N$ incoherently combined data segments), $S_n(f)$ is the one-sided detector noise power spectral density, and $CR_{\rm thr}$ is the threshold on the critical ratio, an SNR-like statistic that is used to select signal candidates. In the figure, we adopt $T_{\rm obs}=\qty{1}{year}$, $T_{\rm FFT}=\qty{1}{day}$,\footnote{The choice $T_{\rm FFT}=\qty{1}{day}$ is much longer than the segment durations currently used in LIGO--Virgo--KAGRA all-sky CW searches with the \textsc{FrequencyHough} pipeline, which are at most a few hours at frequencies of a few tens of hertz. Achieving $T_{\rm FFT}=\qty{1}{day}$ would require substantial methodological and computational developments to avoid a significant increase in computational cost.} $CR_{\rm thr}=4.5$, and assume a single detector with $T_{\mathrm{obs}}=\qty{1}{year}$ and a duty cycle of $\qty{80}{\%}$.%\footnote{Although current searches typically use a shorter data segment duration, values as large as \qty{1}{day} are very reasonable, and likely conservative, for the Post-O5 era, given the algorithmic developments and the increase in available computing power.}

For example, a source located at \qty{5}{\kilo\pc} and emitting a signal at \qty{500}{\Hz} would be detectable for ellipticity $\epsilon \gtrsim 2 \times 10^{-7}$.
At a fixed frequency, the required ellipticity scales linearly with distance, so a nearby source at \qty{500}{\pc} emitting a signal near \qty{500}{\Hz} could be detected with $\epsilon \gtrsim 2 \times 10^{-8}$. 
The required ellipticity also scales linearly with the detector noise amplitude spectral density, such that the \Asharp{} configurations improve the sensitivity to the ellipticity by approximately a factor of two relative to A+ in a wide frequency range.
Although the results presented here are obtained using the \textsc{FrequencyHough} pipeline, other semi-coherent methods are expected to achieve broadly comparable sensitivities; see the review in Ref.~\cite{Riles2022}.

%%%%%%%%%%%%%%%%%%%%%%%%%%%%%%%%%%
%\subsection{Burst searches and accretion disk instabilities}
%\label{subsec:burst}
%%%%%%%%%%%%%%%%%%%%%%%%%%%%%%%%%%
%\input{sections/burst}

%%%%%%%%%%%%%%%%%%%%%%%%%%%%%%%%%%
\subsection{Stochastic background from binary mergers}
\label{subsec:background}
\begin{figure}[h]
    \centering
    \includegraphics[width=0.7\textwidth]{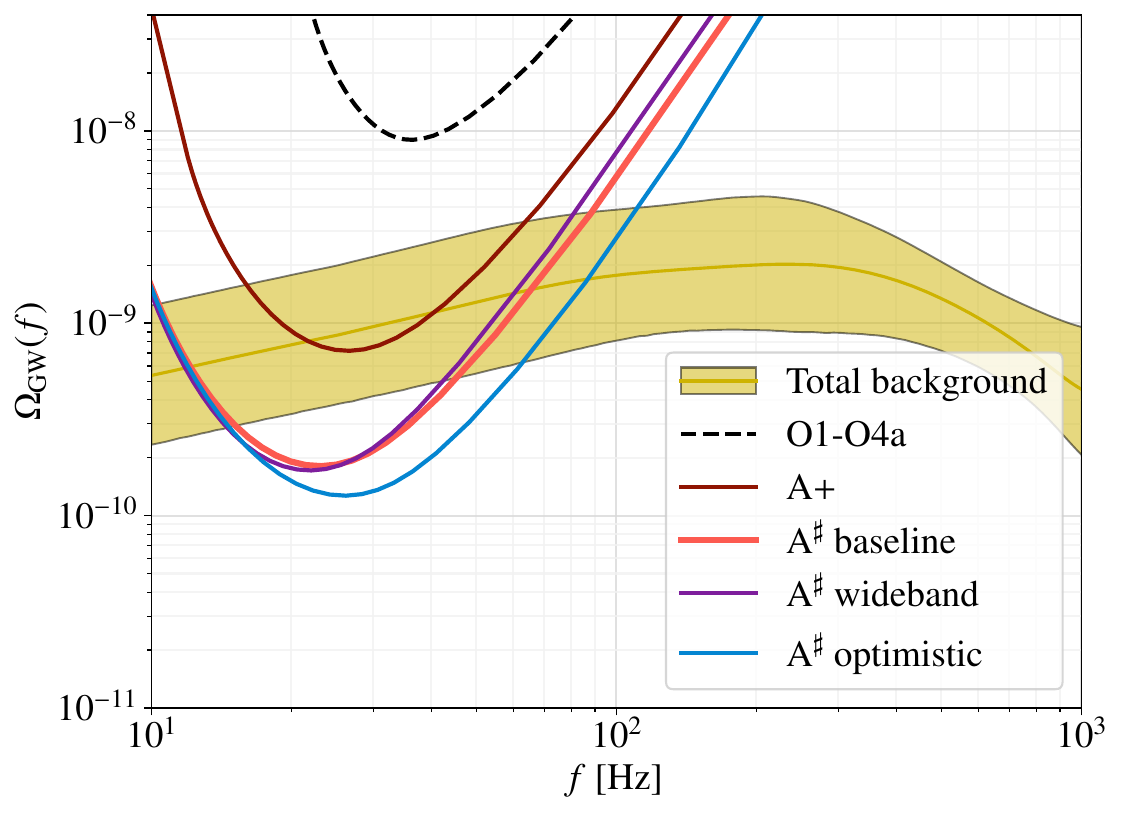}
    \caption[Stochastic Background]
    {Forecast of the astrophysical GW background from compact binary mergers based on observations through O4a. The yellow band shows current estimates for the total dimensionless energy-density spectrum across BNS, NSBH, and BBH sources, compared with the projected sensitivities of detector configurations considered (assuming two LIGO detectors and one year of observation). While A+ is expected to enable detection of this background, all \Asharp{} configurations provide sufficient sensitivity to ensure detection of the astrophysical background.}
    \label{fig:background}
\end{figure}

%%%%%%%%%%%%%%%%%%%%%%%%%%%%%%%%%%
In addition to detecting coherent signals from relatively nearby binaries, the detectors are also expected to be sensitive to the stochastic GW background generated by unresolved compact binary mergers. 
We use the inferred astrophysical merger rates of BNS, NSBH, and BBH systems from GWTC-4 to estimate the corresponding dimensionless energy-density spectra, $\Omega_{\mathrm{GW}}(f)$, radiated by each class (see Refs.~\cite{GWTC-4_pop,GWTC-4_stochastic} for more details). 
\Cref{fig:background} shows current estimates of the total energy-density spectrum from all compact binaries, compared with the projected sensitivities of the detector configurations considered, assuming two LIGO detectors operating with identical configurations and one year of observation.  
Any of the proposed upgrades to the A+ design would provide sufficient sensitivity to ensure detection of the stochastic background from binary mergers, should it remain undetected at the end of O5. 
Such a detection would provide an independent probe of the cosmic merger history and populations of compact binaries, while improved characterization of this astrophysical foreground would benefit searches for weaker stochastic backgrounds of cosmological or other astrophysical origin.

%%%%%%%%%%%%%%%%%%%%%%%%%%%%%%%%%%
\subsection{Beyond-Standard-Model particles and dark matter}
\label{subsec:dm}
%%%%%%%%%%%%%%%%%%%%%%%%%%%%%%%%%%

The CW searches discussed in \cref{subsubsec:cw} are not limited to spinning NSs. 
More broadly, long-duration, narrowband signals in ground-based interferometers also provide a sensitive probe of ultralight beyond-Standard-Model particles and dark matter.  A particularly well-motivated science case is BH superradiance~\cite{Arvanitaki2010,Arvanitaki2011,Brito2020}: if an ultralight bosonic field exists, rapidly spinning BHs can transfer rotational energy to the field and form macroscopic boson clouds when the boson Compton wavelength is comparable to the BH size.  Subsequently, the boson clouds can emit long-duration quasi-monochromatic GWs in the LIGO band, with frequencies set primarily by the boson mass and signal evolution determined by the BH mass, spin, and cloud dynamics.  Searches for these signals therefore connect the GW observations to particle-physics constraints~\cite{O3_all-sky_scalar_bosons,LIGOScientific:2025csr,O2_CygnusX1_scalar_bosons, Collaviti2024}, complementing limits inferred from BH spin measurements and population studies~\cite{Arvanitaki2015, Ng2021, Cardoso2018, Brito2017_2, aswathi2025, Baryakhtar2017, Abac_2025_GW241011_241110,Tsukada2019, Yuan2022,Tsukada2021}. 

\begin{figure}[htb]
	\centering
	\includegraphics[width=\columnwidth]{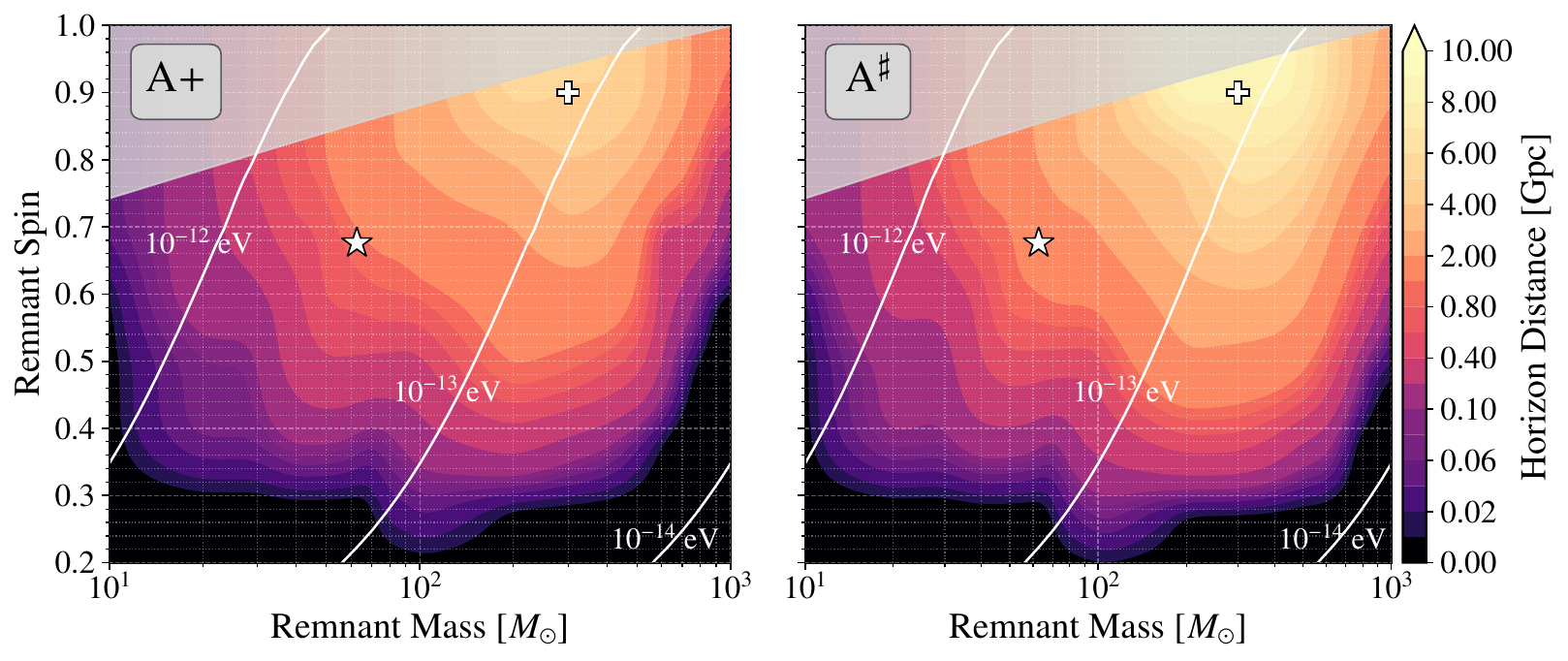}
	\caption[]{Horizon distance (colored contours) as a function of remnant BH mass and spin for a network of two A+ detectors (left) and two \Asharp{} baseline detectors (right). The white contours show the boson masses optimally matched to each BH, for which the superradiant instability is maximized, and therefore approximately indicate the boson-mass range accessible to these detectors through follow-up searches targeting merger remnants. The star and plus markers indicate a GW250114-like system and an optimal high-mass, high-spin system, respectively. 
	The gray region indicates the parameter space in which the signal evolves too rapidly to be tracked with the existing method. }
	\label{fig:sr_vector}
\end{figure}

Follow-up searches targeting the remnant BHs of compact binary mergers are especially interesting, because the remnant age, mass, spin, sky position, and orientation can be constrained from the merger signal itself, enabling a directed search for the putative boson cloud~\cite{Arvanitaki2017,Isi2019}.  For vector boson clouds~\cite{East2017,East2018}, existing observing runs can already reach astrophysically interesting distances for favourable merger remnants~\cite{Jones2023,Jones2025,LIGOScientific:2025csr}. 
The improved sensitivity of \Asharp{} relative to A+ extends this reach further.  As a representative example, for a system like GW250114\_082203 (henceforth
GW250114)~\cite{GW250114:2025oiz,GW250114com:2025wao} with remnant BH mass \qty{63}{\solarmass} and dimensionless spin 0.7, the horizon distance for a vector-boson superradiance search increases from \qty{0.9}{\giga\pc} with A+ to \qty{1.3}{\giga\pc} with \Asharp{}, corresponding to an enhancement of approximately
\qty{45}{\%} in distance and a factor of $\sim$3 in accessible volume. The gain is even larger for more optimal systems: for a remnant BH with mass ${\sim}\qty{300}{\solarmass}$ and spin 0.9, the horizon distance increases from approximately \qty{5.0}{\giga\pc} with A+ to \qty{8.2}{\giga\pc} with \Asharp{}, corresponding to a factor of $\sim$4 increase in accessible volume. 
\Cref{fig:sr_vector} illustrates this improvement across the remnant BH mass-spin plane for vector-boson clouds, with the white contours indicating the corresponding boson masses probed. These improvements substantially boost the discovery potential for GWs from vector
boson clouds and strengthen the role of future ground-based detectors as probes of ultralight particles.

GW interferometers can also act as direct detectors of ultralight dark matter~\cite{Miller_DM_review, Piccinni_review,LVK_DM_O4a}.  In this case the signal does not arise from an astrophysical GW, but from dark-matter fields coupling to components of the instrument, such as the mirrors, beam splitter, or laser light, producing an oscillatory differential strain.  Examples include scalar fields that induce time-dependent variations of fundamental constants~\cite{Vermeulen2021, Gottel2024, Aiello2022}, vector dark matter that exerts oscillatory forces on the test masses~\cite{direct_dark_matter_LIGO_Virgo, direct_dark_matter_KAGRA}, and axion-like fields that modify the propagation of polarized light~\cite{Nagano2019,Nagano2021}. 
Although the physical origin differs from that of boson clouds around BHs, both classes of signals are expected to be long-lived and narrowband, with finite coherence times, and can therefore be searched for using analysis techniques closely related to those developed for CW searches.  The broadband sensitivity improvement of \Asharp{} relative to A+ extends the reach of LIGO to a wider region of ultralight-particle parameter space and strengthens the role of ground-based GW detectors as probes of dark-sector physics.

\begin{figure}[h]
	\centering
	\includegraphics[width=0.7\textwidth]{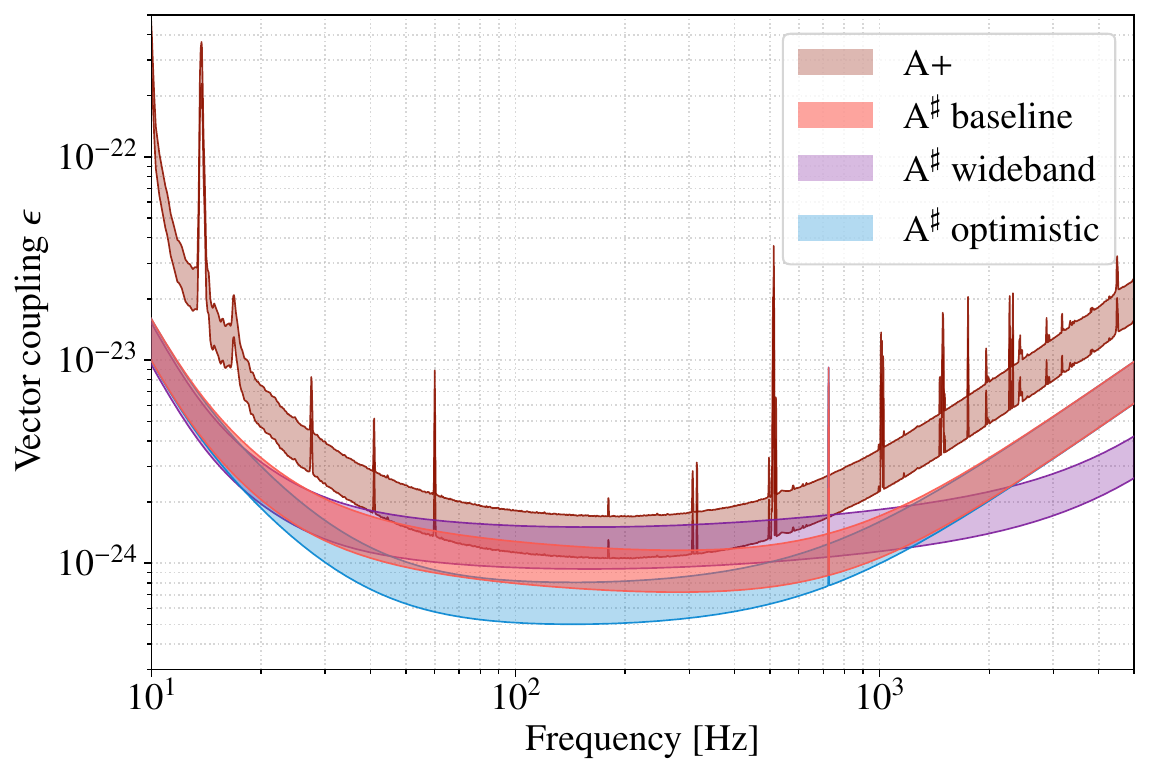}
	\includegraphics[width=0.7\textwidth]{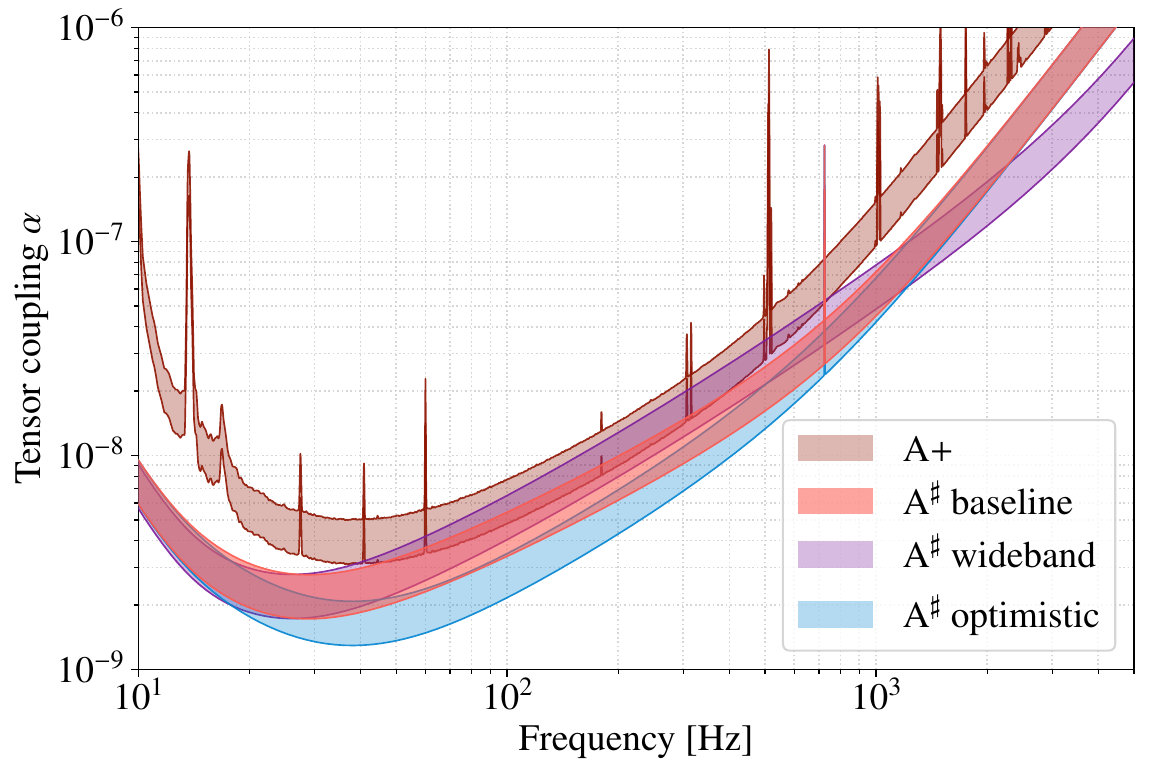}
	\caption[]{Projected sensitivity to ultralight dark matter for the detector configurations considered in this work, assuming one-year observation with two detectors and a duty cycle of 80\%. The top and bottom panels show the projected constraints on vector and tensor dark matter, respectively. Each color corresponds to a different detector configuration. For each colored band, the upper and lower bounds correspond to the BSD-excess-power and BSD-cross-correlation methods, respectively.}
	\label{fig:dm}
\end{figure}

Here we focus on direct searches for spin-1 and spin-2 ultralight dark matter as illustrative examples~\cite{Manita:2023mnc,Armaleo:2020efr}.
Spin-1 dark matter can couple to either baryon number or baryon-minus-lepton number, producing oscillatory forces on the test masses, while spin-2 dark matter couples to the Standard Model energy-momentum tensor and appears, at linear order, as an effective tensor strain.  We forecast the \Asharp{} sensitivity to these signals using two complementary semi-coherent approaches.  Excess-power searches adapt the coherent integration time to the expected dark-matter coherence time in each frequency band and identify statistically significant time-frequency excesses.
Cross-correlation searches instead exploit the strong inter-detector correlation expected for ultralight dark matter, while suppressing uncorrelated instrumental artifacts.  Assuming one year of observing time and a duty cycle of 80\% with two LIGO detectors, the projected strain sensitivities from the BSD-excess-power and BSD-cross-correlation~\cite{DM_sid_Delgado,LSC_sft_2022} implementations can be converted into expected constraints on the spin-1 coupling
$\epsilon_D$ and the spin-2 coupling $\alpha$.
The colored bands in \cref{fig:dm} indicate the estimated sensitivity ranges spanned by the two methods for each detector configuration, with the upper and lower bounds corresponding to the BSD-excess-power and BSD-cross-correlation projections, respectively~\cite{DM_sid_Delgado, peakmap_sensi,LVK_DM_O4a}.

\section{Discussion}
\label{sec:discussion}

The \Asharp{} concept provides a path for substantially improving the scientific reach of the LIGO observatories beyond O5 while remaining within the existing 4-km facilities.
%The design is deliberately broadband: rather than optimizing for a single source class or frequency band, it combines low-frequency improvements from heavier test masses, upgraded suspensions, and improved seismic isolation, with mid- and high-frequency gains from lower coating thermal noise, higher arm power, and enhanced frequency-dependent squeezing. 
The design is deliberately broadband: rather than being tuned for a single source class or frequency band, it combines low-frequency improvements from heavier test masses, upgraded suspensions, and improved seismic isolation, with mid- and high-frequency gains from lower coating thermal noise, higher arm power, and improved frequency-dependent squeezing performance.
This balanced approach enables a broad science program, including increased compact-binary detection rates, improved population inference, stronger access to the properties of astrophysical sources, and enhanced sensitivity to stochastic backgrounds, bursts, continuous waves, and new physics.

Realizing this performance will require addressing several key technical challenges. At low frequencies, the target sensitivity depends not only on reducing fundamental suspension and Newtonian noise, but also on suppressing technical noise associated with seismic motion, suspension damping, and global angular control.
%Achieving the required performance will therefore require the successful development of a new test mass suspension design to support heavier test masses, integration of local interferometric sensors, robust suspension control for heavier optics, improved rotation sensing for seismic isolation and the implementation of Newtonian-noise mitigation.
Achieving the required performance will require the successful realization of several developments, including a new suspension design to support heavier test masses, the integration of local interferometric sensors, robust suspension control systems, improved rotation sensing for seismic isolation, and the implementation of Newtonian-noise mitigation.
At mid-frequencies, improved coating thermal noise remains a major challange, making continued progress in low-loss optical coatings essential.
%The baseline \Asharp{} design assumes achievable improvements in amorphous coatings, while more ambitious options such as crystalline coatings could provide additional scientific gains if their large-scale fabrication, optical absorption and birefringence challenges can be resolved.
The baseline \Asharp{} design assumes continued improvements in amorphous coating performance, while crystalline coatings could provide further scientific gains provided that challenges associated with large-scale fabrication, controlling optical absorption, and mitigating birefringence are successfully addressed.
%At high frequencies, operation with higher circulating power and stronger squeezing will require improved control of thermal distortions, optical losses, mode matching, angular and parametric instabilities, and squeezed-light degradation.
At high frequencies, achieving stable operation at higher circulating optical power and with improved squeezing performance will require further reductions in optical loss, improved mode matching, better control of thermal distortions, and robust suppression of angular and parametric instabilities.
Research and development is underway across these areas to mature the required technologies and assess their readiness for implementation in the LIGO facilities.

The configurations considered in this work illustrate the trade-offs that will shape the final detector design. The baseline \Asharp{} configuration provides the largest overall broadband improvement and is therefore well matched to a wide range of astrophysical goals. A wideband configuration can enhance sensitivity at high frequencies, improving access to some BNS post-merger signals and kHz science, but at the cost of reduced mid-frequency sensitivity and hence lower compact-binary range. A lower-coating-noise configuration provides significant gains for many compact-binary measurements, especially those limited by mid-band sensitivity, but depends on coating technologies that require further development. 

Beyond its immediate scientific return, \Asharp{} serves an important strategic role in the global GW roadmap. Many of its core technologies, including heavier test masses, improved suspensions, low-noise coatings, higher optical power, improved squeezed-light injection, and advanced seismic and Newtonian-noise mitigation, are directly relevant to next-generation observatories such as Cosmic Explorer~\cite{evans2021CEHS,evans2023} and the Einstein Telescope~\cite{Maggiore2020,ET_Abac_2026}. Implementing these technologies in the existing LIGO facilities would therefore provide a critical opportunity to mature hardware, control schemes, commissioning strategies, and data-analysis methods before they are required at larger scale in future detectors, while also sustaining the specialized expertise essential for commissioning and operating next-generation observatories. In this sense, \Asharp{} is both a major scientific upgrade for the 2030s and a technology pathfinder toward next-generation GW observatories.

However, the \Asharp{} design remains constrained by the existing 4-km LIGO infrastructure and therefore cannot deliver the same cosmological volume or broad transformational science capability expected from new observatories such as Cosmic Explorer and the Einstein Telescope. 
Rather, \Asharp{} occupies a natural place between A+ and the next generation of GW detectors: it would provide an important intermediate step by delivering substantial science gains within the existing infrastructure before new facilities come online, while maturing technologies, commissioning experience, and data-analysis methods needed for the observatories that follow. By combining near-term feasibility with broad scientific reach, \Asharp{} provides a compelling route for advancing GW astronomy in the post-A+ era.

\section{Acknowledgments}
% LIGO
This material is based upon work supported by NSF's LIGO Laboratory, which is a major facility fully funded by the National Science Foundation.
LIGO was constructed by the California Institute of Technology and Massachusetts Institute of Technology with funding from the National Science Foundation, and operates under Cooperative Agreement PHY--1764464. Advanced LIGO was built under grant No. PHY--0823459.
% Computing
The authors are grateful for computational resources provided by the LIGO Laboratory and supported by NSF Grants PHY--0757058, PHY--0823459, 
and by the University of Birmingham's BlueBEAR HPC service, which provides a High Performance Computing service to the University of Birmingham's research community.
% NSF (USA) -- all award numbers listed together
The authors also acknowledge support from the NSF under Grants PHY--2146528, PHY--2309064, and PHY--2309267.
%, PHY--2110348, PHY--2409496, PHY--1912380, PHY--1912514, PHY--2011719, PHY--2309290, PHY--2409603, PHY--2309293, PHY--2409601, PHY--2320711, PHY--2309292, PHY--2208079, and PHY--1920023, and under Award Numbers 2207858, 2207998, and 2513439.
% USA -- other
%Research at the Massachusetts Institute of Technology was supported in part by the MIT Undergraduate Research Opportunities Program (UROP). The authors also acknowledge the Gordon and Betty Moore Foundation for support through the Center for Coatings Research. Researchers at California State University, Fullerton acknowledge support from Nicholas and Lee Begovich and from Dan Black and Family.
% ARC (Australia)
This research is supported by the Australian Research Council Centre of Excellence for Gravitational Wave Discovery (OzGrav), Project Number CE230100016, Linkage Infrastructure, Equipment and Facilities, Project Numbers LE210100002 and LE260100008, and Discovery Early Career Researcher Award, Project Number DE240100206.
% UK
The authors also acknowledge the Science and Technology Facilities Council (STFC) for support through grants ST/V005618/1, ST/Y004272/1, ST/V005677/1, ST/Y00423X/1; %ST/Y004256/1, ST/Y004213/1, and ST/V005634/1, and UKRI grant UKRI2487; 
the UK Space Agency through grant ST/Y004922/1; and the Royal Society through University Research Fellowship URF\textbackslash R1\textbackslash 221500 and grant RF\textbackslash ERE\textbackslash 221015.
% Germany
%Researchers in Hannover and Hamburg acknowledge support from the Deutsche Forschungsgemeinschaft (DFG, German Research Foundation) under Germany's Excellence Strategy -- EXC-2123 QuantumFrontiers (Project 390837967) and EXC 2121 ``Quantum Universe'' (Project 390833306).
% Spain
The authors also acknowledge the support of the Spanish Ministerio de Ciencia, Innovacion y Universidades Ramon y Cajal, RYC2023-044489-I funded by MCIN/AEI/10.13039 /501100011033 and the FSE+ and cofinanced by the Universitat de les Illes Balears (UIB). This work was supported by UIB with funds from the Programa de Foment de la Recerca i la Innovació de la UIB 2024-2026 (supported by the yearly plan of the Tourist Stay Tax ITS2023-086); the Spanish Agencia Estatal de Investigación grants RED2024-153978-E, RED2024-153735-E, PID2025-170644NA-I00, funded by MICIU/AEI/10.13039/501100011033 and the ERDF/EU; and the Comunitat Autònoma de les Illes Balears through the Conselleria d'Educació i Universitats with funds from the ERDF (SINCO2022/18146).
% Italy
Researchers in Italy acknowledge support from the Italian National Institute for Nuclear Physics (INFN, Istituto Nazionale di Fisica Nucleare), through Commissione Scientifica Nazionale II (CSN2), under the VIRGO-Italia and ET-Italia projects.
% Canada
%Research performed at U. Montréal and U. Sherbrooke is supported by the NSERC, the CFI, and the FRQNT through the RQMP.
% Korea
%Research performed at SKKU is supported by the National Research Foundation of Korea (NRF) grant funded by the Korean government (MSIT) (No. RS-2024-00455482).
% Japan
The authors acknowledge the support of JST ASPIRE JPMJAP2320.
%
% LVC acknowledgments from P2000488_v30
%
The authors also gratefully acknowledge the support of
the Science and Technology Facilities Council (STFC) of the
United Kingdom, the Max-Planck-Society (MPS), and the State of
Niedersachsen/Germany for support of the construction of Advanced LIGO 
and construction and operation of the GEO\,600 detector. 
Additional support for Advanced LIGO was provided by the Australian Research Council,
%The authors gratefully acknowledge the Italian Istituto Nazionale di Fisica Nucleare (INFN),  
the French Centre National de la Recherche Scientifique (CNRS) and
the Netherlands Organization for Scientific Research (NWO).
%for the construction and operation of the Virgo detector
%and the creation and support  of the EGO consortium. 
The authors also gratefully acknowledge research support
%from these agencies as well as 
by 
the Council of Scientific and Industrial Research of India, 
the Department of Science and Technology, India,
the Science \& Engineering Research Board (SERB), India,
the Ministry of Human Resource Development, India,
the Spanish Agencia Estatal de Investigaci\'on (AEI),
the Spanish Ministerio de Ciencia, Innovaci\'on y Universidades,
%the European Union NextGenerationEU/PRTR (PRTR-C17.I1),
%the ICSC - CentroNazionale di Ricerca in High Performance Computing, Big Data
%and Quantum Computing, funded by the European Union NextGenerationEU,
%the Comunitat Auton\`oma de les Illes Balears through the Conselleria d'Educaci\'o i Universitats,
%the Conselleria d'Innovaci\'o, Universitats, Ci\`encia i Societat Digital de la Generalitat Valenciana and
the CERCA Programme Generalitat de Catalunya, Spain,
%the Polish National Agency for Academic Exchange,
%the National Science Centre of Poland and the European Union - European Regional
%Development Fund;
%the Foundation for Polish Science (FNP),
%the Polish Ministry of Science and Higher Education,
the Swiss National Science Foundation (SNSF),
the Deutsche Forschungsgemeinschaft (DFG, German Research Foundation),
the Russian Science Foundation,
%the European Commission,
%the European Social Funds (ESF),
%the European Regional Development Funds (ERDF),
the Royal Society, 
the Scottish Funding Council, 
the Scottish Universities Physics Alliance, 
the Hungarian Scientific Research Fund (OTKA),
%the French Lyon Institute of Origins (LIO),
%the Belgian Fonds de la Recherche Scientifique (FRS-FNRS), 
%Actions de Recherche Concert\'ees (ARC) and
%Fonds Wetenschappelijk Onderzoek - Vlaanderen (FWO), Belgium,
%the Paris \^{I}le-de-France Region, 
the National Research, Development and Innovation Office of Hungary (NKFIH), 
the National Research Foundation of Korea,
the Natural Sciences and Engineering Research Council of Canada (NSERC),
the Canadian Foundation for Innovation (CFI), the FRQNT through the RQMP, 
the Brazilian Ministry of Science, Technology, and Innovations,
the International Center for Theoretical Physics South American Institute for Fundamental Research (ICTP-SAIFR), 
the Research Grants Council of Hong Kong,
the National Natural Science Foundation of China (NSFC),
the Israel Science Foundation (ISF),
the US-Israel Binational Science Fund (BSF),
the Leverhulme Trust, 
the Research Corporation,
the National Science and Technology Council (NSTC), Taiwan,
the United States Department of Energy,
and
the Kavli Foundation.
%The authors gratefully acknowledge the support of the NSF, STFC, INFN and CNRS for provision of computational resources.

\appendix
\addtocontents{toc}{\fixappendix}
% iopart sets \thesection = "Appendix A" in appendix mode, which causes
% cleveref to produce "appendix Appendix A".
% So store only the letter in
% the label, and add "Appendix " manually to the heading and TOC display.
\makeatletter
\renewcommand\thesection{\Alph{section}}
\def\@sect#1#2#3#4#5#6[#7]#8{%
  \ifnum #2>\c@secnumdepth
    \let\@svsec\@empty
  \else
    \refstepcounter{#1}%
    \edef\@svsec{Appendix \csname the#1\endcsname. }%
  \fi
  \@tempskipa #5\relax
  \ifdim \@tempskipa>\z@
    \begingroup #6\relax
      \noindent{\hskip #3\relax\@svsec}{\interlinepenalty \@M #8\par}%
    \endgroup
    \csname #1mark\endcsname{#7}%
    \addcontentsline{toc}{#1}{\ifnum #2>\c@secnumdepth \else
      \protect\numberline{Appendix~\csname the#1\endcsname}\fi #7}%
  \else
    \def\@svsechd{#6\hskip #3\relax
      \@svsec #8\csname #1mark\endcsname
      {#7}\addcontentsline{toc}{#1}{\ifnum #2>\c@secnumdepth \else
        \protect\numberline{Appendix~\csname the#1\endcsname}\fi #7}}%
  \fi
  \@xsect{#5}%
}
\makeatother

%\section{Detector parameters}
%\input{sections/detector_parameters}

%\section{Noise budgets}
%\input{sections/noise_budgets}

\medskip

\section*{References}
\bibliographystyle{iopart-num}
\bibliography{references}

\end{document}